%% file: PR_spin.tex
\pdfoutput=1
\newcommand*{\ATLASLATEXPATH}{latex/}
\documentclass[atlasdraft=false, texlive=2016, UKenglish, cernpreprint]{\ATLASLATEXPATH atlasdoc}
\pdfoutput=1

\usepackage{\ATLASLATEXPATH atlaspackage}
\usepackage{\ATLASLATEXPATH atlasbiblatex}

\usepackage{\ATLASLATEXPATH atlasphysics}

\graphicspath{{logos/}{figures/}}

\usepackage{PR_spin-defs}

\input{PR_spin-metadata}

\hypersetup{pdftitle={ATLAS document},pdfauthor={The ATLAS Collaboration}}

\begin{document}

\maketitle

\tableofcontents

\input{sections/introduction}
\input{sections/atlasDetector}
\input{sections/dataMc}
\input{sections/eventSelectionAndReconstruction}
\input{sections/unfoldingProcedure}
\input{sections/systematicsUncertainties}
\input{sections/results_diff}
\input{sections/results_spin}
\input{sections/results_susy}
\input{sections/conclusion}
\section*{Acknowledgements}

\input{acknowledgements/Acknowledgements}


\clearpage


\printbibliography
\clearpage
\input{atlas_authlist}



\end{document}

%% file: PR_spin-metadata.tex
\AtlasTitle{Measurements of top-quark pair spin correlations in the $e\mu$
channel at $\sqrt{s} = 13$ TeV using $pp$ collisions in the ATLAS detector}

\AtlasVersion{3.5}

\AtlasAbstract{%
  A measurement of observables sensitive to spin correlations in \ttbar production is presented, using 36.1 fb$^{-1}$ of $pp$ collision data at $\sqrt{s} = 13$ TeV recorded with the ATLAS detector at the Large Hadron Collider.  Differential cross-sections are measured in events with exactly one electron and one muon with opposite-sign electric charge as a function of the azimuthal opening angle and the absolute difference in pseudorapidity between the electron and muon candidates in the laboratory frame. The azimuthal opening angle is also measured as a function of the invariant mass of the \ttbar system. The measured differential cross-sections are compared to predictions by several NLO Monte Carlo generators and fixed-order calculations. The observed degree of spin correlation is somewhat higher than predicted by the generators used. The data are consistent with the prediction of one of the fixed-order calculations at NLO, but agree less well with higher-order predictions. Using these leptonic observables, a search is performed for pair production of supersymmetric top squarks decaying into Standard Model top quarks and light neutralinos. Top squark masses between 170 and 230~\GeV\ are largely excluded at the 95\% confidence level for kinematically allowed values of the neutralino mass.}

\AtlasRefCode{TOPQ-2016-10}

\PreprintIdNumber{CERN-EP-2019-034}

\AtlasJournal{EPJC}
\AtlasJournalRef{\EPJC 80 (2020) 754}
\AtlasDOI{10.1140/epjc/s10052-020-8181-6}

\AtlasCoverEgroupEditors{atlas-TOPQ-2016-10-editors@cern.ch}

\AtlasCoverEgroupEdBoard{atlas-TOPQ-2016-10-editorial-board@cern.ch}

%% file: sections/introduction.tex
\clearpage
\section{Introduction} \label{sec:intro}

The lifetime of the top quark is shorter than the timescale for hadronisation (${\sim}10^{-23}$~s) and is much shorter than the spin decorrelation time (${\sim}10^{-21}$~s)~\cite{PhysRevD.98.030001}. As a result, the spin information of the top quark is transferred directly to its decay products. Top quark pair production (\ttbar) in QCD is parity invariant and hence the top quarks are not expected to be polarised in the Standard Model (SM); however, the spins of the top and the anti-top quarks are predicted to be correlated. This correlation has been observed experimentally by the ATLAS and CMS collaborations in proton--proton collision data at the Large Hadron Collider (LHC) at centre-of-mass energies of $\sqrt{s}=7$~\TeV\ \cite{TOPQ-2011-11,TOPQ-2013-01,TOPQ-2013-10,CMS-TOP-13-003} and $\sqrt{s}=8$~\TeV\ \cite{TOPQ-2014-07,CMS-TOP-13-015,CMS-TOP-14-023,TOPQ-2015-13}. It has been also studied in proton--antiproton collisions at the Tevatron collider \cite{Abazov:2011ka,Abazov:2011qu,Abazov:2011gi,Abazov:2015psg,Aaltonen:2010nz}. This paper presents measurements of spin correlation at a centre-of-mass energy of $\sqrt{s}=13$~\TeV\ in proton--proton collisions using the ATLAS detector and data collected in 2015 and 2016.

Due to the unstable nature of top quarks, their spin information is accessed through their decay products. However, not all decay particles carry the spin information to the same degree, with charged leptons arising from leptonically decaying $W$ bosons carrying almost the full spin information of the parent top quark~\cite{Bernreuther:2001rq,Jezabek:1988ja,Czarnecki:1990pe,Parke:2012jb}. This feature, along with the fact that charged leptons are readily identified and reconstructed by collider experiments, means that observables to study spin correlation in \ttbar events are often based on the angular distributions of the charged leptons in events where both $W$ bosons decay leptonically (referred to as the dilepton channel). The simplest observable is the absolute azimuthal opening angle between the two charged leptons~\cite{Mahlon:2010gw}, measured in the laboratory frame in the plane transverse to the beam line. This opening angle is denoted by $\Delta\phi$. Non-vanishing spin correlation was observed by the ATLAS experiment using the $\Delta\phi$ observable and $\sqrt{s}=7$~\TeV\ data~\cite{TOPQ-2011-11}. Since that time, spin correlation in \ttbar pairs has been extensively studied by both ATLAS and CMS using many observables and techniques. Spin correlation measurements have also been used to search for physics beyond the Standard Model (BSM) either directly, by searching for decreases in the expected SM spin correlation induced by scalar supersymmetric top squarks (stops)~\cite{TOPQ-2014-07}, or indirectly by setting limits on effective field theory operators, such as the chromo-magnetic and chromo-electric dipole operators~\cite{CMS-TOP-14-023}. 
Previous measurements by ATLAS~\cite{TOPQ-2011-11,TOPQ-2013-01,TOPQ-2014-07} and CMS~\cite{CMS-TOP-13-003,CMS-TOP-14-023} using $\Delta\phi$ show slightly stronger spin correlation than expected in the SM, but with experimental uncertainties large enough that the results are still consistent with the SM expectation. In this paper, improved Monte Carlo (MC) generators are employed relative to previous spin correlation results from ATLAS to better control the systematic uncertainties. The spin correlation is measured as a function of the invariant mass of the \ttbar system, as well as inclusively.

Charged-lepton observables can be used to search for the production of supersymmetric top squarks with masses close to that of the SM top quark. Such a scenario is difficult to constrain with conventional searches; however, observables such as $\Delta\phi$ and the absolute difference between the pseudorapidities of the two charged leptons, $\Delta\eta$, are highly sensitive in this regard. The $\Delta\phi$ distribution was previously used in such a search by ATLAS~\cite{TOPQ-2014-07} and this new paper also includes $\Delta\eta$ for this purpose. Although this observable is only mildly sensitive to the SM spin correlation, it is sensitive to different supersymmetry (SUSY) hypotheses; the two observables are therefore used together in this paper to set limits on SUSY top squark production.

This paper is organised as follows. The ATLAS detector is described in Section 2. Section 3 describes the data and Monte Carlo (MC) used in the analysis and Section 4 describes the object definitions and event selection requirements. The unfolding procedure is described in Section 5 and the systematic uncertainties that are considered are described in Section 6. The differential cross-section results are presented in Section 7, the spin correlation extraction is described in Section 8, and the SUSY limits are presented in Section 9. Finally, the conclusions of the paper are summarised in Section 10.

\FloatBarrier

%% file: sections/atlasDetector.tex
\section{ATLAS detector}
\label{sec:detector}

The ATLAS detector~\cite{PERF-2007-01} at the LHC covers nearly the entire solid angle\footnote{ATLAS uses a right-handed coordinate system with its origin at the nominal interaction point (IP) in the centre of the detector and the $z$-axis along the beam pipe. The $x$-axis points from the IP to the centre of the LHC ring, and the $y$-axis points upwards. Cylindrical coordinates $(r, \phi)$ are used in the transverse plane, $\phi$ being the azimuthal angle around the $z$-axis. The pseudorapidity is defined in terms of the polar angle $\theta$ as $\eta = -\ln \tan(\theta / 2)$. Angular distance is measured in units of $\Delta R \equiv \sqrt{(\Delta\eta)^{2} + (\Delta\phi)^{2}}$.} around the interaction point. It consists of an inner tracking detector surrounded by a thin superconducting solenoid, electromagnetic and hadronic calorimeters, and a muon spectrometer incorporating three large superconducting toroidal magnet systems. The inner-detector system is immersed in a \SI{2}{\tesla} axial magnetic field and provides charged-particle tracking in the range $|\eta| < 2.5$.

The high-granularity silicon pixel detector surrounds the collision region and provides four measurements per track. The innermost layer, known as the insertable B-Layer~\cite{PIX-2018-001,ATLAS-TDR-2010-19}, was added in 2014 and provides high-resolution hits at small radius to improve the tracking performance. The pixel detector is followed by the silicon microstrip tracker, which provides four three-dimensional measurement points per track. These silicon detectors are complemented by the transition radiation tracker, which enables radially extended track reconstruction up to $|\eta| = 2.0$. The transition radiation tracker also provides electron identification information based on the number of hits (typically 30 in total) passing a higher charge threshold indicative of transition radiation.

The calorimeter system covers the pseudorapidity range $|\eta| < 4.9$. Within the region $|\eta|< 3.2$, electromagnetic calorimetry is provided by barrel and endcap high-granularity lead/liquid-argon (LAr) sampling calorimeters, with an additional thin LAr presampler covering $|\eta| < 1.8$ to correct for energy loss in material upstream of the calorimeters. Hadronic calorimetry is provided by the steel/scintillator-tile calorimeter, segmented into three barrel structures within $|\eta| < 1.7$, and two copper/LAr hadronic endcap calorimeters that cover $1.5 < |\eta| < 3.2$. The solid angle coverage is completed with forward copper/LAr and tungsten/LAr calorimeter modules optimised for electromagnetic and hadronic measurements respectively, in the region $3.1 < |\eta| < 4.9$.

The muon spectrometer comprises separate trigger and high-precision tracking chambers measuring the deflection of muons in a magnetic field generated by superconducting air-core toroids. The precision chamber system covers the region $|\eta| < 2.7$ with three layers of monitored drift tubes, complemented by cathode strip chambers in the forward region, where the background is highest. The muon trigger system covers the range $|\eta| < 2.4$ with resistive-plate chambers in the barrel, and thin-gap chambers in the endcap regions.

A two-level trigger system is used to select interesting events~\cite{TRIG-2016-01}. The level-1 trigger is hardware-based and uses a subset of detector information to reduce the event rate to a design value of at most \SI{100}{\kHz}. This is followed by the software-based high-level trigger, which reduces the event rate to around \SI{1}{\kHz}.

\FloatBarrier

%% file: sections/dataMc.tex
\section{Data and Monte Carlo simulation}
\label{sec:data}

The $pp$ collision data used in this analysis were collected during 2015 and 2016 by the ATLAS experiment at a centre-of-mass energy of $\sqrt{s}=13$~TeV and correspond to an integrated luminosity of 36.1 fb$^{-1}$. The data considered in this analysis were recorded under stable beam conditions and required all sub-detectors to be operational. Each selected event included additional interactions from, on average, 24 inelastic $pp$ collisions in the same proton bunch crossing, as well as residual detector signals from previous and subsequent bunch crossings, collectively referred to as ``pile-up''. Events were required to pass either a single-electron or single-muon trigger.
Multiple triggers were used to select events: the lowest-threshold triggers utilised isolation requirements to reduce the trigger rate, and had transverse momentum (\pt) thresholds of 24~GeV for electrons and 20~GeV for muons in 2015 data, or 26~GeV for both lepton types in 2016 data. These triggers were complemented by others with higher \pt thresholds and no isolation requirements to increase event acceptance.

MC simulations were used to model background processes and to correct the data for detector acceptance and resolution effects. The ATLAS detector was simulated~\cite{SOFT-2010-01} using \GEANT4~\cite{Agostinelli:2002hh}. A faster detector simulation~\cite{SOFT-2010-01}, utilising parameterised showers in the calorimeter, but with full simulation of the inner detector and muon spectrometer, was used in the samples generated to estimate certain \ttbar modelling uncertainties. Additional $pp$ interactions were generated with \PYTHIA\!8 (v8.186)~\cite{pythia8} and overlaid onto signal and background processes in order to simulate the effect of pile-up. The simulated events were weighted to match the distribution of the average number of interactions per bunch crossing that are observed in data.  The same reconstruction algorithms and analysis procedures were applied to both data and MC events. Corrections derived from dedicated data samples were applied to the MC simulation to improve agreement with data.

The primary \ttbar sample used in this result (hereafter referred to as \emph{nominal}) was simulated using the next-to-leading order (NLO) \POWHEG-Box~(v2) matrix-element (ME) event generator \cite{powheg1,powheg2,powheg3} interfaced to \PYTHIA\!8 (v8.210) for the parton shower (PS) and fragmentation. The NNPDF3.0 NLO parton distribution function (PDF) set~\cite{NNPDF} was used in the matrix element (ME) generation and the NNPDF2.3 PDF set was used in the PS. Non-perturbative QCD effects were modelled using a set of tuned parameters called the A14 tune~\cite{ATL-PHYS-PUB-2014-021}. The ``$h_{\text{damp}}$'' parameter, which controls the \pt of the first additional gluon emission beyond the Born configuration, was set to 1.5 times the mass of the top quark ($m_t$) of 172.5 GeV. The main effect of this was to regulate the high-\pt emission against which the \ttbar system recoils. The choice of this $h_{\text{damp}}$ value was found to improve the modelling of the \ttbar system kinematics in previous analyses~\cite{ATL-PHYS-PUB-2016-004}. The renormalisation and factorisation scales were set to $\mu_{\textrm{F}} = \mu_{\textrm{R}} = \sqrt{(m_t^2 + p_{\textrm{T}}(t)^2)}$, where the $p_{\textrm{T}}$ of the top quark is evaluated before radiation. The \ttbar contribution was normalised using the predicted cross-section, $\sigma_{t\bar{t}} = 832^{+20}_{-29}\left(\mathrm{scale}\right)\pm 35\left(\mathrm{PDF}\right)^{+23}_{-22}\left(\mathrm{mass}\right)$~{pb} as calculated with the Top++2.0 program at next-to-next-to-leading (NNLO) order in perturbative QCD, including soft-gluon resummation to next-to-next-to-leading-log order~\cite{Czakon:2013goa} and assuming a top quark mass of $172.5 \pm 1.0$~GeV. The top quark mass was set to 172.5~GeV in all simulated top quark samples. An alternative \ttbar sample was simulated with the same settings but with the top quarks decayed using \MADSPIN~\cite{Artoisenet:2012st} and with spin correlations between the $t$ and $\bar{t}$ disabled. This sample was used, along with the nominal sample, as a template in the extraction of spin correlation, described in Section~\ref{sec:results_spin}. A further \POWHEG+\PYTHIA\!8 sample was generated with the spin correlations enabled in \MADSPIN, to allow a comparison of the simulation of \POWHEG+\PYTHIA\!8 with and without the use of \MADSPIN. In order to facilitate comparisons to predictions from fixed-order calculations or from other MC generators, the primary spin correlation coefficients as measured in the nominal \POWHEG-Box sample, using the formalism described in Ref.~\cite{Bernreuther:2015yna}, are: $C(k,k)= 0.314 \pm 0.002$, C$(n,n) = 0.320 \pm 0.002$, C$(r,r) = 0.050 \pm 0.002$, under the assumption that the spin-analysing power of the leptons is equal to unity. The uncertainties quoted are purely statistical.

In order to investigate the effects of initial- and final-state radiation, an alternative \POWHEG-Box + \mbox{\PYTHIA\!8} sample was generated with the renormalisation and factorisation scales varied by a factor of 2, using the low radiation variation of the A14 tune and an $h_{\text{damp}}$ value of $1.5\times m_t$, corresponding to reduced parton-shower radiation~\cite{ATL-PHYS-PUB-2016-004}. The A14 Var3c~\cite{ATL-PHYS-PUB-2014-021} tune variation corresponded to varying $\alpha_\textrm{s}$, which impacts the initial-state radiation in the A14 tune, and covered the size of the other available A14 variations. In order to estimate the effect of the choice of ME event generator, a sample was generated with \MGMCatNLO~(v2.2.1)~\cite{Alwall:2014hca}, interfaced to \PYTHIA\!8. The choice of PS algorithm is evaluated using a sample generated using \POWHEG-Box interfaced to \HERWIG\!7~\cite{Bellm:2015jjp}. An additional \SHERPA~(v2.2.1)~\cite{sherpanlo} sample was used in which events were generated with up to one additional parton simulated at NLO and two, three and four partons at LO with the CT10~\cite{CT10} PDF set for comparison purposes.

Background processes were simulated using a variety of MC event generators. Single top quark production in association with a $W$ boson ($tW$) was simulated at NLO using the \POWHEG-Box~(v1)~\cite{powheg1} ME event generator with CT10 as the PDF. It was interfaced to \PYTHIA\!6 (v6.428)~\cite{pythia6} for the PS, fragmentation and underlying event with the CTEQ6L1~\cite{CT10} NLO PDF set, and a set of tuned parameters called the Perugia 2012 tune~\cite{perugia2012}. The sample was normalised to the theoretical cross-section $\sigma_{tW} = 71.7 \pm 1.8\left(\mathrm{scale}\right) \pm 3.4\left(\mathrm{PDF}\right)$~{pb}~\cite{Kidonakis:2010ux}. The higher-order overlap with \ttbar production was addressed according to the ``diagram removal'' (DR) generation scheme~\cite{diagram_subtraction}. A sample generated with an alternative ``diagram subtraction'' (DS) method was used to evaluate systematic uncertainties~\cite{diagram_subtraction}.

\SHERPA (v2.2.1) with the NNPDF3.0 PDF set was used to model Drell--Yan production. For the $Z/\gamma^*\rightarrow\tau^+\tau^-$ process, \SHERPA calculated matrix elements at NLO for up to two partons and at LO for up to two additional partons using the OpenLoops~\cite{openloops} and Comix~\cite{comix} ME event generators. The MEs were merged with the \SHERPA PS~\cite{Schumann:2007mg} using the ME + PS@NLO prescription~\cite{sherpanlo}. The simulation was normalised using the total cross-section from NNLO predictions~\cite{ATL-PHYS-PUB-2017-006}.

Electroweak diboson production~\cite{ATL-PHYS-PUB-2016-002}, with both bosons decaying leptonically, was simulated with the same \SHERPA version and PDF settings as Drell--Yan production. \SHERPA calculated the MEs for diboson samples at NLO for zero or one additional partons and at LO for two to three additional partons.  The \SHERPA PS was used for all parton multiplicities of four or more. The number of simulated events was normalised using the cross-section computed by the event generator.  Electroweak and loop-induced diboson processes were simulated using \SHERPA (v2.1.1)~\cite{Sherpa,sherpanlo} with the CT10 PDF set.

Events with \ttbar\ production in association with a vector boson or a Higgs boson were simulated using \MGMCatNLO + \PYTHIA\!8~\cite{madgraph}, using the NNPDF2.3 PDF set and the A14 tune, as described in Ref.~\cite{ATL-PHYS-PUB-2016-005}. The $t$-channel production of a single top quark in association with a $Z$ boson ($tZ$) was generated using \MGMCatNLO interfaced with \PYTHIA\!6~\cite{pythia6} with the CTEQ6L1 PDF~\cite{Pumplin:2002vw} set and the Perugia 2012 tune~\cite{perugia2012}. The $tW$ channel production of a single top quark together with a $Z$ boson ($tWZ$) was generated with \MGMCatNLO and showered with \PYTHIA\!8, using the PDF set NNPDF3.0NLO and the A14 tune.  The production of \ttbar\!$WW$ and $\ttbar\ttbar$ were simulated at LO using \MGMCatNLO + \PYTHIA\!8, using the NNPDF2.3 PDF set and the A14 tune.

\textsc{EvtGen} (v1.2.0)~\cite{EvtGen} was used for the heavy-flavour hadron decays in all samples, with the exception of \SHERPA, which performed these decays internally.

Backgrounds also arise from events containing one prompt lepton from the decay of a $W$ or $Z$ boson and either a non-prompt lepton or a particle misidentified as a lepton.  These ``fake leptons'' can arise from heavy-flavour hadron decays, photon conversions, jet misidentification or light-meson decays, and were estimated using MC simulations. The history of the stable particles in the generator-level record was used to identify fake leptons from these processes. The majority ({\raise.17ex\hbox{$\scriptstyle\mathtt{\sim}$}}90\%) of events containing a fake lepton originated from the single-lepton \ttbar process, with smaller contributions arising from $W$ boson production in association with jets, $t$-channel single top quark production, and \ttbar production in association with a vector boson. \SHERPA (v2.2.1) with the NNPDF3.0 PDF set was used to simulate $W$~boson production in association with jets. The $t$-channel single-top quark process was generated using \POWHEG-Box~v1 + \PYTHIA\!6 with the same parameters and PDF sets as those used for the $tW$ sample. Other possible processes with fake leptons, such as multi-jet and Drell--Yan production, were negligible for the event selection used in this analysis.  The fake-lepton contribution derived from MC simulation was verified using a same-charge lepton control region in the data; the MC distributions were scaled up by a small amount as a consequence.

Fully simulated samples involving the SUSY decays $\tilde{t}\rightarrow t \tilde{\chi}^0_1$ with left-handed top squarks were generated using \MGMCatNLO + \PYTHIA8 interfaced to EvtGen and \MADSPIN, with the A14 tune and the LO PDF set NNPDF2.3. The samples contained dilepton $e\mu$ final states only, and covered a range of $170.0 < m({\tilde{t}}) < 300.0$~GeV and $0.5 < m({\tilde{\chi}^0_1}) < 142.5$~GeV. The top quark mass was set to $172.5$~GeV but was allowed to be off-shell by $2\cdot\Gamma_{t}$ and therefore decays of top squarks to top quarks with a mass of $170$~GeV were permitted.

\FloatBarrier

%% file: sections/eventSelectionAndReconstruction.tex
\section{Event selection and reconstruction}
\label{sec:selection}

\subsection{Object and event selection}

This analysis utilises reconstructed electrons, muons, jets, and missing transverse momentum. Jets are reconstructed with the anti-$k_t$ algorithm~\cite{Cacciari:2005hq,Cacciari:2008gp}, using a radius parameter of $R = 0.4$, from topological clusters of energy deposits in the calorimeters~\cite{PERF-2014-07}. Jets are accepted within the range $\pt > 25$~GeV and $|\eta|<2.5$ and are calibrated using simulation with corrections derived from data~\cite{PERF-2016-04}. Jets likely to originate from pile-up are suppressed using a multivariate jet-vertex-tagger (JVT)~\cite{PERF-2014-03} for candidates with $\pt < 60$~GeV and $|\eta| < 2.4$. Additionally, pile-up effects on all jets are corrected using a jet area method~\cite{Cacciari:2007fd,PERF-2016-04}. Jets are identified as containing $b$-hadrons using a multivariate discriminant~\cite{PERF-2016-05}, which uses track impact parameters, track invariant mass, track multiplicity, and secondary vertex information to discriminate $b$-jets from light-quark or gluon jets (light jets). The average $b$-tagging efficiency is 77\%, with a purity of 95\% for $b$-tagged jets in simulated dileptonic \ttbar events with the selection used in this analysis.

Electron candidates are identified by matching an inner-detector track to an isolated energy deposit in the electromagnetic calorimeter, within the fiducial region of transverse momentum $\pt > 25$~GeV and $|\eta| < 2.47$. Electron candidates are excluded if the pseudorapidity of the calorimeter cluster is within the transition region between the barrel and the endcap of the electromagnetic calorimeter, $1.37 < |\eta| < 1.52$. Electrons are selected using a multivariate algorithm and are required to satisfy a \emph{Tight} likelihood-based quality criterion in order to provide high efficiency and good rejection of fake electrons~\cite{PERF-2017-01}. Electron candidates must have tracks that pass the requirements of transverse impact parameter significance with respect to the primary vertex\footnote{The transverse impact parameter significance is defined as $d_0^\text{sig} = d_0 / \sigma_{d_0}$, where $\sigma_{d_0}$ is the uncertainty in the transverse impact parameter $d_0$.} $|d_0^\text{sig}|<5$ and longitudinal impact parameter $|z_0 \sin\theta| < 0.5$~mm. Electrons must pass \pt- and $\eta$-dependent isolation requirements based on inner-detector tracks and topological clusters in the calorimeter. These requirements have an efficiency of 95\% for an electron \pt of 25~GeV and 99\% for an electron \pt above 60~GeV, when determined in simulated $Z\rightarrow e^{+}e^{-}$ events. 

Electrons that share a track with a muon are discarded. Double counting of electron energy deposits as jets is prevented by removing the closest jet within $\Delta R=0.2$ of a reconstructed electron. Following this, the electron is discarded if a jet exists within $\Delta R = 0.4$ of the electron to ensure sufficient separation from nearby jet activity, where in this case $\Delta R$ was calculated using the rapidity of the jets.

Muon candidates are identified from muon-spectrometer tracks that match tracks in the inner detector, with \pt > 25~GeV and $|\eta| < 2.5$~\cite{PERF-2015-10}. The tracks of muon candidates are required to have a transverse impact parameter significance $|d_0^\text{sig}|<3$ and a longitudinal impact parameter $|z_0 \sin\theta| < 0.5$~mm. Muons must satisfy quality criteria and isolation requirements based on inner-detector tracks and topological clusters in the calorimeter which depend on $\eta$ and \pt. These requirements reduce the contributions from fake muons and provide the same efficiency as for electrons. The criteria used for the muons in this analysis is the \emph{Medium} working point. Muons may leave energy deposits in the calorimeter that could be misidentified as a jet, so jets with fewer than three associated tracks are removed if they are within $\Delta R = 0.4$ of a muon. Muons are discarded if they are separated from the nearest jet by $\Delta R < 0.4$ to reduce the background from muons from heavy-flavour hadron decays inside jets.

The missing transverse momentum (with magnitude \met) is defined as the negative vector sum of the transverse momenta of reconstructed, calibrated objects in the event.  It is computed using calibrated electrons, muons, and jets~\cite{PERF-2016-07} and includes contributions from soft tracks associated with the primary vertex but not forming the lepton or jet candidates.  The primary vertex of an event is defined as the vertex for which the associated tracks have the highest sum of $\pt^2$, where each track has $\pt > 400$~\MeV.

Two types of signal events are considered, depending on whether a full reconstruction of the $t\bar{t}$ system is performed, denoted here as \emph{inclusive} and \emph{reconstructed} selections. The inclusive selection is used for the $\Delta\phi$ and $\Delta\eta$ differential cross-sections.  It is defined by requiring exactly one electron and one muon of opposite electric charge, where at least one of them has $\pt > 27$~GeV, and at least two jets, at least one of which must be $b$-tagged. The reconstructed selection is used for the measurement of $\Delta\phi$ as a function of the \ttbar invariant mass.  It has a more stringent $b$-tagging requirement of at least two $b$-tagged jets and also requires that at least one solution was found for the reconstruction of the \ttbar system (described in detail later in this section). The tighter $b$-tagging requirement is imposed in the reconstructed selection to improve the performance of the $t\bar{t}$ reconstruction by removing light jets that are erroneously assigned to the top-quark or top-antiquark decay. A less strict $b$-tagging selection requirement of only one or more $b$-tagged jets is used in the inclusive selection in order to increase the event selection efficiency. Only events with exactly one electron and one muon are considered as this decay mode provides the highest signal purity as well as more than sufficient data statistics. The dielectron and dimuon decay modes are not considered due to their enhanced Drell--Yan and heavy flavour backgrounds, while the increase in statistical power would not improve the overall uncertainty on the results.

Using the inclusive selection, 93\% of selected events are expected to be \ttbar events. The other processes that pass the signal selection are Drell--Yan ($Z/\gamma^{*}\rightarrow\tau^{+}\tau^{-}$), diboson, single top quark ($tW$) production, boson production in association with a $t\bar{t}$ pair ($t\bar{t}V$ and others), and fake-lepton events. The reconstructed selection gives a subset of these events, in which 96\% of selected events are expected to be \ttbar events. This is higher than the inclusive selection because of the tighter $b$-tagging requirement and because the \ttbar reconstruction procedure tends to succeed more often for \ttbar events than for background processes.

\begin{figure}
\centering
\begin{tabular}{cc}
\subfloat[\label{fig:emu_electron_pt}]{\includegraphics[width=0.45\textwidth]{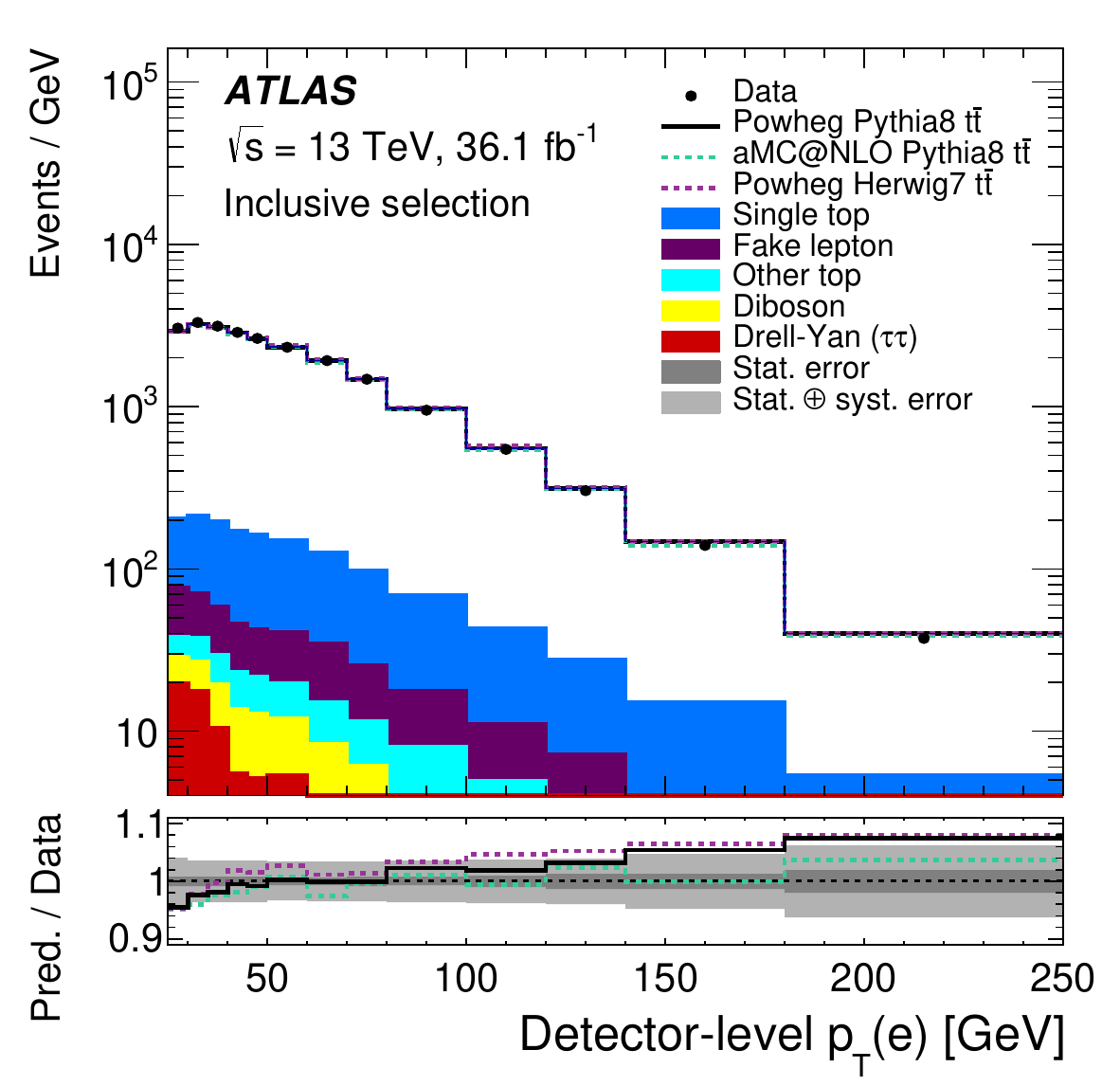}}&
\subfloat[\label{fig:emu_muon_pt}]{\includegraphics[width=0.45\textwidth]{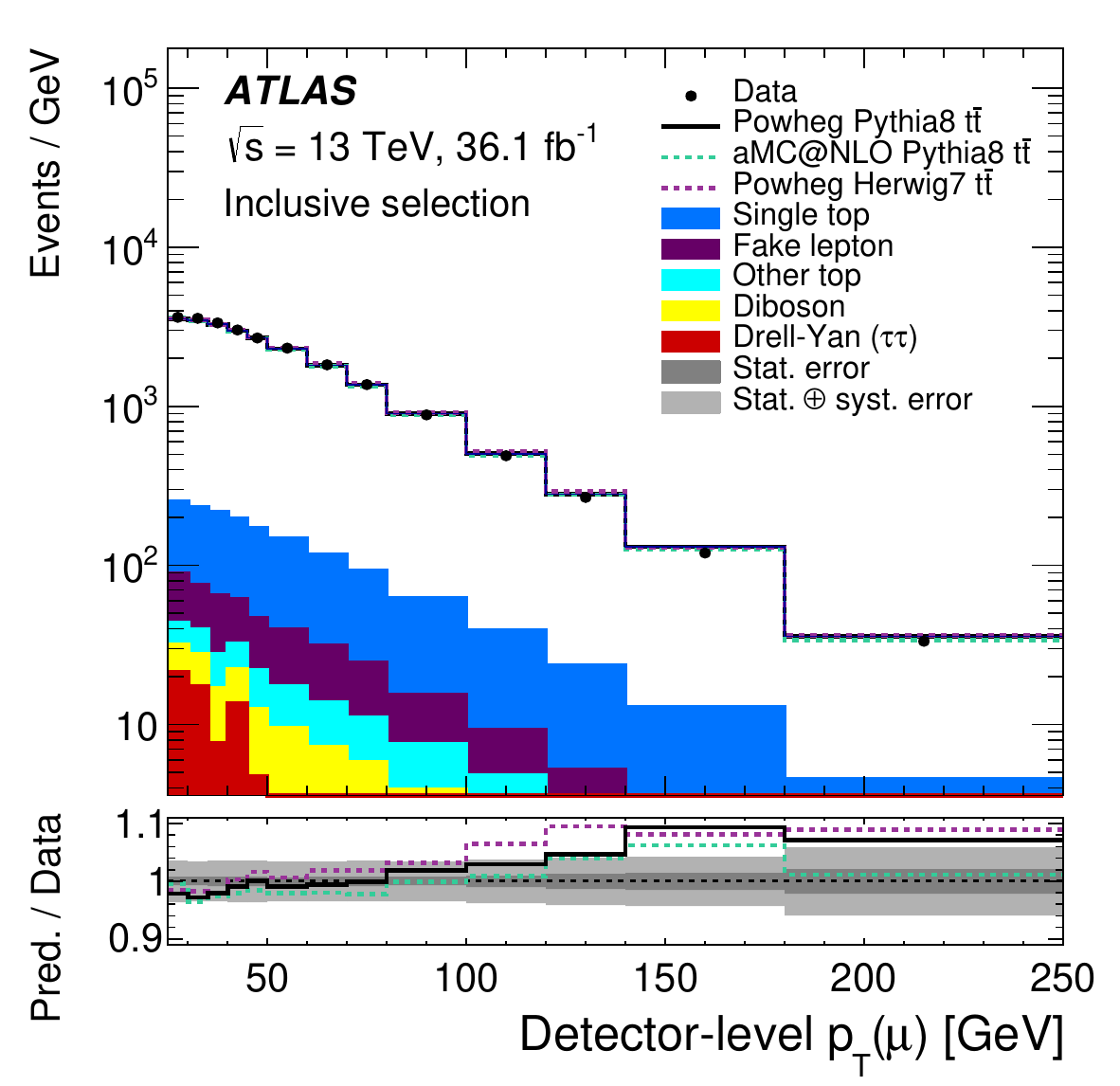}} \\
\subfloat[\label{fig:emu_bjet_pt}]{\includegraphics[width=0.45\textwidth]{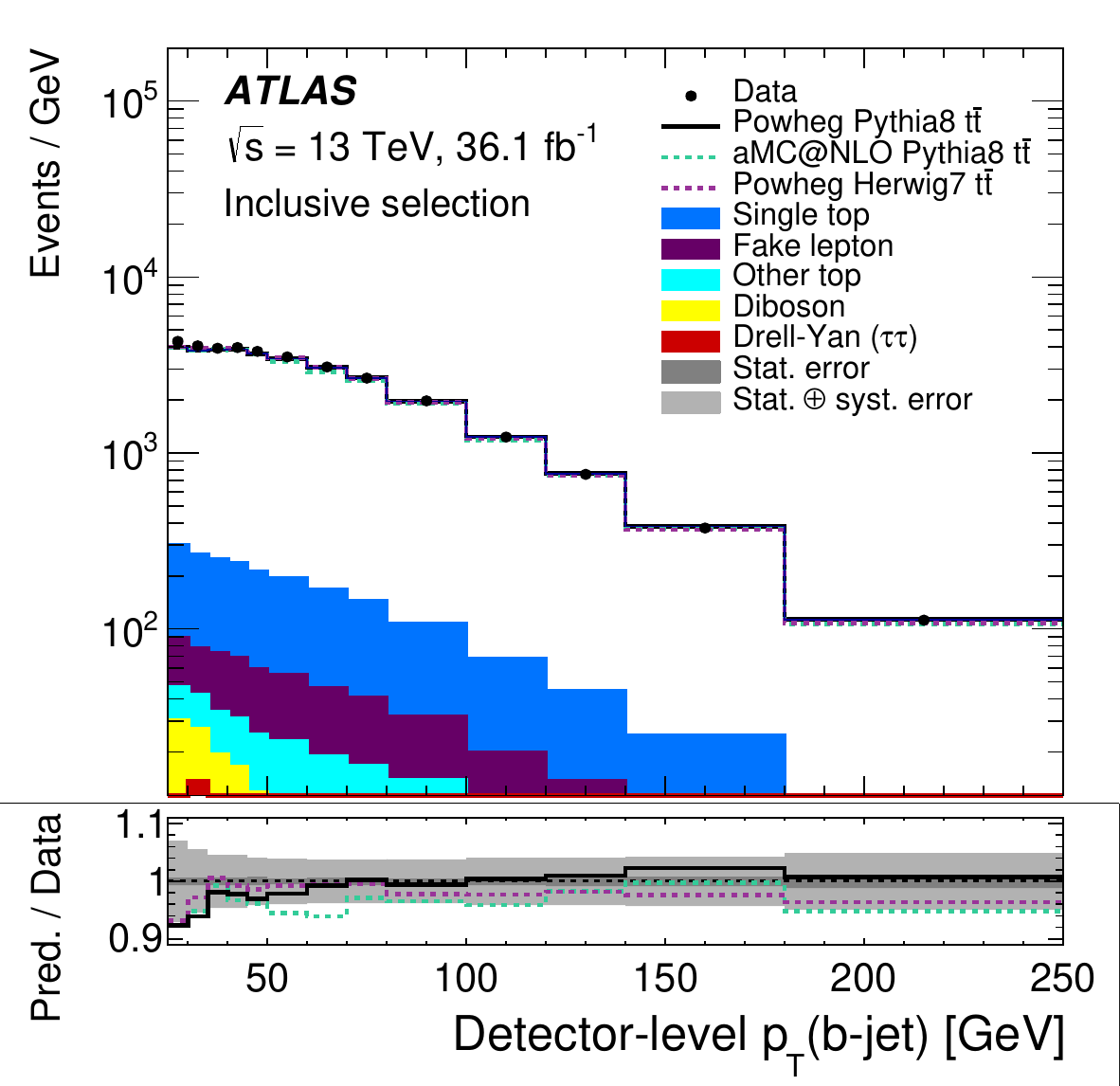}} &
\subfloat[\label{fig:emu_met}]{\includegraphics[width=0.45\textwidth]{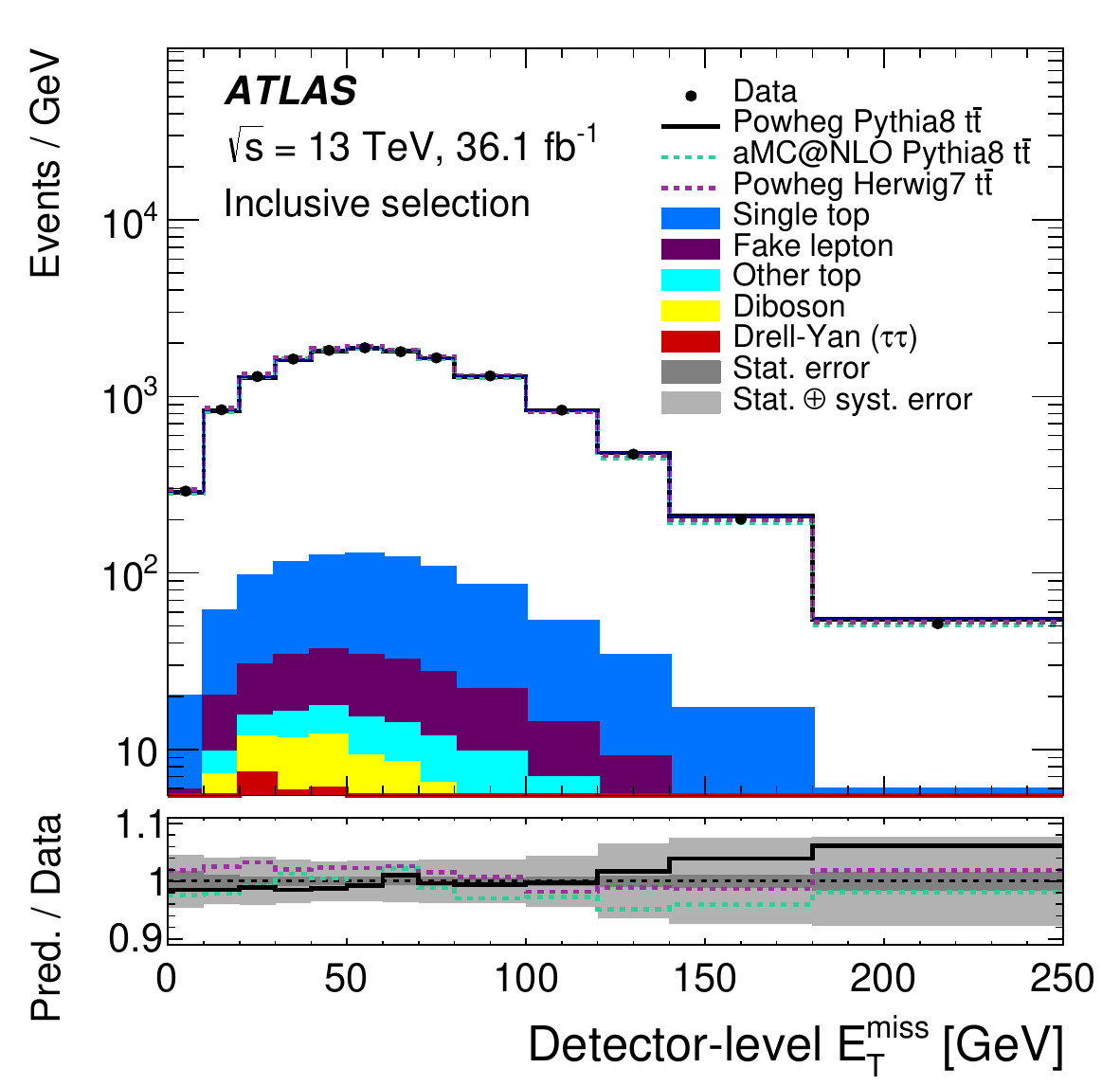}}\\
\end{tabular}
\caption{Kinematic distributions for the (a) electron \pt, (b) muon \pt, (c)  leading $b$-jet \pt,  and (d) \met for the $e^{\pm}\mu^{\mp}$ inclusive selection. In all figures, the rightmost bin also contains events that are above the $x$-axis range. The dark uncertainty bands in the ratio plots represent the statistical uncertainties while the light uncertainty bands represent the statistical and systematic uncertainties added in quadrature. The systematic uncertainties include contributions from leptons, jets, missing transverse momentum, background modelling, pile-up modelling and luminosity, but not PDF or signal \ttbar modelling uncertainties. The observed distribution is compared to the sum of signal and background using three different \ttbar signal models: \POWHEG+\PYTHIA\!8, \POWHEG+\HERWIG\!7 and \MGMCatNLO+\PYTHIA\!8, and the ratio panel compares the summed prediction to data for the three models.}
\label{fig:reco}
\end{figure}

\begin{figure}[htb]
\centering 
\subfloat[\label{fig:emu_delta_phi_nomNoNW}]{\includegraphics[width=0.4\textwidth]{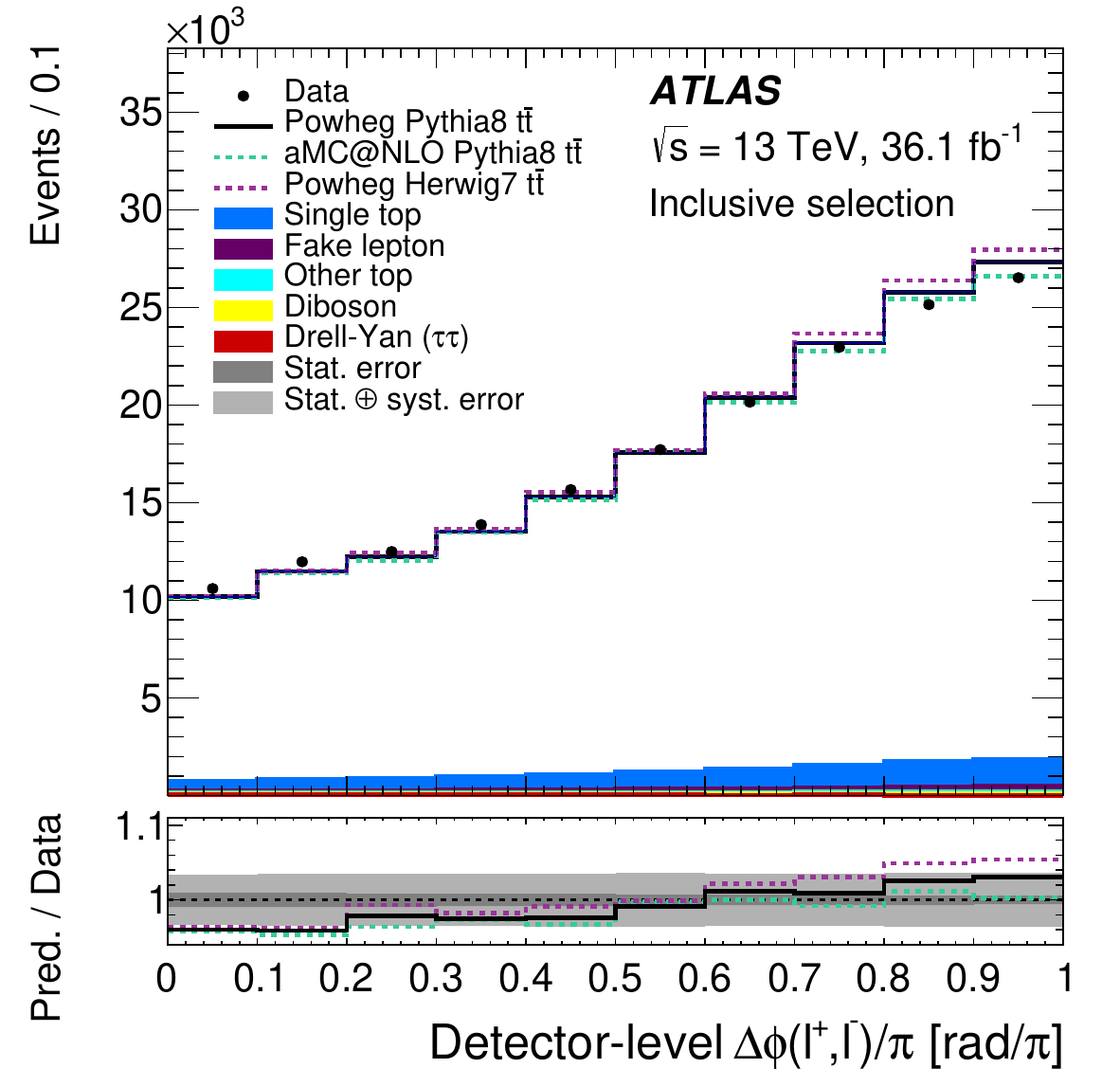}}
\subfloat[\label{fig:emu_delta_eta_nomNoNW}]{\includegraphics[width=0.4\textwidth]{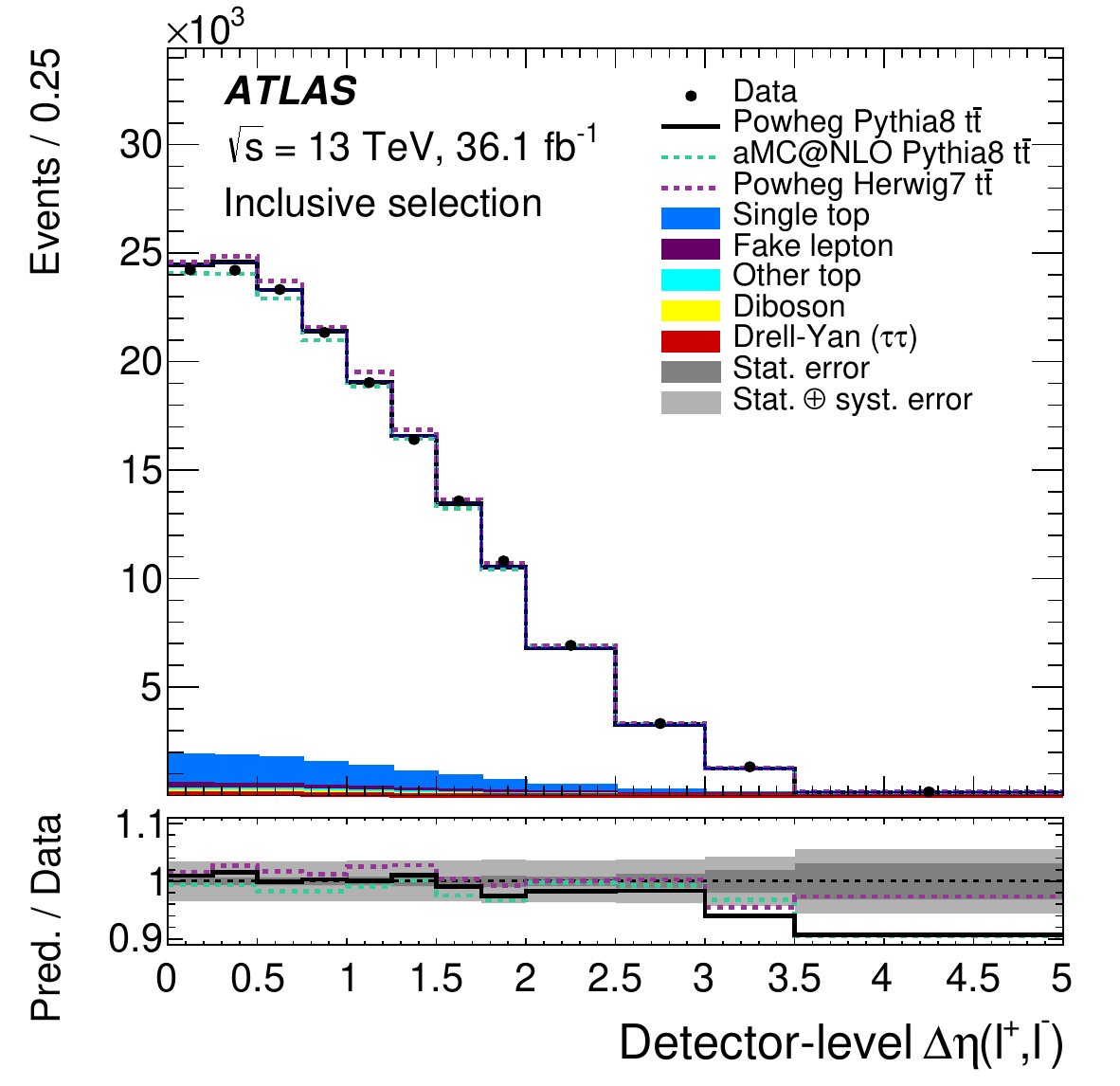}}
\caption{Distribution of (a) the $\Delta\phi$ and (b) $\Delta\eta$ observables for the $e\mu$ selection after the requirement of at least one $b$-tagged jet (inclusive selection). The highest bin for $\Delta\eta$ also contains events that are above the $x$-axis range. 
The dark uncertainty bands in the ratio plots represent the statistical uncertainties while the light uncertainty bands represent the statistical and systematic uncertainties added in quadrature. The systematic uncertainties include contributions from leptons, jets, missing transverse momentum, background modelling, pile-up modelling and luminosity, but not PDF or signal \ttbar modelling uncertainties. The observed distribution is compared to the sum of signal and background using three different \ttbar signal models: \POWHEG+\PYTHIA\!8, \POWHEG+\HERWIG\!7 and \MGMCatNLO+\PYTHIA\!8, and the ratio panel compares the summed prediction to data for the three models.} 
\label{fig:emu_selection3} 
\end{figure}

The event yields after both selections are listed in Table~\ref{tab:yields}. The expected yields are in agreement with the observed number of events in both cases. Distributions of the lepton and jet \pt and \met are shown in Figure~\ref{fig:reco} for the inclusive selection. The data and prediction agree within the total uncertainty for all of these kinematic observables. The trends observed in the lepton and jet \pt arise from the well-documented limitations of the modelling of the top quark's \pt spectrum at NLO~\cite{TOPQ-2015-06,TOPQ-2016-01,TOPQ-2016-04}. The systematic uncertainties included in both the table and the figures are described in Section~\ref{sec:systematics}. The azimuthal opening angle of the electron and muon, $\Delta\phi$, and the absolute value of the separation of the leptons in pseudorapidity, $\Delta\eta$, are shown in Figure~\ref{fig:emu_selection3} for the inclusive selection.  The observed distribution is compared to the sum of signal and background using three different signal models: \POWHEG+\PYTHIA\!8, \POWHEG+\HERWIG\!7, and \MGMCatNLO+\PYTHIA\!8, and the ratio panel compares the combined signal plus background to data for the three models.

\begin{table}[h]
\caption{Event yields in the inclusive and reconstructed selections for the observed data, expected signal and expected background. The uncertainties quoted include contributions from leptons, jets, missing transverse momentum, luminosity, background modelling, and pile-up modelling. They do not include uncertainties from PDF or signal \ttbar\ modelling. The ``$t\bar{t}V$ and others'' entries contain events from $t\bar{t}Z$, $t\bar{t}W$, $t\bar{t}WW$, $t\bar{t}H$, and the $t\bar{t}t\bar{t}$ processes.}
\vspace{3mm}
\centering
\begin{tabular}{ l | r c r | r c r }
\toprule
Process                                   & \multicolumn{3}{c|}{Inclusive selection}  & \multicolumn{3}{c}{Reconstructed selection} \\
                                          & \multicolumn{3}{c|}{$\ge 1$ $b$-tag}  & \multicolumn{3}{c}{$\ge 2$ $b$-tags} \\
\midrule
$t\bar{t}$                                &  165~000   & $\pm$ &  5000     &   75~000   & $\pm$ &  4000   \\
      $tW$                                &    8900    & $\pm$ &  1400     &    1550    & $\pm$ &   170   \\
      $t\bar{t}V$ and others              &     670    & $\pm$ &    60     &     233    & $\pm$ &    22   \\
   Diboson                                &     580    & $\pm$ &    60     &      15.1  & $\pm$ &     2.8 \\
$Z/\gamma^{*}\rightarrow\tau^{+}\tau^{-}$ &     420    & $\pm$ &    70     &      26    & $\pm$ &    17   \\
Fake Lepton                               &    1800    & $\pm$ &   700     &     630    & $\pm$ &   250   \\
\midrule                                                                                             
Expected                                  &  177~000   & $\pm$ &   6000     &  78~000   & $\pm$ & 4000  \\
Observed                                  &  177~113   &       &            &  75~885   &       &       \\
\bottomrule
\end{tabular}
\label{tab:yields}
\end{table}

\subsection{Reconstruction of the \ttbar system}

In order to measure spin correlations as a function of the \ttbar invariant mass at detector level, the kinematic properties of the event must be reconstructed from the identified leptons, jets, and missing transverse momentum. The top quark, top antiquark, and reconstructed \ttbar system are built using the Neutrino Weighting (NW) method~\cite{NeutrinoWeighting}. While the individual four-momenta of the two neutrinos in the final state are not directly measured in the detector, the sum of their transverse momenta is measured as \met. The absence of the measured four-momenta of the two neutrinos leads to an under-constrained system that cannot be solved analytically. 
The following invariant mass constraints were applied to each event:
\begin{equation}
 \label{eq:nus}
 \begin{aligned}
(\ell_{1,2} + \nu_{1,2})^2 &= m^2_W = (80.4~\text{\GeV})^2\text{,} \\
(\ell_{1,2} + \nu_{1,2} + b_{1,2})^2 &= m_t^2 = (172.5~\text{\GeV})^2\text{,}\\
\end{aligned}
\end{equation}
where $\ell_{1,2}$, $\nu_{1,2}$ and $b_{1,2}$ represent the four-momenta of the charged leptons, neutrinos and $b$-quarks, respectively. Since the neutrino pseudorapidities ($\eta(\nu)$ and $\eta(\bar{\nu}$)) required for $\nu_{1,2}$ are unknown, their values are scanned, in steps of 0.2, between $-5$ and 5.

With the assumptions about $m_t$, $m_W$ and values for $\eta(\nu)$ and $\eta(\bar{\nu})$, Eq.~(\ref{eq:nus}) can now be solved, leading to two possible solutions for each assumption of $\eta(\nu)$ and $\eta(\bar{\nu})$. Only real solutions without an imaginary component are considered. An ``inferred'' \met value, resulting from the neutrinos for each solution, is compared to the \met observed in the event. A weight is introduced in order to quantify this agreement:
\begin{equation}
 \label{eq:weight}
w = \exp\Bigg(\frac{-\Delta E^2_x}{2\sigma_x^2}\Bigg) \cdot \exp \Bigg( \frac{-\Delta E^2_y}{2\sigma^2_y} \Bigg) \text{,} \nonumber
\end{equation}
where $\Delta E_{x,y}$ is the difference between the ($x$,$y$) component of the missing transverse momentum computed from the neutrino four momenta in Eq.~(\ref{eq:nus}) and the observed missing transverse momentum, and $\sigma_{x,y}$ is a fixed scale related to the resolution of the observed \met in the detector in ($x,y$), based on studies in $Z$ boson events~\cite{PERF-2016-07}. The assumption for $\eta(\nu)$ and $\eta(\bar{\nu})$ that gives the highest weight is used to reconstruct the $t$ and $\bar{t}$ quarks for that event. 

In each event, there may be more than two $b$-tagged jets (on average there are 2.04 $b$-tagged jets per event) and therefore several possible combinations of jets to use in the kinematic reconstruction. In addition, there is an ambiguity in assigning a jet to the $t$ or $\bar{t}$ quark candidate. To reduce this ambiguity, the two $b$-tagged jets with the highest weight from the $b$-tagging algorithm are used to reconstruct the $t$ and $\bar{t}$ quarks and the assignment which produces the solution with highest weight in the NW is taken as the correct assignment. 

Equation~(\ref{eq:nus}) cannot always be solved for a particular assumption of $\eta(\nu)$ and $\eta(\bar{\nu})$. This can be caused by mis-assignment of the input objects or through mis-measurement of the input object four-momenta. It is also possible that the assumed $m_t$ is sufficiently different from the true value to prevent a valid solution for a particular event, or the event is from a background process, and therefore cannot be solved. To mitigate these effects, the assumed value of $m_t$ is scanned between the values of 171 and 174 GeV, in steps of 0.5 GeV, and the \pt of the measured jets are smeared using a Gaussian function with a \pt-dependent width between 14\% and 8\% of their measured \pt. This smearing is repeated 5 times.

This procedure allows the NW algorithm to shift the four-momenta of the two jets and the $m_t$ hypothesis to see if a solution can be found. The solution which produces the highest $w$ gives the kinematics of the reconstructed event. Solutions which provide an invariant mass of the \ttbar system below 300 GeV, or which provide $t$ or $\bar{t}$ quarks with negative energies, are rejected. For around 5\% of events, no solution can be found, even after smearing. Only events with at least one solution with a weight above 0.4 are considered, where this criterion was chosen to optimise the angular resolution in the top quark reconstruction.
The efficiency for \ttbar reconstruction is {\raise.17ex\hbox{$\scriptstyle\mathtt{\sim}$}}80\%. Due to the implicit assumptions about $m_t$ and $m_W$, the reconstruction efficiency found in simulated background samples is much lower ({\raise.17ex\hbox{$\scriptstyle\mathtt{\sim}$}}60\% for $tW$ and Drell--Yan processes) and leads to a suppression of background events. Table~\ref{tab:yields} shows the event yields before and after reconstruction in the signal region. The different effects of the systematic uncertainties on each type of selection are discussed in greater detail in Section \ref{sec:results_diff}.

\begin{figure}[htb]
\centering 
\begin{tabular}{cc}
\subfloat[\label{fig:emu_dphi}]{\includegraphics[width=0.4\textwidth]{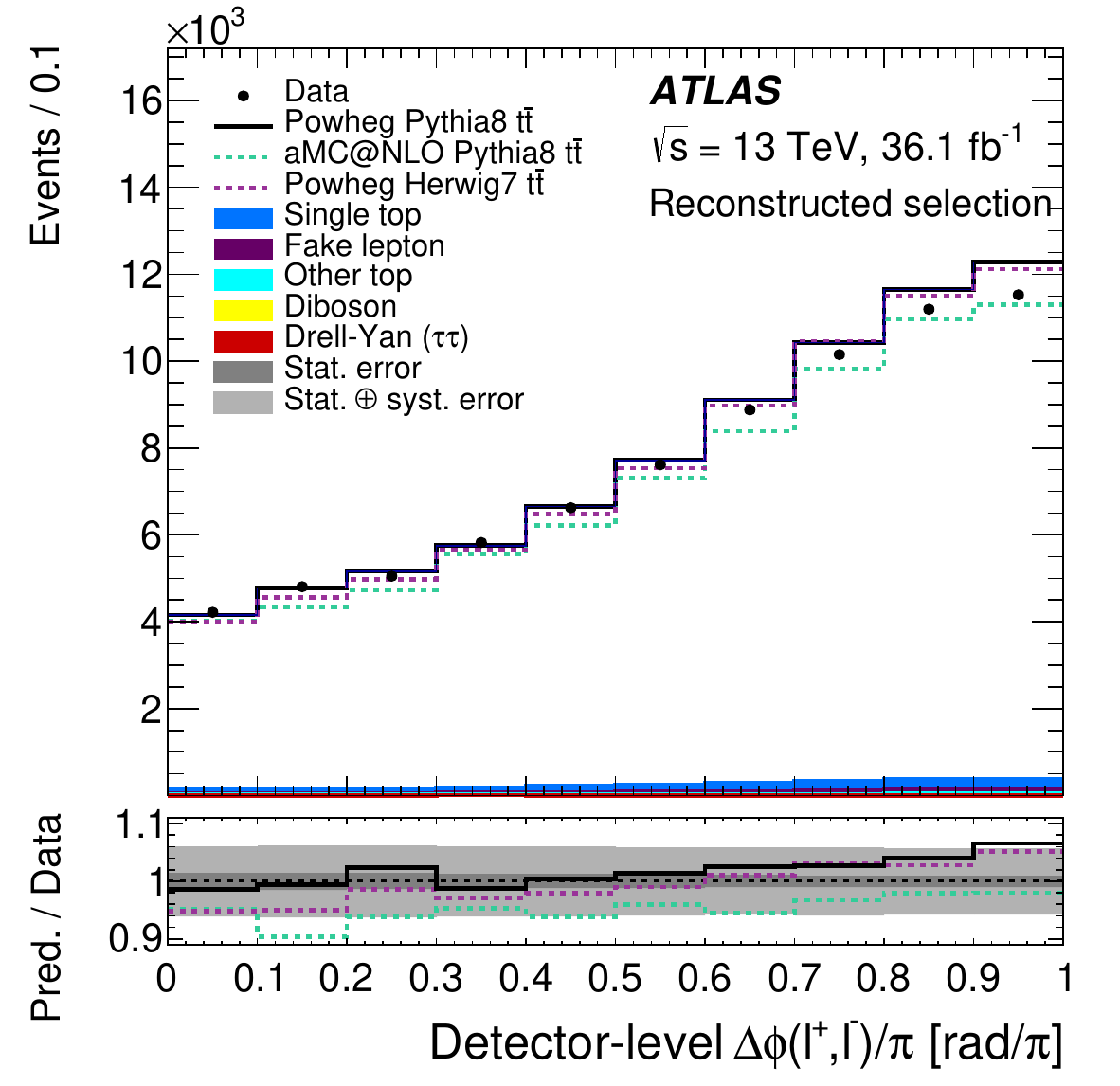}}&
\subfloat[\label{fig:emu_ttbar_m}]{\includegraphics[width=0.4\textwidth]{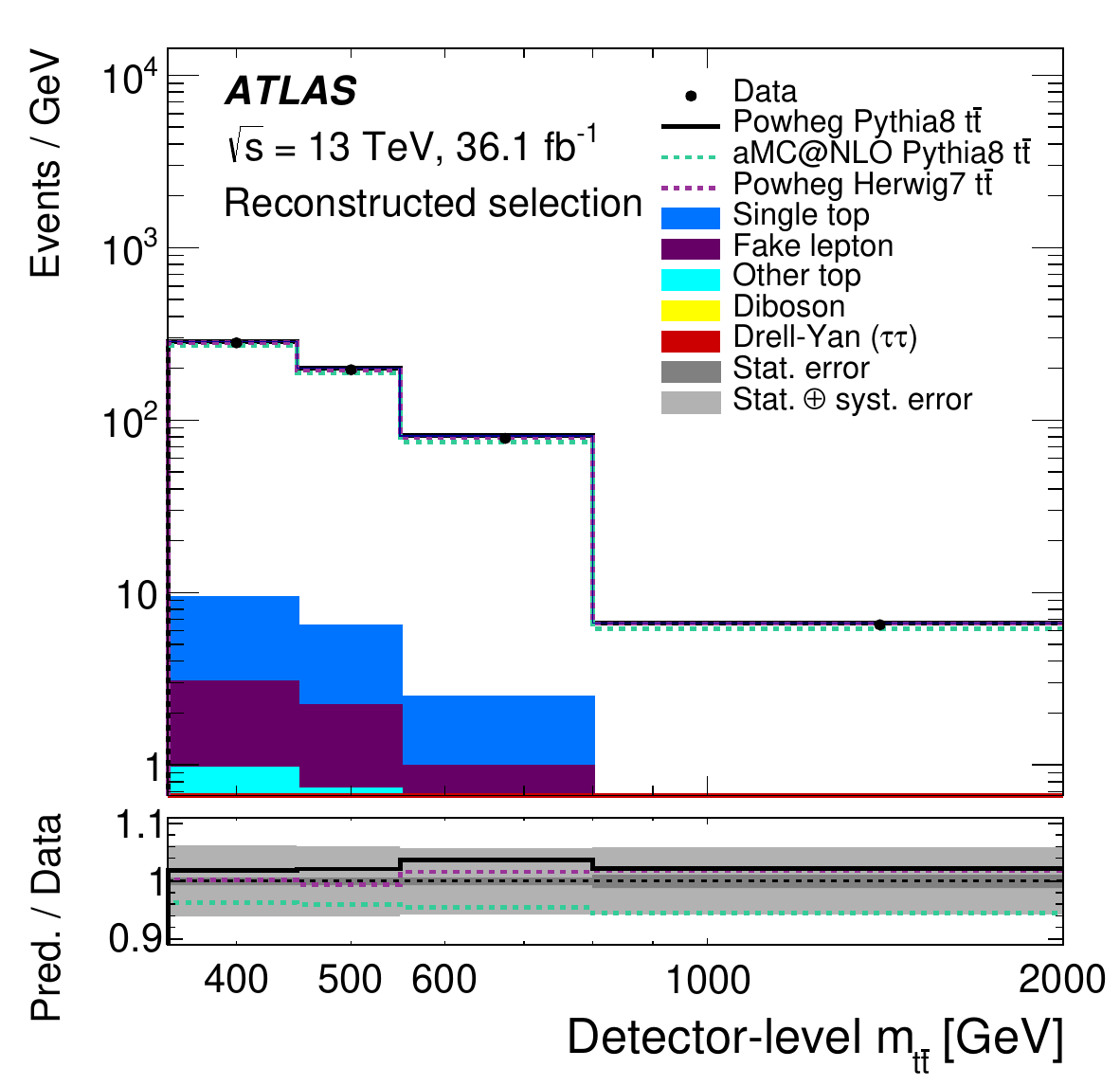}}\\
\end{tabular}
\caption{Kinematic distributions for (a) $\Delta \phi$ and (b) $m_{\ttbar}$ after the requirement of at least two $b$-tagged jets and Neutrino Weighting (reconstructed selection). The highest bin in~Figure~\ref{fig:emu_ttbar_m} also contains events that are above the $x$-axis range. The dark uncertainty bands in the ratio plots represent the statistical uncertainties while the light uncertainty bands represent the statistical and systematic uncertainties added in quadrature. The systematic uncertainties include contributions from leptons, jets, missing transverse momentum, background modelling, pile-up modelling and luminosity, but not PDF or signal \ttbar modelling uncertainties. The observed distribution is compared to the sum of signal and background using three different \ttbar signal models: \POWHEG+\PYTHIA\!8, \POWHEG+\HERWIG\!7, and \MGMCatNLO+\PYTHIA\!8, and the ratio panel compares the summed prediction to data for the three models.}
\label{fig:emu_selection_v2_nuW_delta_phi_and_mttbar} 
\end{figure}

\begin{figure}[htb]
\centering 
\begin{tabular}{cc}
\subfloat[\label{fig:emu_dphi_v2_nuW1}]{\includegraphics[width=0.4\textwidth]{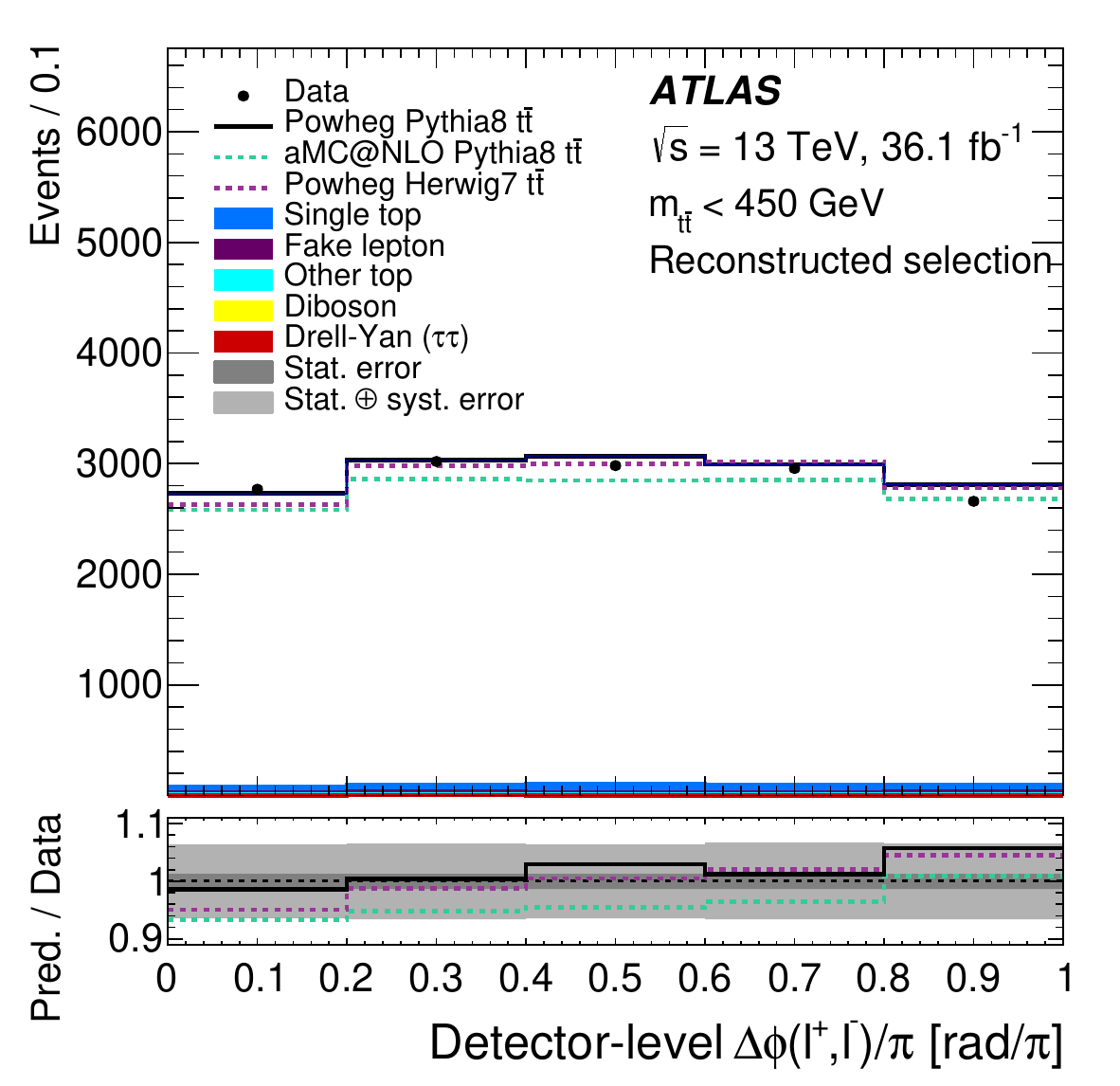}}&
\subfloat[\label{fig:emu_dphi_v2_nuW2}]{\includegraphics[width=0.4\textwidth]{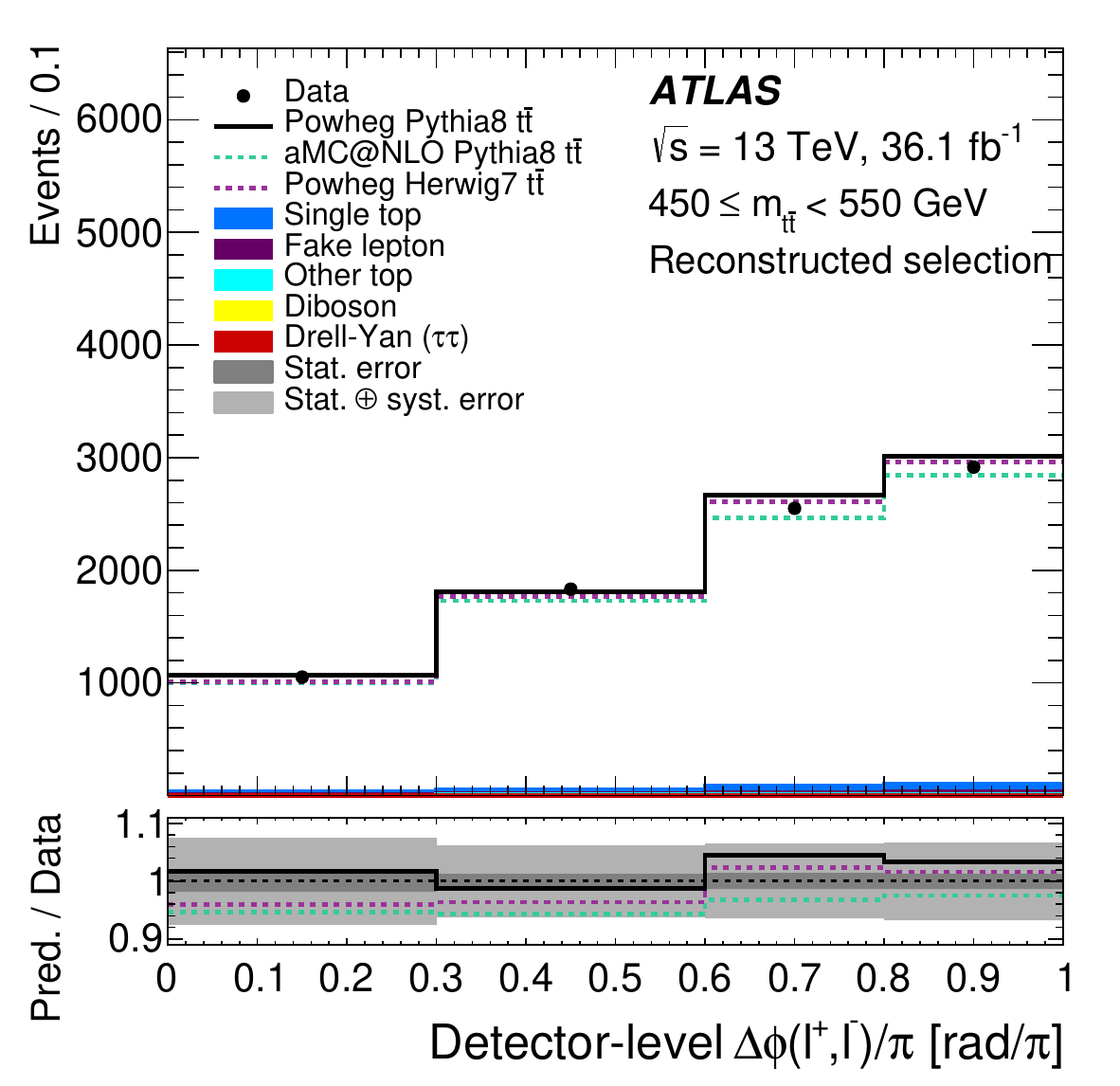}}\\
\subfloat[\label{fig:emu_dphi_v2_nuW3}]{\includegraphics[width=0.4\textwidth]{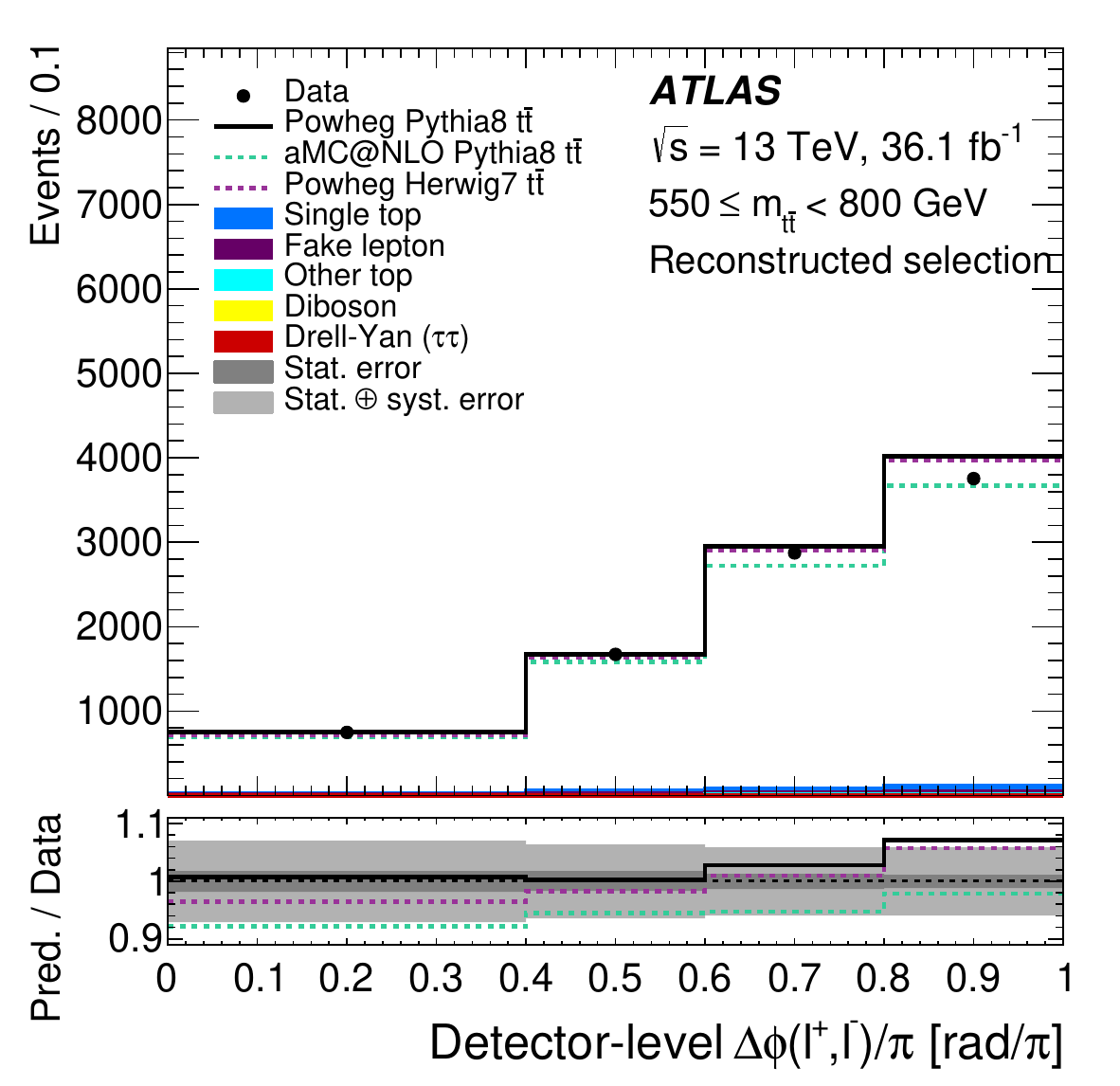}}&
\subfloat[\label{fig:emu_dphi_v2_nuW4}]{\includegraphics[width=0.4\textwidth]{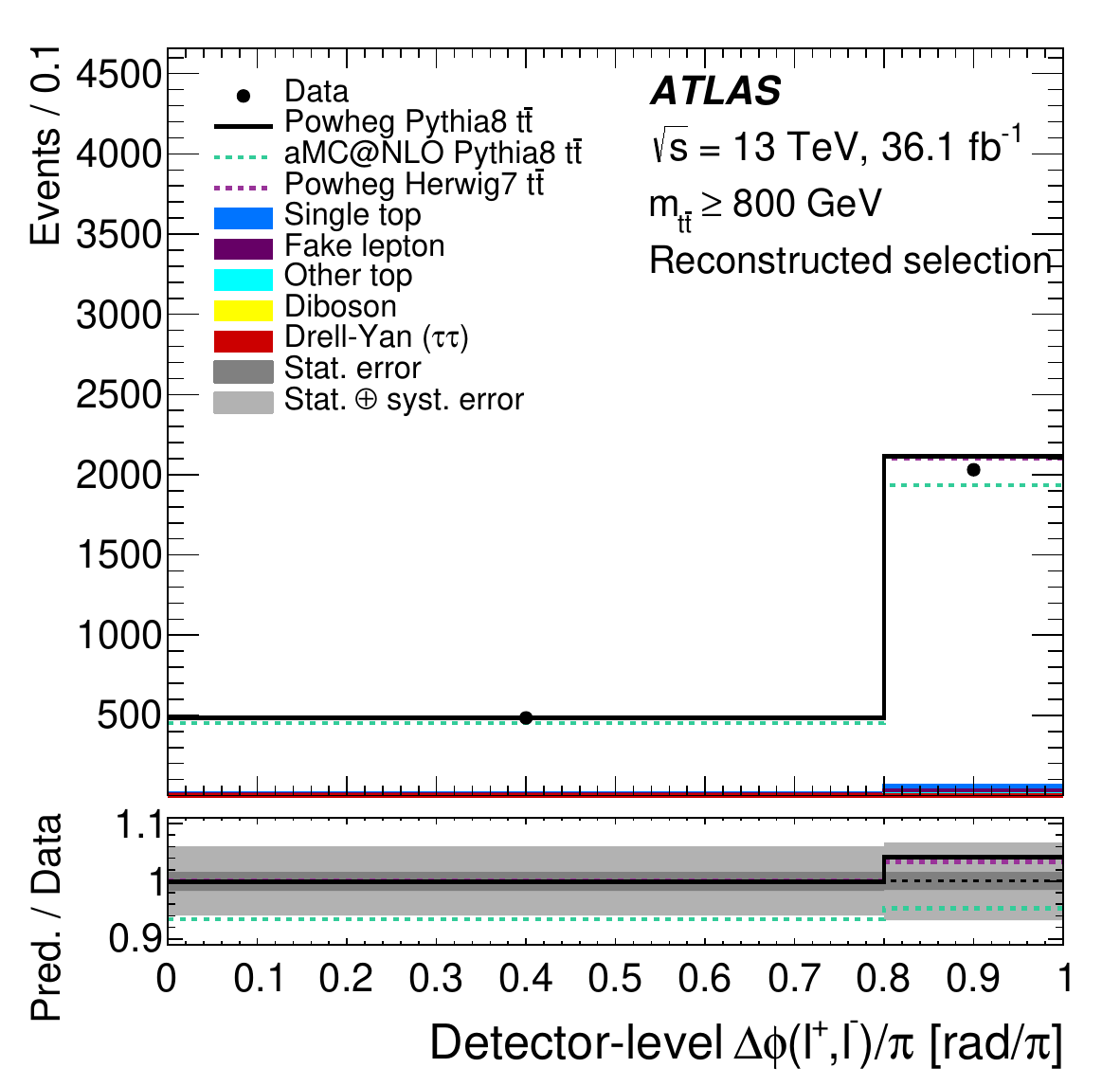}}\\
\end{tabular}
\caption{Kinematic distributions after the requirement of at least two $b$-tagged jets and Neutrino Weighting (reconstructed selection). The plots display $\Delta\phi/\pi$ in individual mass ranges: 
(a) $m_{\ttbar}<450$~\GeV,
(b) $450\leq m_{\ttbar}<550$~\GeV,
(c) $550\leq m_{\ttbar}<800$~\GeV, and
(d) $m_{\ttbar}\geq800$~\GeV. The dark uncertainty bands in the ratio plots represent the statistical uncertainties while the light uncertainty bands represent the statistical and systematic uncertainties added in quadrature. The systematic uncertainties include contributions from leptons, jets, missing transverse momentum, background modelling, pile-up modelling and luminosity, but not PDF or signal \ttbar modelling uncertainties. The observed distribution is compared to the sum of signal and background using three different \ttbar signal models: \POWHEG+\PYTHIA\!8, \POWHEG+\HERWIG\!7 and \MGMCatNLO+\PYTHIA\!8, and the ratio panel compares the summed prediction to data for the three models.}
\label{fig:emu_selection_v2_nuW_delta_phi_vs_mttbar} 
\end{figure}

Figure~\ref{fig:emu_selection_v2_nuW_delta_phi_and_mttbar} shows the distributions of $\Delta\phi$ and $m_{\ttbar}$ after reconstruction and with a requirement of at least two $b$-tagged jets (reconstructed selection). The four plots in Figure~\ref{fig:emu_selection_v2_nuW_delta_phi_vs_mttbar} show the $\Delta\phi$ distribution split into four mass regions: 
$m_{\ttbar}<450$~\GeV; 
$450\leq m_{\ttbar}<550$~\GeV; 
$550\leq m_{\ttbar}<800$~\GeV; and 
$m_{\ttbar}\geq800$~\GeV. These bins in $m_{\ttbar}$ were determined to have the finest possible granularity whilst maintaining an unbiased and stable unfolding procedure for the $\Delta\phi$ observable (described further in Section \ref{sec:unfolding}).

\subsection{Definitions of partons and particles}

 In the measurements presented in this paper, events are corrected for detector effects using two definitions of particles in the generator-level record of the simulation: parton level and particle level. 
Parton-level objects are taken from the MC simulation history. Top quarks are taken after radiation but before decay (this is the last top quark in a decay chain) whereas leptons are taken before radiation (i.e. \emph{Born} level leptons). The measurement corrected to parton level is extrapolated to the full phase-space, where all generated dilepton events are considered. However, events with leptons originating from an intermediate $\tau$-lepton in the $t \rightarrow bW \rightarrow b \ell \nu$ decay chain are not considered as their subsequent decays do not carry the full spin information of their parent top quark and hence, dilute the spin correlation information. Fiducial requirements are not made on the partonic objects so that the results at parton level can be more easily compared to fixed-order predictions.

Particle-level objects are constructed using a procedure intended to correspond as closely as possible to the detector-level object and event selection. Only objects in the MC simulation considered stable (with lifetimes longer than $3 \times 10^{-11}$~s) in the generator-level information are used. Particle-level leptons are identified as those originating from a $W$ boson decay. The four-momentum of each electron or muon is summed with the four-momenta of all radiated photons within a cone of size $\Delta R = 0.1$ about its direction, excluding photons from hadron decays. The resulting leptons are required to have $\pt > 25$~GeV and $|\eta|<2.5$. Particle-level jets are constructed using stable particles, with the exception of selected particle-level electrons and muons, photons that are summed into the electrons or muons, and particle-level neutrinos originating from $W$ boson decays. The jets are constructed using the anti-$k_t$ algorithm with a radius parameter of $R=0.4$, and selected if they pass the requirements of $\pt>25$~GeV and $|\eta|<2.5$. Intermediate $b$-hadrons in the MC decay chain history are clustered in the stable-particle jets with their energies set to zero. If, after clustering, a particle-level jet contains one or more of these ``ghost'' $b$-hadrons, the jet is said to have originated from a $b$-quark. This technique is referred to as ``ghost matching''~\cite{Cacciari:2007fd}. Particle-level \met is calculated using the vector transverse-momentum sum of all neutrinos in the event, excluding those originating from hadron decays, either directly or via a $\tau$-lepton. 

Events are selected at the particle level in a fiducial phase-space region with similar requirements to the phase-space region in the detector. They must contain exactly one particle-level electron and one particle-level muon of opposite electric charge, at least one of which must have $\pt > 27$~GeV, and at least two particle-level jets. The particle-level requirement on the number of jets that must be ghost-matched to a $b$-hadron mimics the \emph{inclusive} and \emph{reconstructed} selections at detector-level: for the inclusive selection, at least one particle-level jet must be ghost-matched, while for the reconstructed case, the particle-level selection requires exactly two ghost-matched jets.  In addition, the \emph{reconstructed} selection excludes particle-level leptons originating from an intermediate $\tau$-lepton in the $t \rightarrow bW \rightarrow b \ell \nu$ decay chain. The particle-level \ttbar\ object is constructed using the sum of the particle-level electron and muon, the two ghost-matched jets, and the two neutrinos that originate from the same $W$ boson decays as the selected particle-level leptons.

\FloatBarrier

%% file: sections/unfoldingProcedure.tex
\section{Unfolding procedure}
\label{sec:unfolding}

The data are corrected for detector resolution and acceptance effects using an iterative Bayesian unfolding procedure~\cite{bayesian} in order to create distributions at particle (parton) level in a fiducial (full) phase-space. The unfolding itself is performed using the \textsc{RooUnfold} package~\cite{Adye:2011gm}.

In the unfolding procedure, background-subtracted data are corrected for detector acceptance and resolution effects as well as for the efficiency to pass the event selection requirements in order to obtain the absolute differential cross-sections:
\begin{equation}
\frac{\text{d}\sigma_{\ttbar}}{\text{d}X^i} = \frac{1}{\mathcal{L} \cdot \Delta X^i \cdot \epsilon^i_{\text{eff}}} \cdot \sum_{j} R^{-1}_{ij} \cdot f^j_{\text{acc}} \cdot (N^j_{\text{obs}}-N^j_{\text{bkg}}) \text{,} \nonumber
\end{equation}
where $j$ is the index for bins of observable $X$ at detector level and $i$ labels the bins at particle or parton level.  $\Delta X^i$ is the width of bin $i$, $N^j_{\text{obs}}$ is the number of observed events in data in bin $j$, $\mathcal{L}$ is the integrated luminosity, $N^j_{\text{bkg}}$ is the estimated number of background events in bin $j$, $R$ is the response matrix and $R^{-1}_{ij}$ symbolises the effective inversion of $R$ in the Bayesian unfolding. The acceptance correction $f^j_{\text{acc}}$ accounts for events that are outside the fiducial phase-space but pass the detector-level selection. The efficiency correction $\epsilon^i_{\text{eff}}$ corrects for events that are in the fiducial phase-space but are not reconstructed in the detector.  

\begin{figure}
\centering
\begin{tabular}{cc}
\subfloat[\label{fig:migration_em_parton_NoNW_dphi}]{\includegraphics[width=0.44\textwidth]{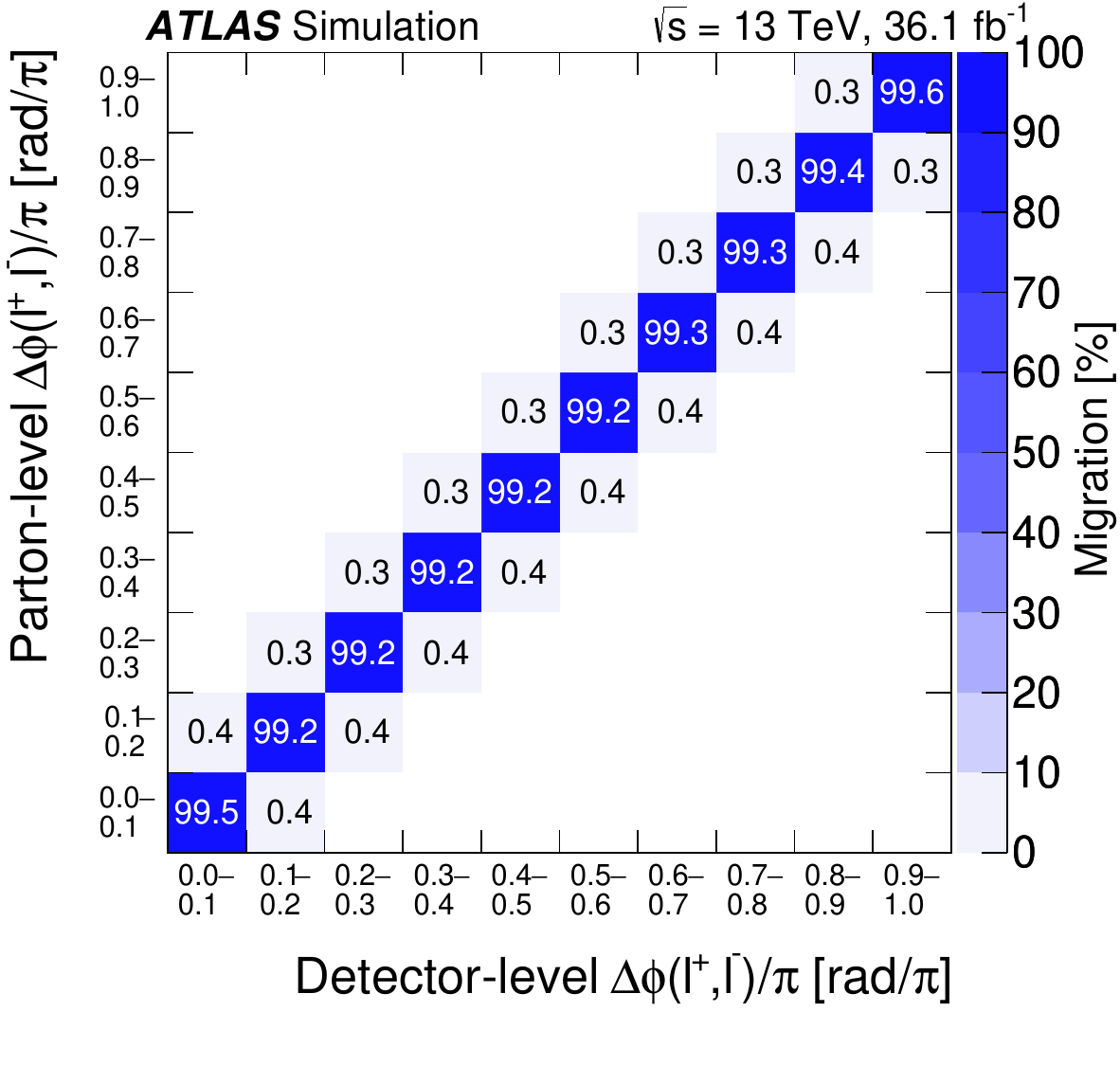}}&
\subfloat[\label{fig:migration_em_parton_NoNW_deta}]{\includegraphics[width=0.44\textwidth]{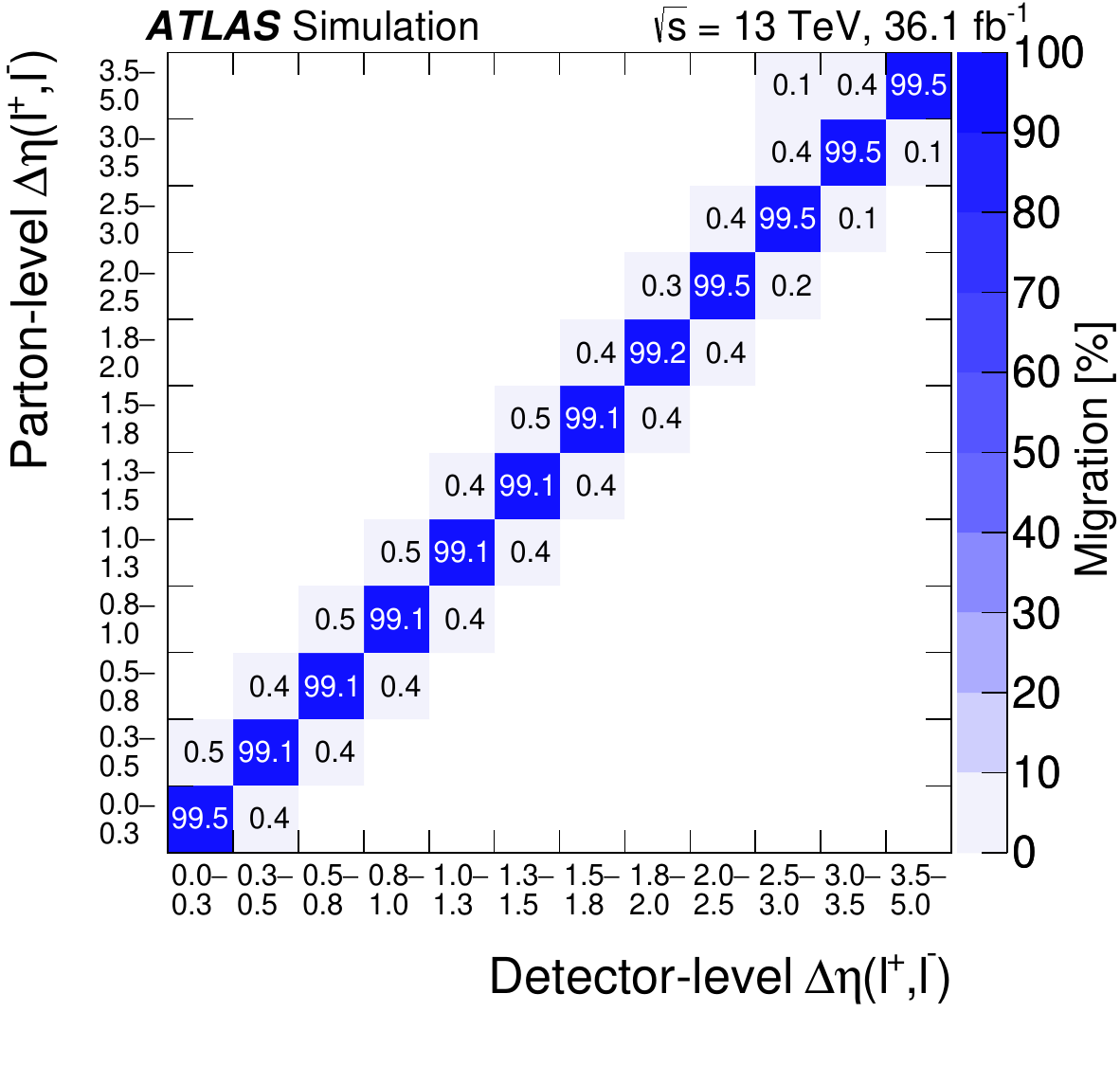}}\\
\subfloat[\label{fig:migration_em_particle_parton_NW_dphi}]{\includegraphics[width=0.44\textwidth]{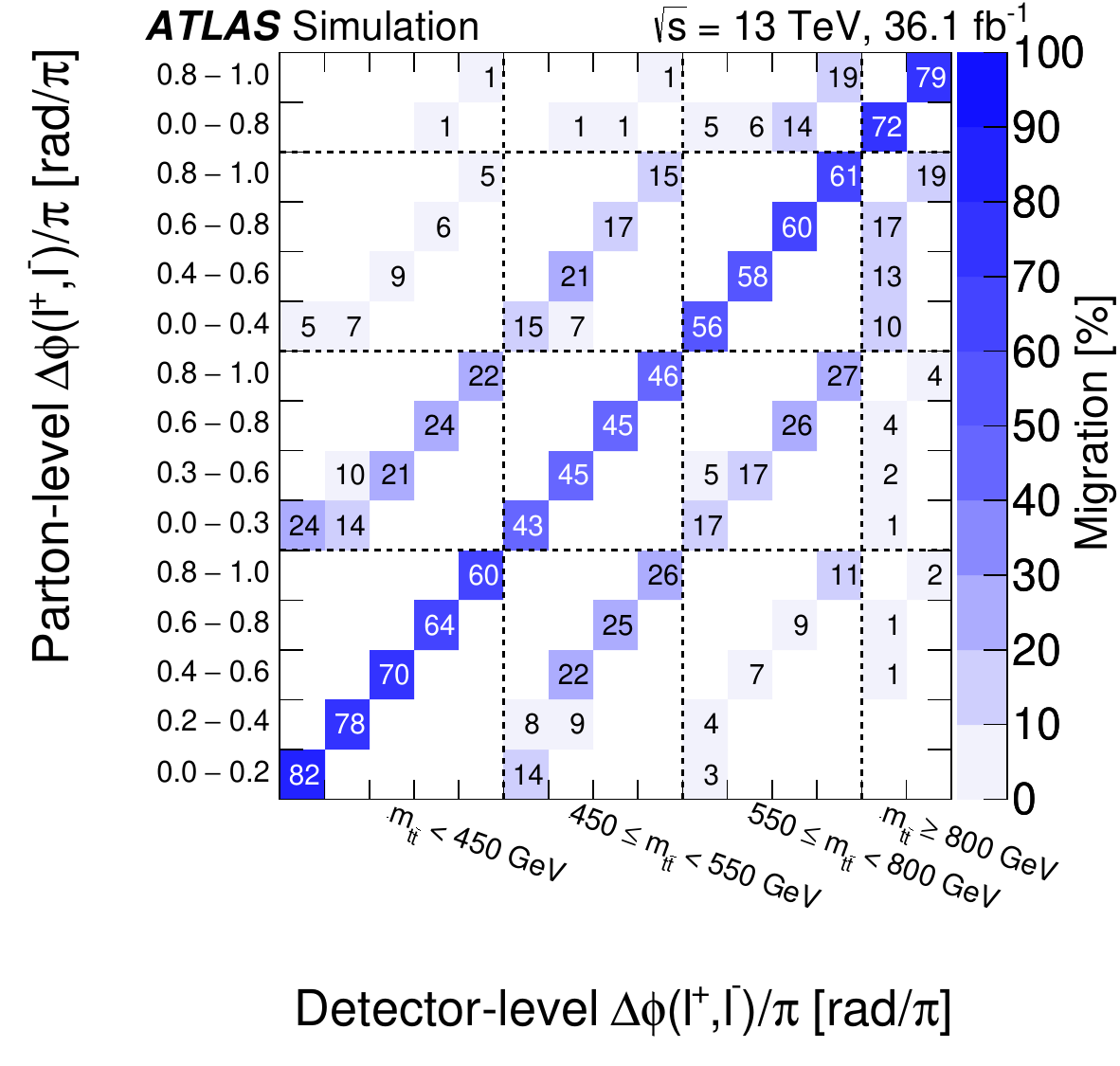}}\\
\end{tabular}
\caption{Parton-level response matrices, normalised by row and shown as percentages, for: (a) $\Delta\phi$, (b) $\Delta\eta$, and (c) $\Delta\phi$ as a function of $m_{t\bar{t}}$, after Neutrino Weighting. For (c), the binning on the horizontal and vertical axes is identical, with each invariant mass region subdivided into $\Delta\phi$ bins. The dotted lines separate different invariant mass regions, while the tick marks indicate the $\Delta\phi$ bins.}
\label{fig:migration_em_parton}
\end{figure}

The fiducial differential cross-sections are divided by the measured total cross-section, obtained by integrating over all bins in the differential distribution, in order to obtain the normalised differential cross-sections. The response matrix, $R$, describes the detector response and is determined by mapping the bin-to-bin migration of events from particle or parton level to detector level in the nominal \ttbar MC simulation. Figure~\ref{fig:migration_em_parton_NoNW_dphi} and~\ref{fig:migration_em_parton_NoNW_deta} illustrate the response matrices that are used for the single-differential $\Delta\phi$ and $\Delta\eta$ observables at parton level. Each response matrix is normalised such that the sum of entries in each row is equal to one. The values represent the fraction of events at either particle or parton level in bin $i$ that are reconstructed in bin $j$ at detector level. Figure~\ref{fig:migration_em_particle_parton_NW_dphi} shows the response matrix for the double-differential distribution of $\Delta\phi$ as a function of $m_{\ttbar}$ at parton level. The $\Delta\phi$ distributions for each $m_{\ttbar}$ region are concatenated into a single one-dimensional distribution, such that the response matrix takes into account the migrations between different $m_{\ttbar}$ regions. As can be observed in the figure, the $\Delta\phi$ observable is diagonal in each region, with the majority of the off-diagonal smearing occurring due to the resolution of the $m_{\ttbar}$ observable.

The binning for each observable is chosen in order to minimise the effect of statistical fluctuations in the data as well as in the alternative \ttbar samples which are used in the systematic prescription (and are a dominant source of systematic uncertainty), as well as to account for the experimental resolution. The size of the chosen bins is usually much larger than the detector resolution on the $\Delta\phi$ observable, which is illustrated by the highly diagonal response matrices in the inclusive selection. In contrast, the resolution of the reconstructed $m_{\ttbar}$ observable is significantly larger and so the binning here is chosen to be the smallest possible binning that reproduces the underlying truth-level distribution without bias, when measured using MC pseudo-experiments. 

The stability of the unfolding procedure is determined by constructing pseudo-data sets by randomly sampling events from the nominal \ttbar MC sample with approximately the same statistical power as the expected data. Pull tests are performed as part of the binning optimisation and are therefore always successful for the chosen observable bins. In addition, the unfolding procedure is tested to see how it responds to various \emph{stresses} introduced into the pseudo-data. Three such stresses are investigated: introducing linear slopes in the observables, the difference between the spin correlated and uncorrelated MC samples, and the observed difference between data and the expectation at detector level. In all cases, the unfolding procedure is able to correct the pseudo-data back to their underlying truth spectra and so a systematic uncertainty for the unfolding procedure is not included.

The number of iterations used in the iterative Bayesian unfolding is also optimised using pseudo-experiments. Iterations are performed until the $\chi^2$ per degree-of-freedom, calculated by comparing the unfolded pseudo-data to the corresponding generator-level distribution for that pseudo-data set, is less than or equal to unity. For the inclusive observables ($\Delta\phi$ and $\Delta\eta$), the optimal number of iterations is determined to be two, whereas for the reconstructed observable ($\Delta\phi$ in bins of $m_{t\bar{t}}$), the optimal number of iterations is determined to be four. All distributions are unfolded to the particle level and to the parton level.

\FloatBarrier

%% file: sections/systematicsUncertainties.tex
\section{Systematic uncertainties}
\label{sec:systematics}

The measured differential cross-sections are affected by systematic uncertainties arising from detector response, signal modelling, and background modelling. The contributions from various sources of uncertainty are described in this section. These individual systematic uncertainties are summed in quadrature to obtain the total systematic uncertainty, and the overall uncertainty is calculated by summing the systematic and statistical uncertainties in quadrature.

\subsection{Signal modelling uncertainties}

The following four systematic uncertainties related to the modelling of the \ttbar\ system in the MC generators are considered: the choice of matrix-element generator, the hadronisation and parton-shower model, the amount of initial- and final-state radiation, and the choice of PDF set.  In each case (except for the PDF uncertainty), alternative MC samples are unfolded with the nominal \ttbar MC response and the difference to their generator-level spectra is taken as the systematic uncertainty.  A fast detector simulation (described in Section~\ref{sec:data}) is used for each of the alternative models and for the response matrix, rather than the full detector simulation used in the nominal unfolding procedure. In most cases, the resulting systematic shift is used to define a symmetric uncertainty, where deviations from the generator-level spectra are also considered to be mirrored in the opposite direction, resulting in equal and opposite symmetric uncertainties (called symmetrising). 

The choice of NLO ME generator affects the invariant mass of the simulated \ttbar events, the observables themselves, and the reconstruction efficiencies. To estimate this uncertainty, \MGMCatNLO~(with \PYTHIA\!8 for the parton-shower simulation) is used, applying the nominal unfolding procedure based on the \POWHEG-Box+\PYTHIA\!8 \ttbar sample. The resulting uncertainty is symmetrised.

To evaluate the uncertainty arising from the choice of parton-shower algorithm and the hadronisation model, the alternative sample generated with \mbox{\POWHEG-Box}~+~\HERWIG\!7 is unfolded with the nominal \ttbar MC response. The resulting uncertainty is symmetrised.

The uncertainty arising from initial- and final-state radiation is evaluated using the reduced radiation sample of \POWHEG-Box~+~\mbox{\PYTHIA\!8}, and is again symmetrised. An enhanced radiation sample was also investigated as this has been used in previous similar analyses. However, it was found to markedly disagree with the data and is therefore not used here.

The uncertainty due to the choice of PDF set is evaluated using the PDF4LHC15 prescription~\cite{Butterworth:2015oua}, utilising 30 eigenvector shifts derived from fits to multiple NLO PDF sets. Each shift is evaluated for each bin added in quadrature and the resulting uncertainty in each bin is symmetrised.

\subsection{Background modelling uncertainties}
The uncertainties in the background processes are assessed by repeating the full analysis using pseudo-data sets and by varying the background predictions by one standard deviation of their nominal values. The difference between the nominal pseudo-data set result and the shifted result is taken as the systematic uncertainty, then the separate background uncertainties are combined in quadrature.

Each background prediction has an uncertainty associated with its theoretical cross-section. The cross-section for the $tW$ process is varied by $\pm$5.3\%~\cite{Kidonakis:2010ux}, the diboson cross-section is varied by $\pm$6\%, and the Drell--Yan $Z/\gamma^*\rightarrow\tau^+\tau^-$ background cross-section is varied by $\pm$5\% based on studies of different MC generators. 
Uncertainties on the remaining  SM backgrounds are taken to be $13\%$ for $\ttbar V$~\cite{Alwall:2014hca,Frixione:2015zaa}, $^{+6.8}_{-9.9}\%$ for $\ttbar H$~\cite{HiggsXS}, $^{+10}_{-28}\%$ for $tWZ$ and $\pm50\%$ for $tZ$, $\ttbar WW$ and $\ttbar\ttbar$~\cite{TOPQ-2015-22}.

An additional scaling factor and uncertainty of $1.07\pm0.12$ is assigned to the $Z/\gamma^*$ background, based on a comparison of data and MC simulation in a region enriched in $Z\rightarrow\ell^{+}\ell^{-}$ decays in association with $b$-jets.

A 40\% uncertainty is assigned to the normalisation of the fake-lepton background based on comparisons between data and MC simulation in a fake-dominated control region, which is selected in the same way as the \ttbar signal region but the leptons are required to have same-sign electric charges. An additional uncertainty is included, to account for slight differences in shapes between the data-driven and MC estimates in $\Delta\phi(\ell^+,\ell^-)$ and $\Delta\eta(\ell^+,\ell^-)$.

An additional uncertainty is evaluated for the $tW$ process by replacing the nominal DR sample with a DS sample, as discussed in Section~\ref{sec:data}, and taking the difference between
the two as the systematic uncertainty. Other background process uncertainties are found to be insignificant and are not discussed further.

\subsection{Detector modelling uncertainties}

Systematic uncertainties due to the modelling of the detector response affect the signal reconstruction efficiency, the unfolding procedure, and the background estimation. In order to evaluate their impact, the full analysis is repeated with variations of the detector modelling and the difference between the nominal and the shifted results is taken as the systematic uncertainty. 

The uncertainties due to lepton isolation, trigger, identification, and reconstruction requirements are evaluated in data using a tag-and-probe method in events with a leptonically decaying $Z$ boson~\cite{PERF-2017-01,PERF-2015-10}. 

The jet energy scale uncertainty is assessed in data~\cite{PERF-2016-04}, using simulation-based corrections and in situ techniques based on jets, photons and $Z$ bosons.  A 21-component breakdown of the uncertainty is used, with contributions from pile-up, jet flavour composition, single-particle response, and punch-through. 
The jet energy resolution uncertainty is parametrised as a function of jet \pT and rapidity~\cite{ATL-PHYS-PUB-2015-015}.  

Uncertainties related to the $b$-jet tagging procedure, summarised under ``$b$-tagging,'' are determined separately for $b$-jets, $c$-jets and light-jets using a 27-component breakdown (6 for $b$-jets, 3 for $c$-jets, 16 for light-jets, and two extrapolation uncertainties)~\cite{PERF-2016-05, ATLAS-CONF-2018-001, ATLAS-CONF-2018-006}.  These uncertainties account for differences between data and simulation.

The systematic uncertainty due to the track-based terms (i.e.\ those tracks not associated with other reconstructed objects such as leptons and jets) used in the calculation of \met\ is evaluated by comparing the \met in $Z\rightarrow\mu\mu$ events, which do not contain prompt neutrinos from the hard process, using different generators. Uncertainties associated with energy scales and resolutions of leptons and jets are propagated to the \met calculation~\cite{PERF-2016-07}.

The uncertainty in the combined 2015+2016 integrated luminosity is 2.1\%. It is derived, following a methodology similar to that detailed in Ref.~\cite{DAPR-2013-01}, and using the LUCID-2 detector for the baseline luminosity measurements \cite{LUCID2}, from calibration of the luminosity scale using $x-y$ beam--separation scans. The uncertainty in the reweighting of the MC pile-up distribution to match the data is evaluated according to the uncertainty on the average number of interactions per bunch crossing.

\FloatBarrier

%% file: sections/results_diff.tex
\section{Differential cross-section results}
\label{sec:results_diff}

\begin{figure}
\centering
\begin{tabular}{cc}
\subfloat[]{\includegraphics[width=0.45\textwidth]{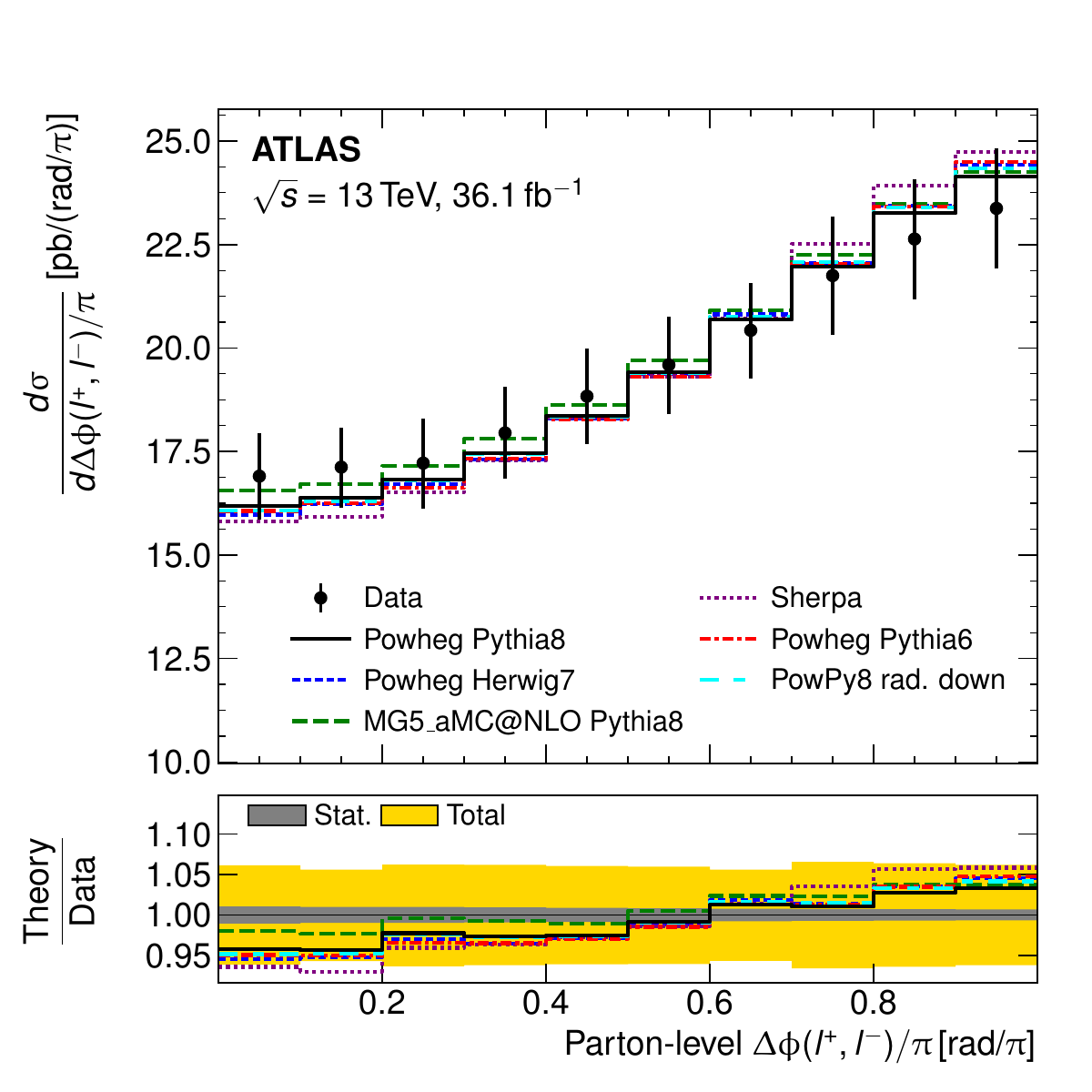}}&
\subfloat[]{\includegraphics[width=0.45\textwidth]{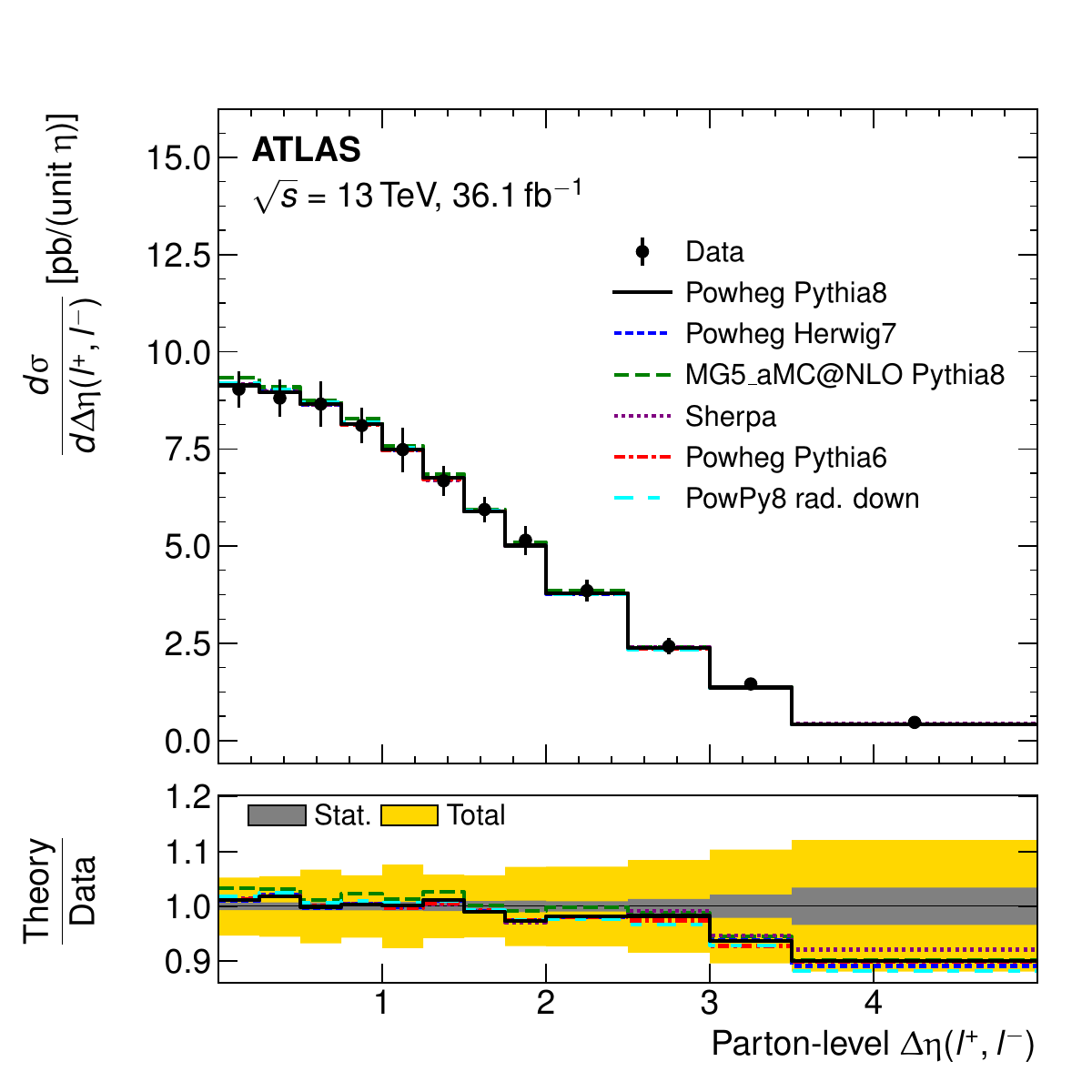}}\\
\subfloat[]{\includegraphics[width=0.45\textwidth]{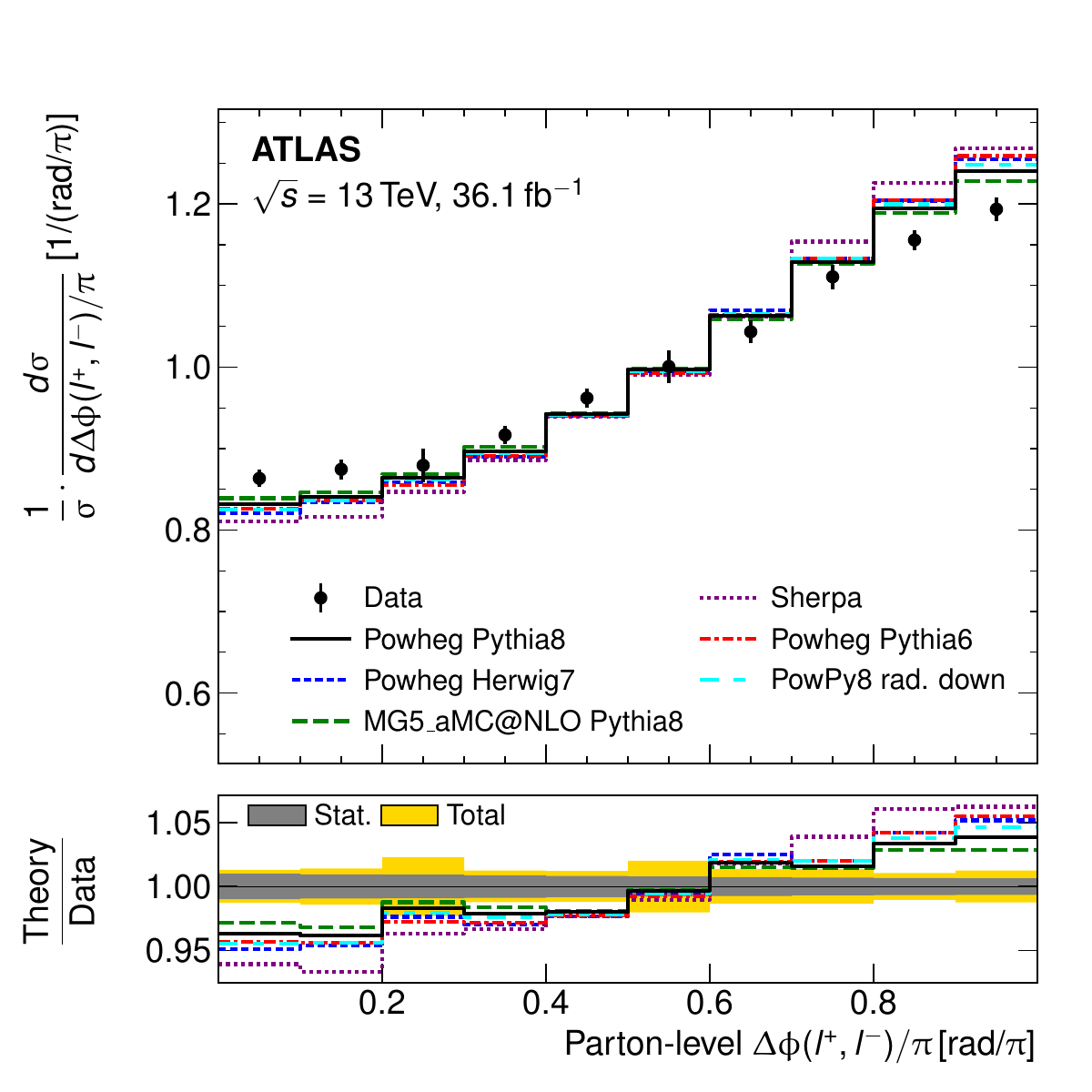}}&
\subfloat[]{\includegraphics[width=0.45\textwidth]{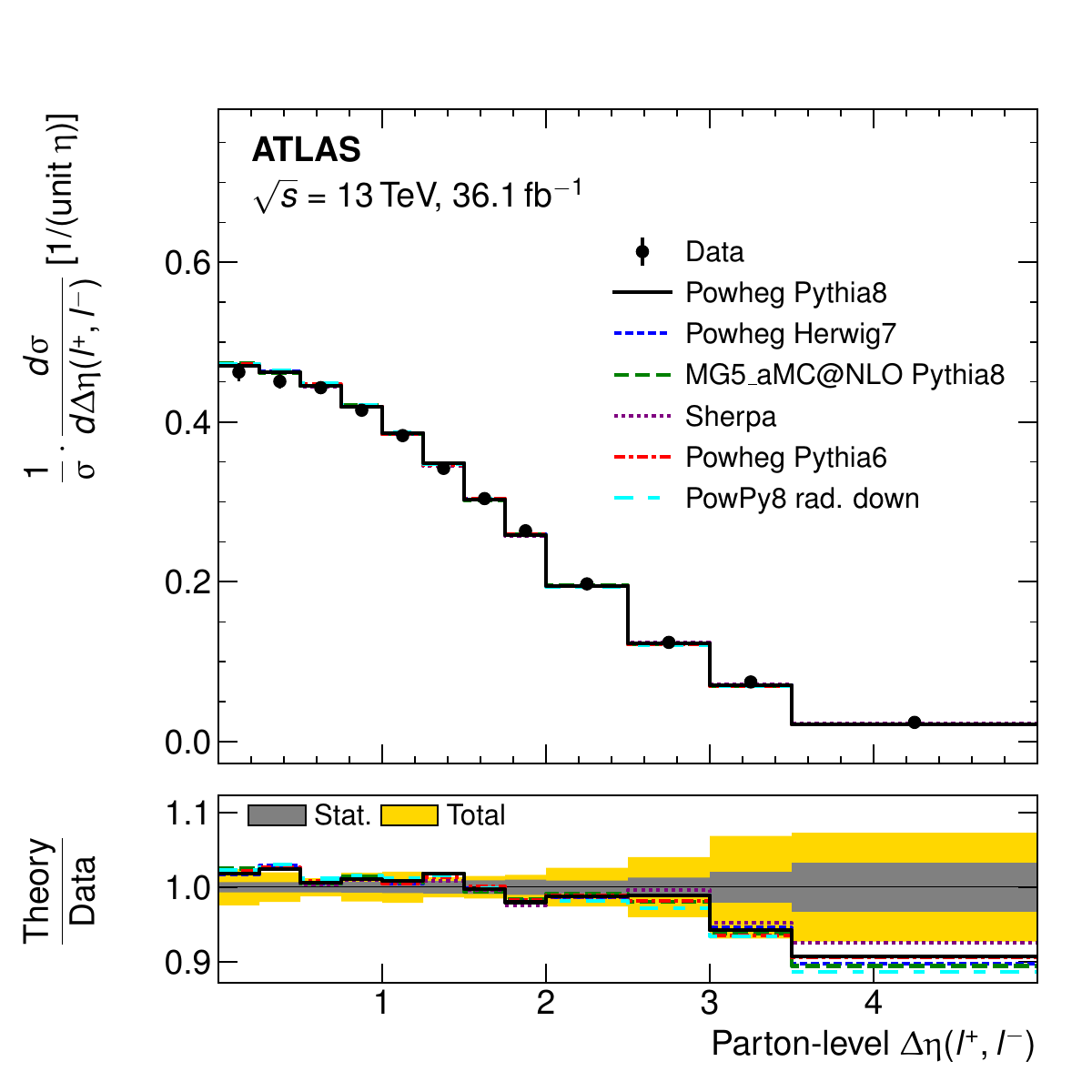}}\\
\end{tabular}
\caption{The parton-level differential cross-sections compared to predictions from \POWHEG, \MGMCatNLO\ and \SHERPA:  \emph{(top)}, absolute (a) $\Delta \phi$ and (b) $\Delta \eta$
 and \emph{(bottom)}, normalised (c) $\Delta \phi$ and (d) $\Delta \eta$, using the inclusive selection.}
\label{fig:results_parton_nlo_NoNW}
\end{figure}

\begin{figure}
\centering
\begin{tabular}{cc}
\subfloat[]{\includegraphics[width=0.45\textwidth]{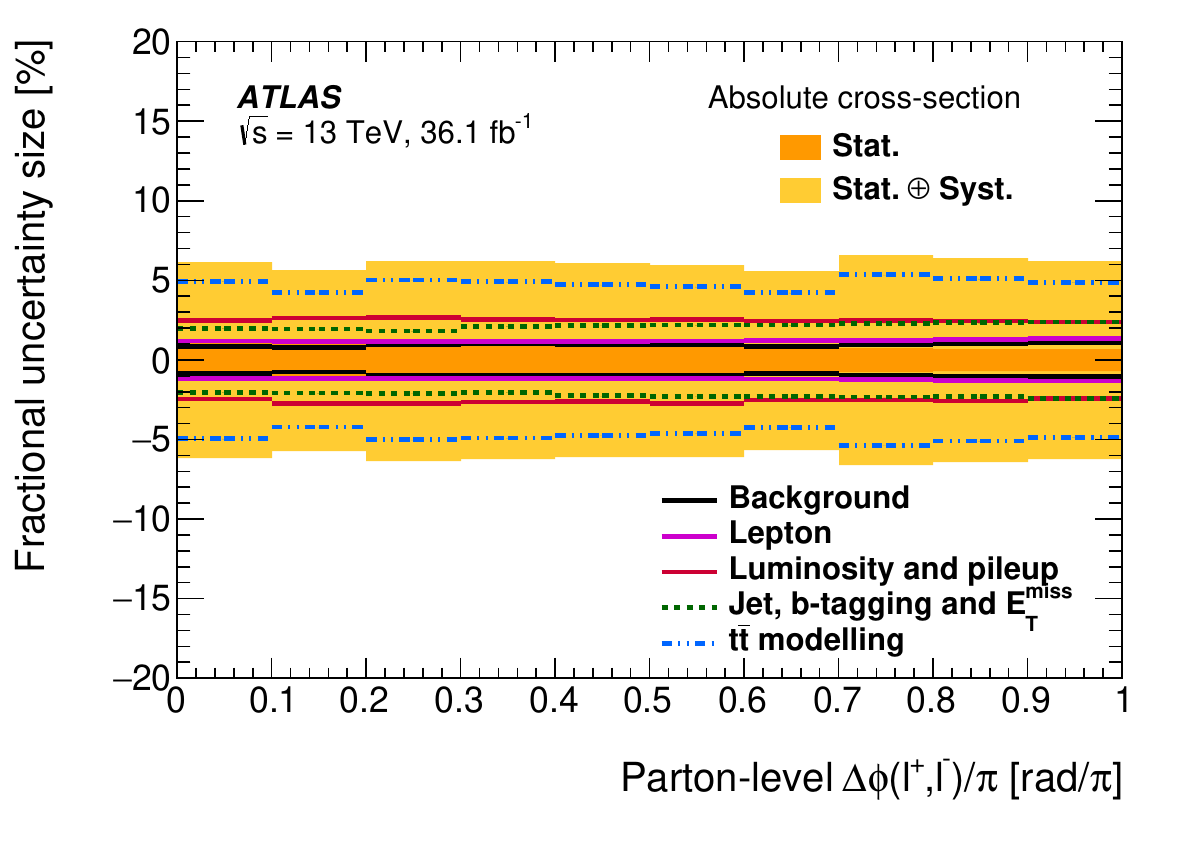}}&
\subfloat[]{\includegraphics[width=0.45\textwidth]{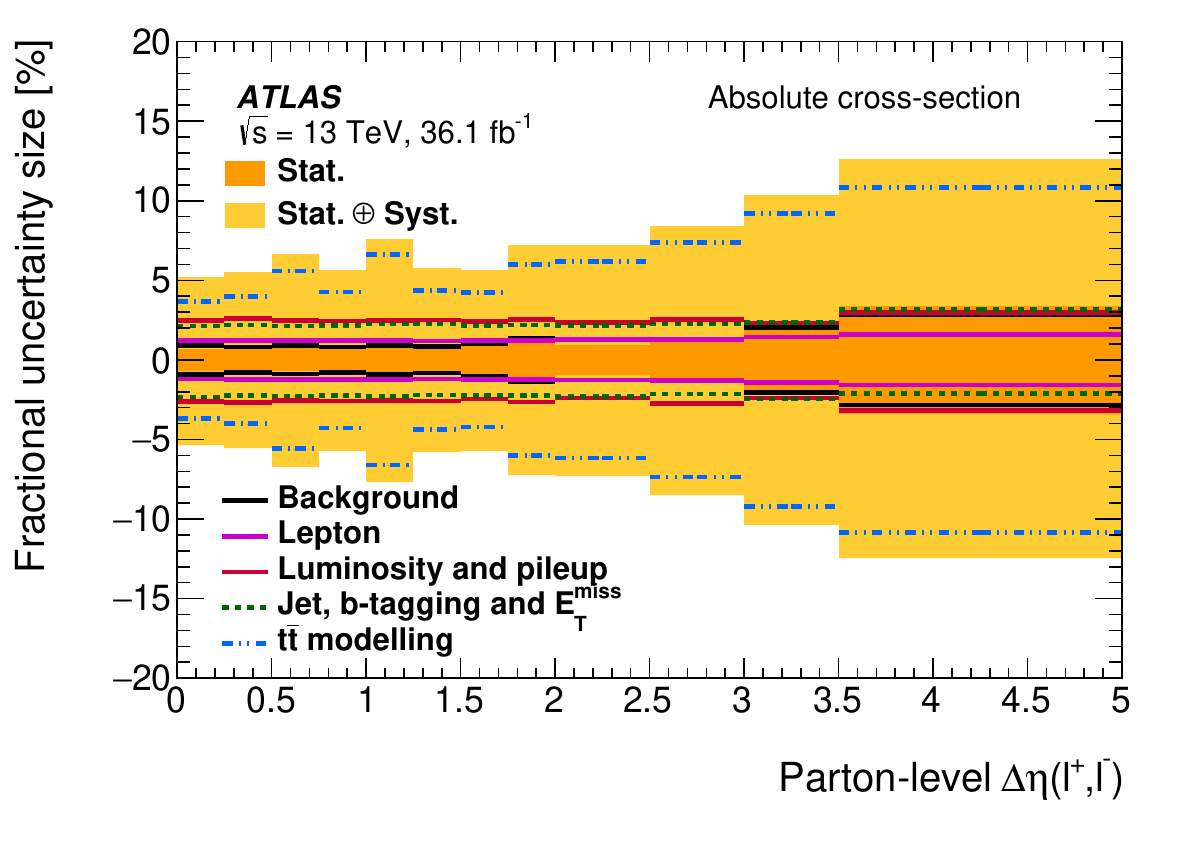}}\\
\subfloat[]{\includegraphics[width=0.45\textwidth]{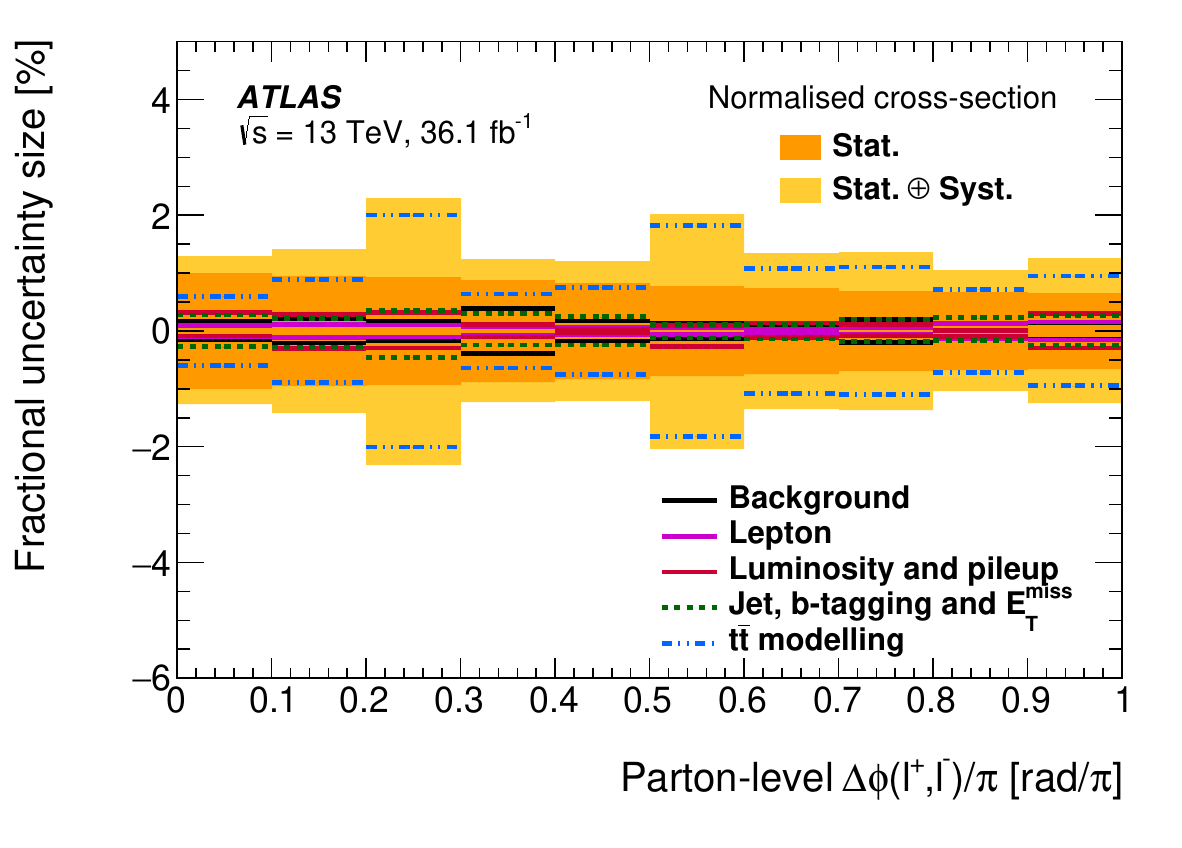}}&
\subfloat[]{\includegraphics[width=0.45\textwidth]{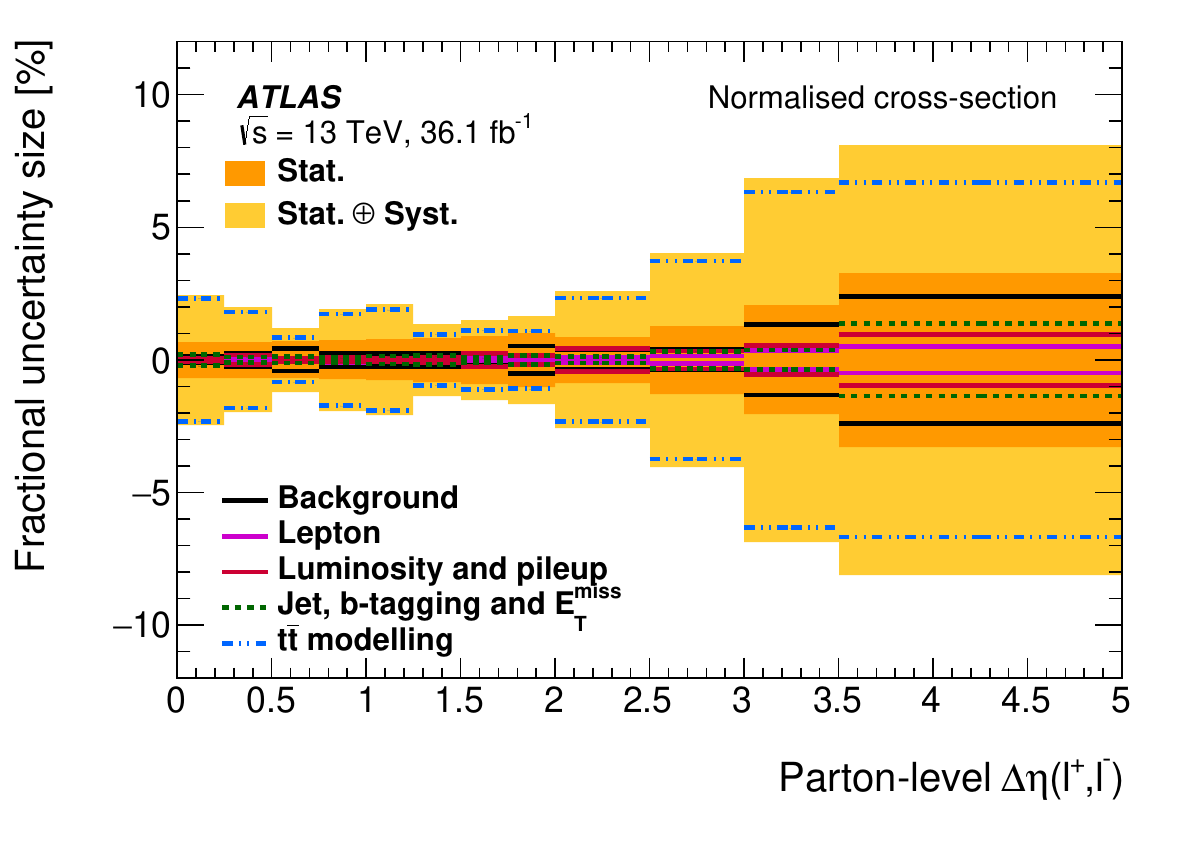}}\\
\end{tabular}
\caption{Systematic uncertainties for the parton-level differential cross-sections:   \emph{(top)}, absolute (a) $\Delta \phi$ and (b) $\Delta \eta$
 and \emph{(bottom)}, normalised (c) $\Delta \phi$ and (d) $\Delta \eta$.  The \ttbar modelling uncertainties refer to the contributions from the NLO matrix-element generator (``Generator''), the PS algorithm (``Shower'') and the variation of initial- and final-state radiation (``Radiation'').}
\label{fig:systematics_parton_nlo_NoNW}
\end{figure}

The absolute and normalised parton-level cross-sections for $\Delta\phi$ and $\Delta\eta$ are presented in Table~\ref{tab:results_dphi_ll_parton} and Table~\ref{tab:results_deta_ll_parton}. These results are compared to several NLO MC generators interfaced to parton showers (described in Section~\ref{sec:data}) in Figure~\ref{fig:results_parton_nlo_NoNW} and the breakdown of the contributions to the systematic uncertainties are shown in Figure~\ref{fig:systematics_parton_nlo_NoNW}. In each case, the total generator cross-section was normalised to the NNLO values described in Section~\ref{sec:data}.

All uncertainties that are normalisation effects but which do not cause large changes in the shape of the observable (luminosity, for example) cancel when performing the normalised cross-sections. Jet and pile-up effects are also significant, but only in the absolute cross-sections. Overall, reasonable agreement is observed in the inclusive cross-section between the data and MC predictions but significant shape effects are apparent, particularly in the normalised observables where the uncertainties are small. Ignoring the differences in the absolute fiducial cross-sections between different MC generators, the shapes predicted by different generators are fairly consistent, except perhaps at very high $\Delta\eta$. In the $\Delta\phi$ observable, an obvious trend is observed, with the data tending to be higher than the expectation at low $\Delta\phi$ and lower than the expectation at high $\Delta\phi$. For $\Delta\eta$, the data and expectation agree well at low values, even in the normalised cross-sections, but there is a slight tension at higher values.

\input{tables/results_dphi_ll_parton.tex}
\input{tables/results_deta_ll_parton.tex}

\begin{figure}
\centering
\begin{tabular}{cc}
\subfloat[]{\includegraphics[width=0.45\textwidth]{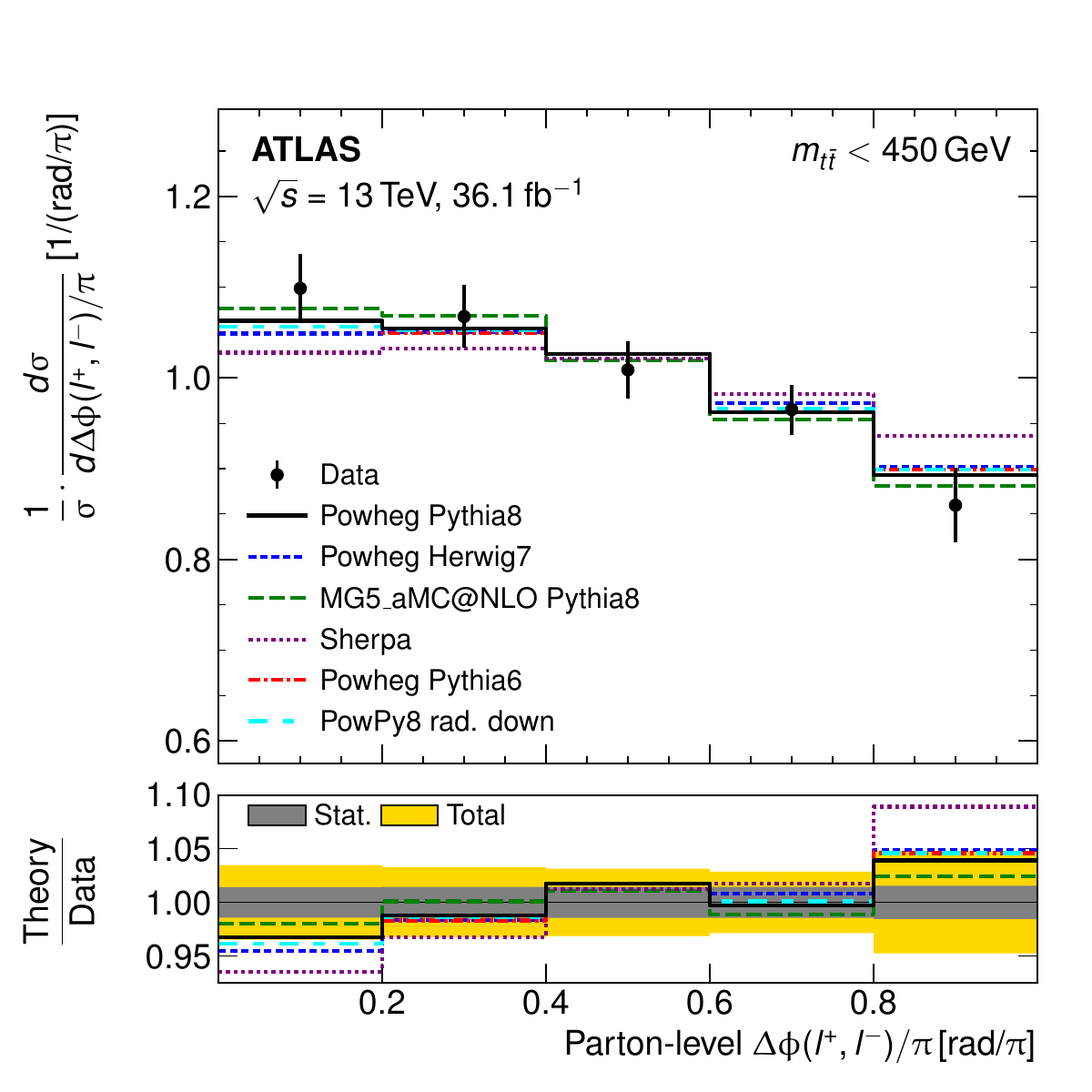}}&
\subfloat[]{\includegraphics[width=0.45\textwidth]{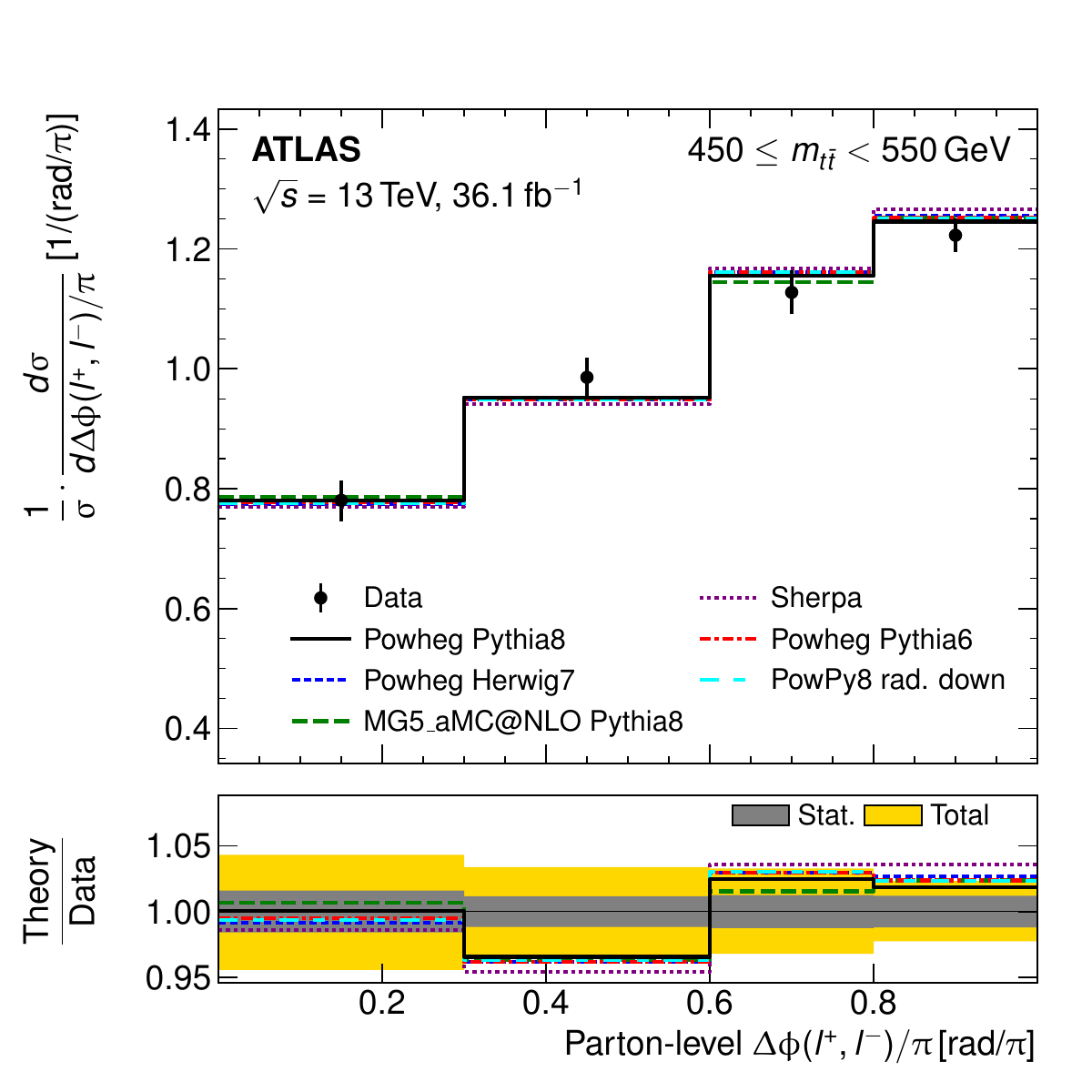}}\\
\subfloat[]{\includegraphics[width=0.45\textwidth]{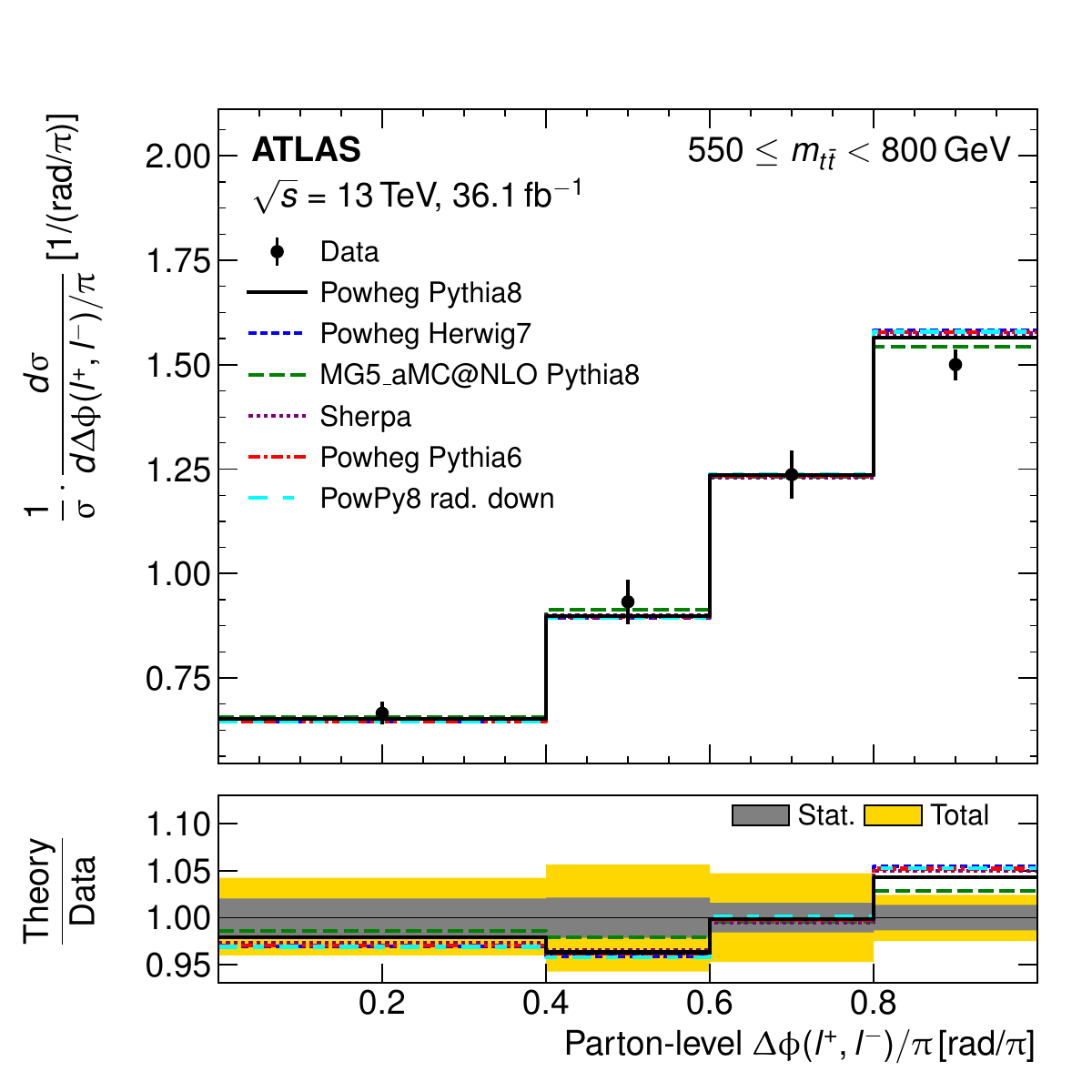}}&
\subfloat[]{\includegraphics[width=0.45\textwidth]{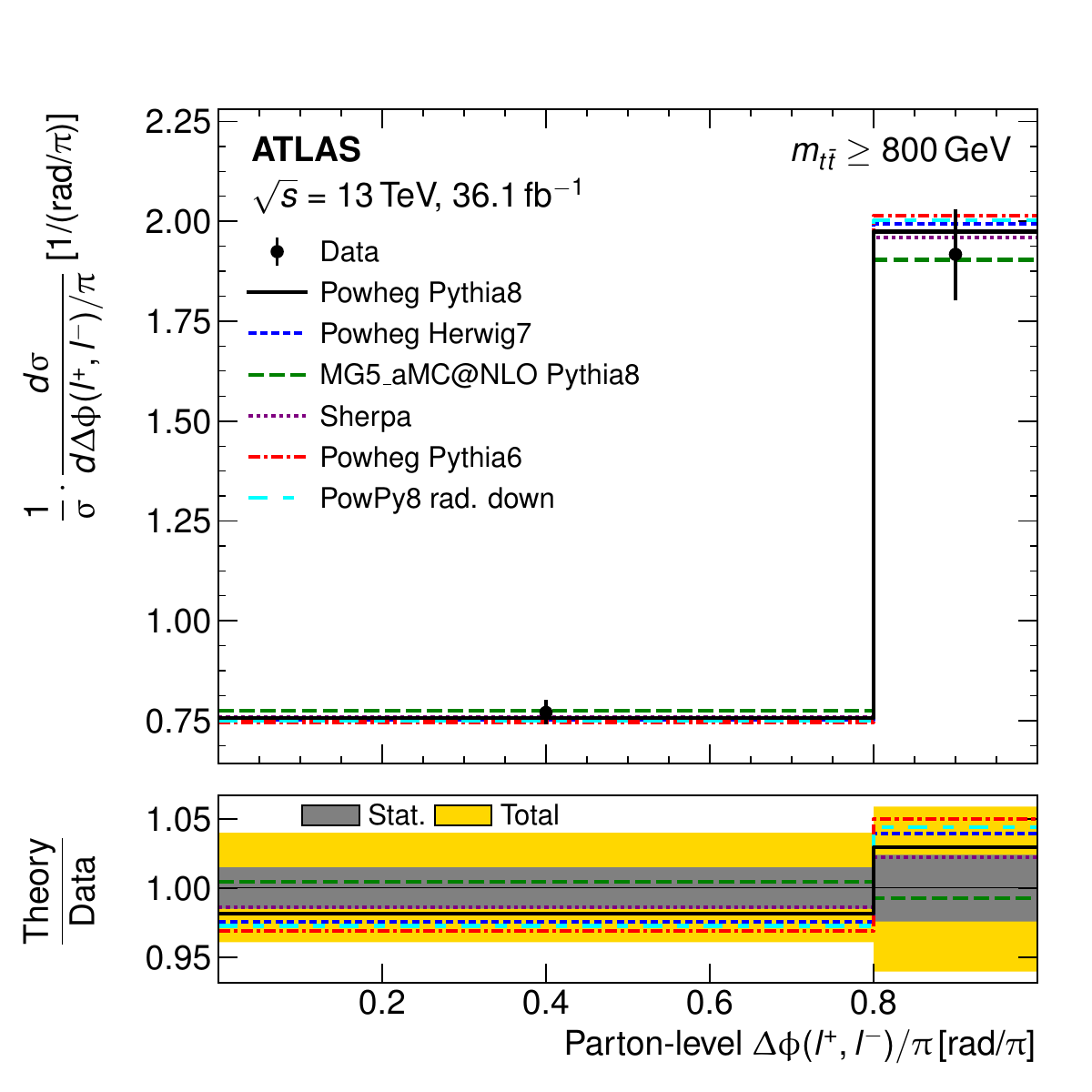}}\\
\end{tabular}
\caption{The normalised parton-level differential cross-sections in four \ttbar\ mass bins compared to
  predictions from \POWHEG, \MGMCatNLO\ and \SHERPA, using the reconstructed selection: 
(a) $m_{\ttbar}<450$~\GeV,
(b) $450\leq m_{\ttbar}<550$~\GeV,
(c) $550\leq m_{\ttbar}<800$~\GeV, and
(d) $m_{\ttbar}\geq800$~\GeV. 
Each differential distribution is normalised to the integrated cross-section within the individual $m_{\ttbar}$ region.}
\label{fig:results_parton_nlo}
\end{figure}

\begin{figure}
\centering
\begin{tabular}{cc}
\subfloat[]{\includegraphics[width=0.45\textwidth]{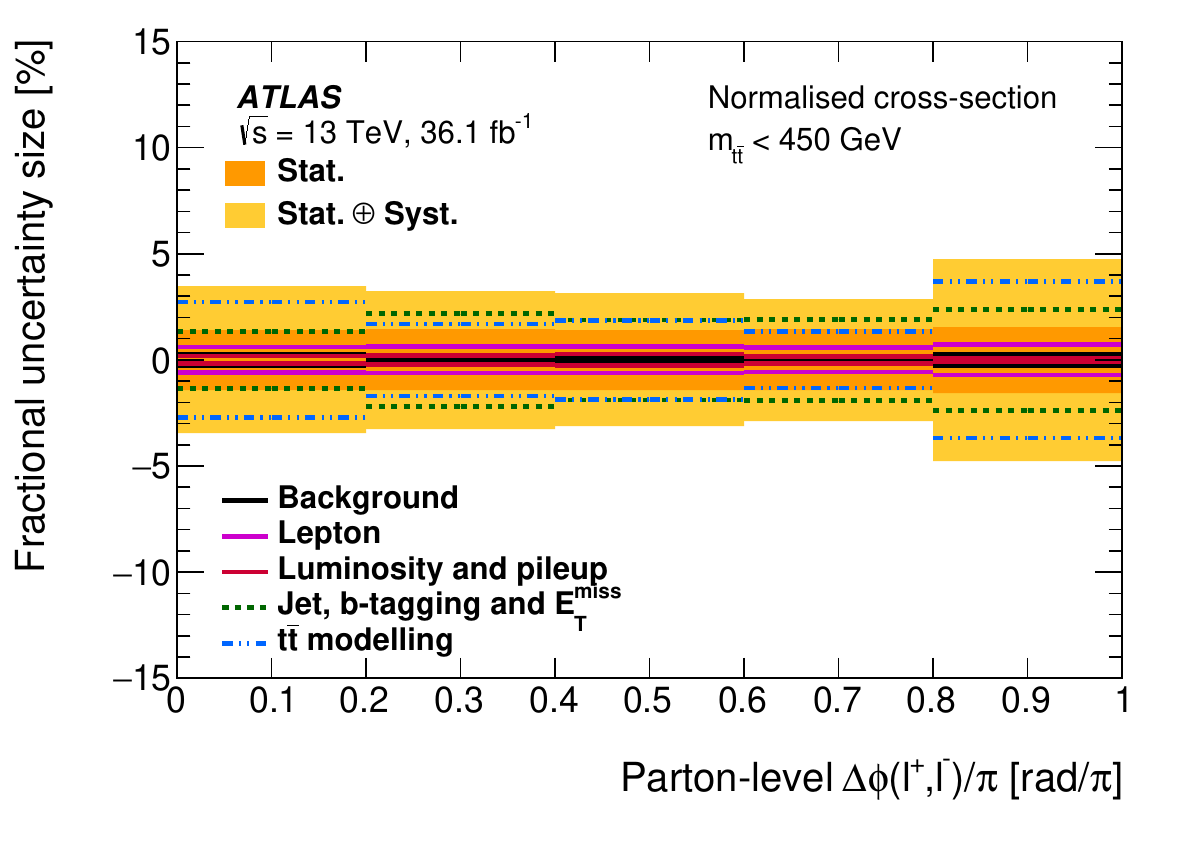}}&
\subfloat[]{\includegraphics[width=0.45\textwidth]{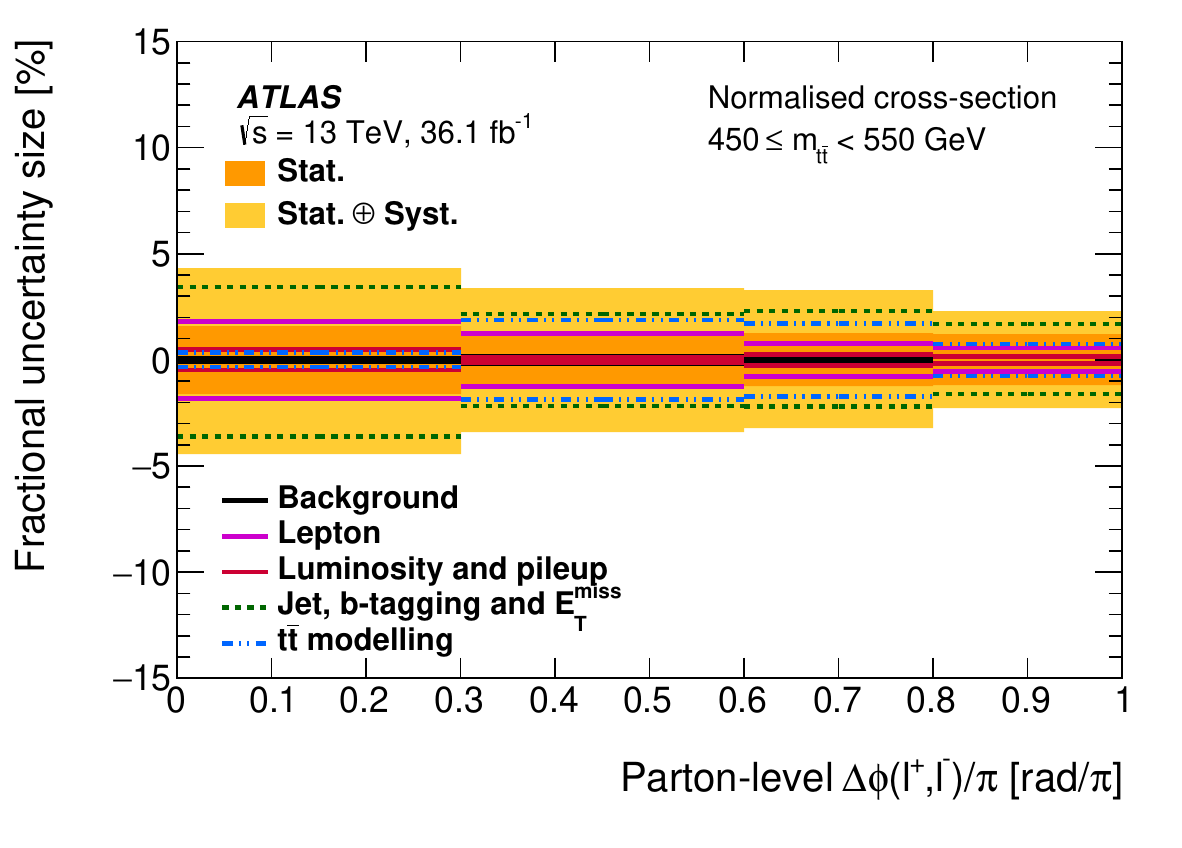}}\\
\subfloat[]{\includegraphics[width=0.45\textwidth]{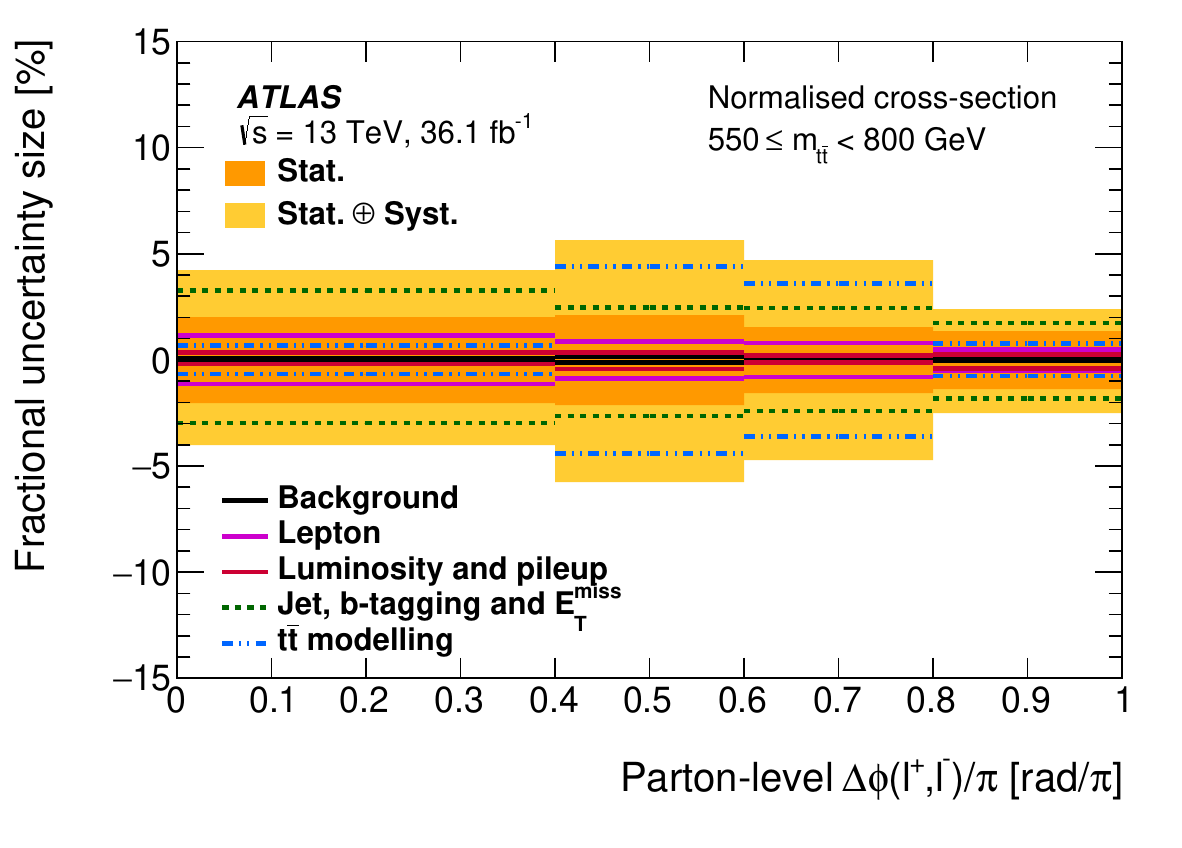}}&
\subfloat[]{\includegraphics[width=0.45\textwidth]{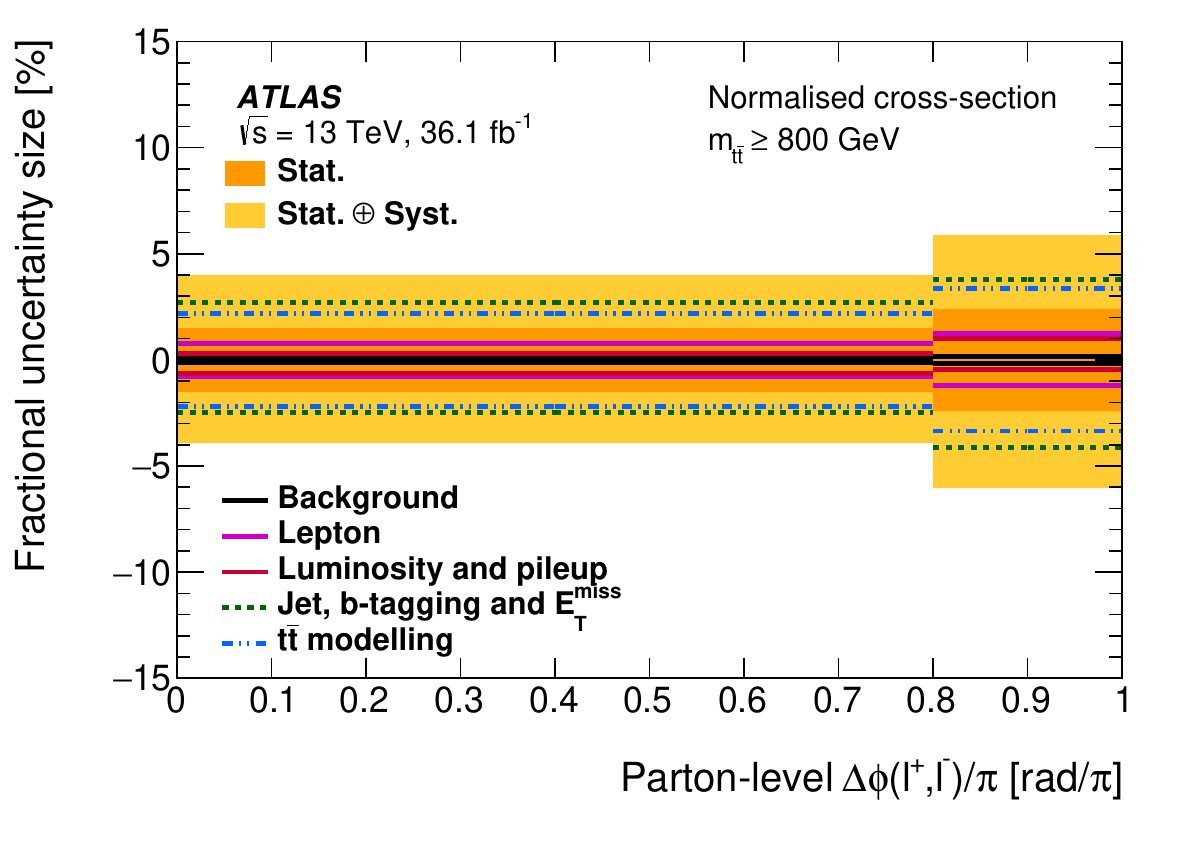}}\\
\end{tabular}
\caption{Systematic uncertainties for the normalised parton-level differential cross-sections in four \ttbar\ mass bins:
(a) $m_{\ttbar}<450$~\GeV,
(b) $450\leq m_{\ttbar}<550$~\GeV,
(c) $550\leq m_{\ttbar}<800$~\GeV, and
(d) $m_{\ttbar}\geq800$~\GeV.
  The \ttbar modelling uncertainties refer to the contributions from the NLO matrix-element generator (``Generator''), the PS algorithm (``Shower'') and the variation of initial- and final-state radiation (``Radiation'').}
\label{fig:systematics_parton_nlo}
\end{figure}

The unfolded, normalised parton-level cross-sections for $\Delta\phi$ in four \ttbar\ invariant mass bins are shown in Table~\ref{tab:results_dphi_ll_vs_ttbar_m_parton}.  They are compared with different NLO ME generators and parton showers in Figure~\ref{fig:results_parton_nlo} and the systematic uncertainties are illustrated in Figure~\ref{fig:systematics_parton_nlo}.
Each differential cross-section is normalised within its $m_{\ttbar}$ range. In all regions of invariant mass, the systematic uncertainties arising from the modelling of the \ttbar and jets are dominant, with statistical uncertainties on the data becoming more important at higher values of invariant mass. In the lowest region of invariant mass, the various NLO predictions differ from each other and from the data. In the other regions of $m_{\ttbar}$ the differences are less pronounced and agree within the uncertainties. 

\input{tables/results_dphi_ll_vs_ttbar_m_parton.tex}

\begin{figure}
\centering
\begin{tabular}{cc}
\subfloat[\label{fig:results_nlo_NoNW_dphi_particle}]{
\includegraphics[width=0.45\textwidth]{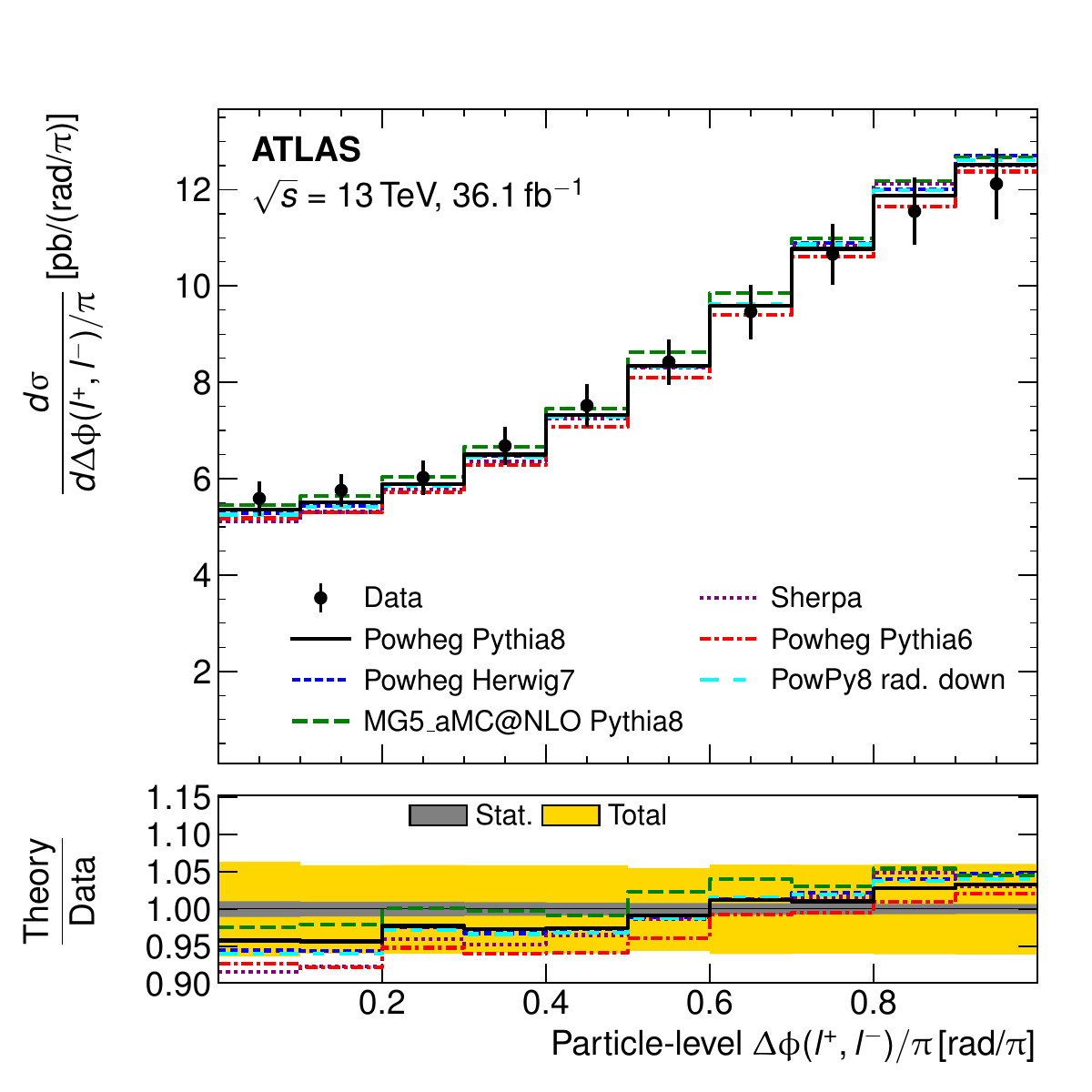}}&
\subfloat[\label{fig:results_nlo_NoNW_deta_particle}]{
\includegraphics[width=0.45\textwidth]{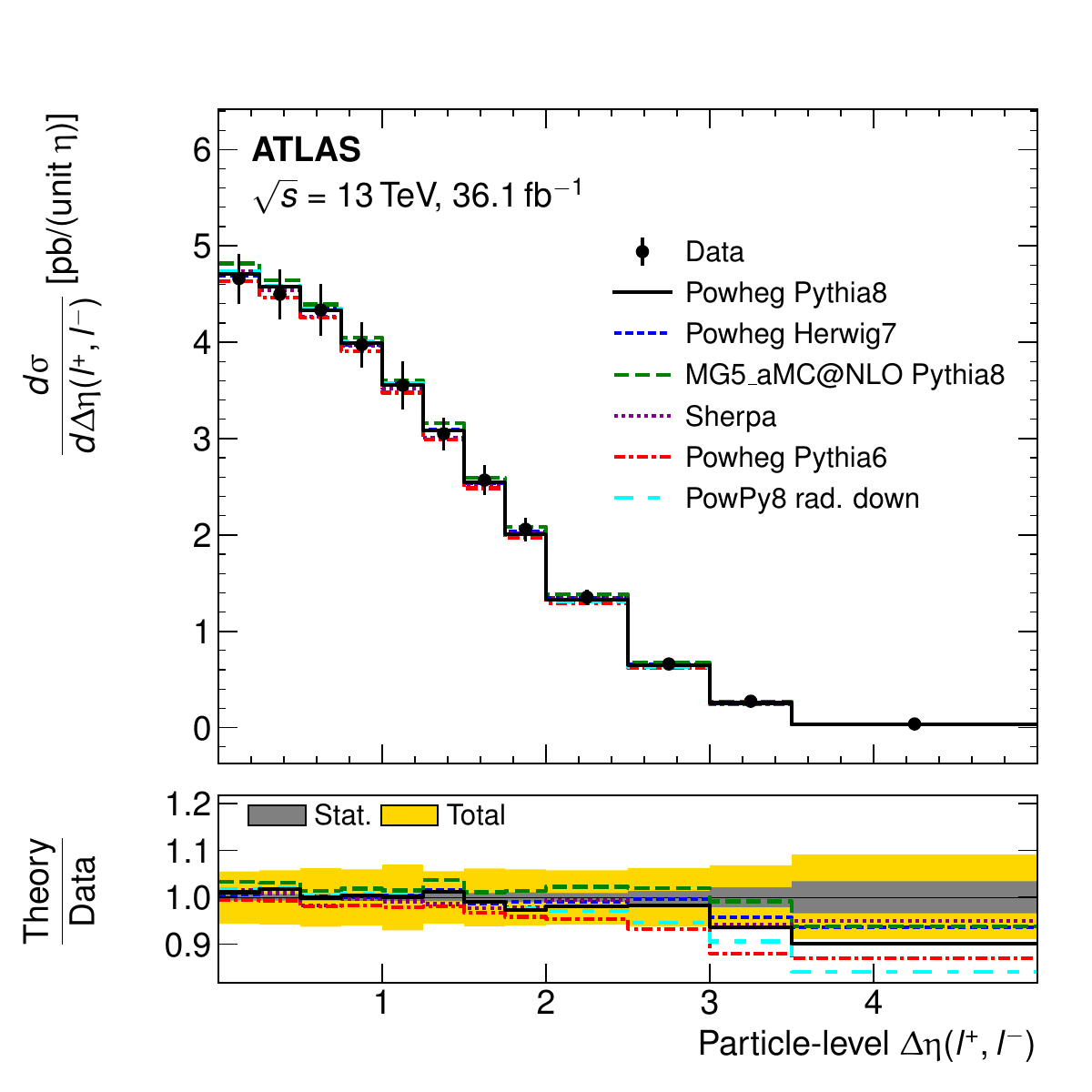}}\\
\subfloat[\label{fig:results_nlo_NoNW_dphi_particle_norm}]{
\includegraphics[width=0.45\textwidth]{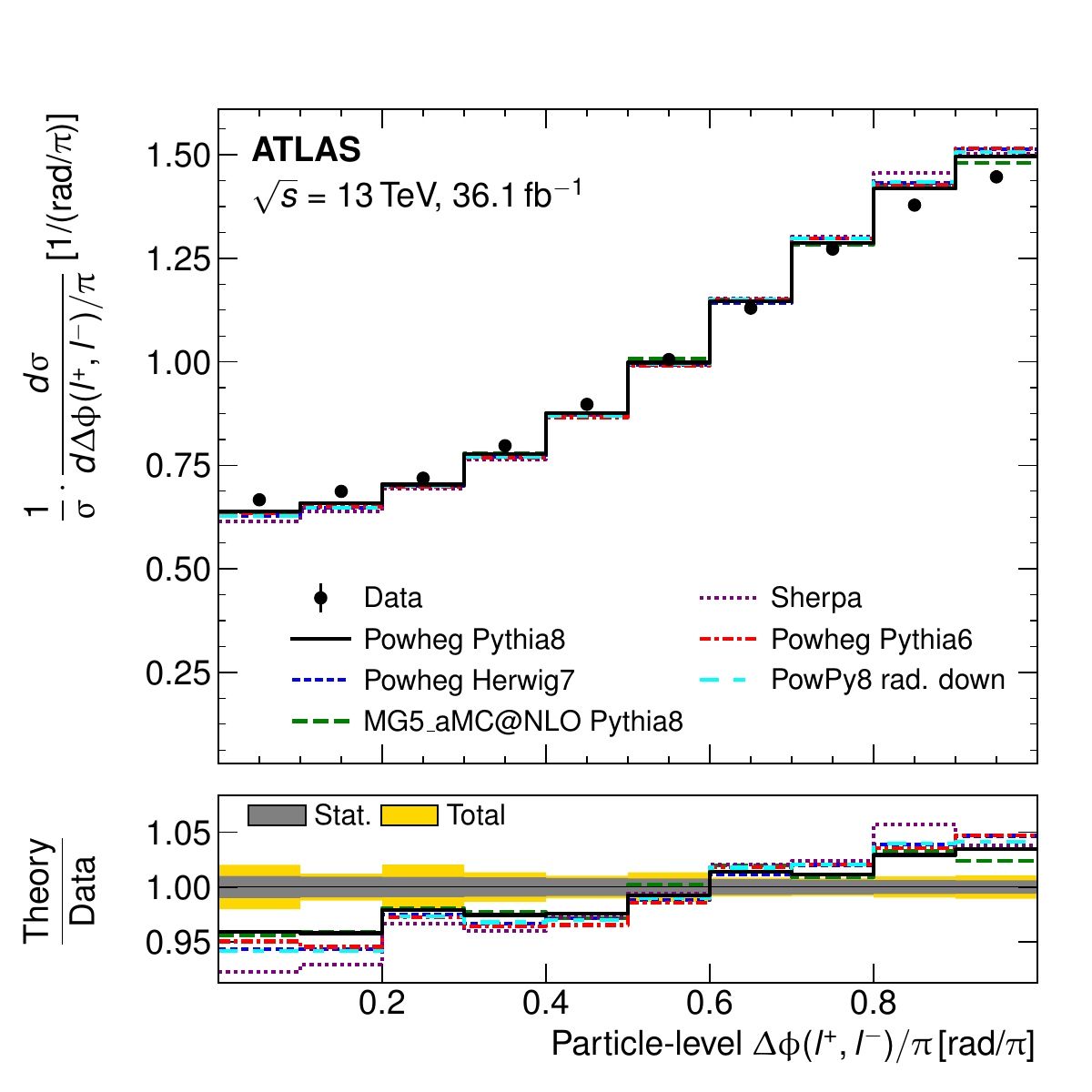}}&
\subfloat[\label{fig:results_nlo_NoNW_deta_particle_norm}]{
\includegraphics[width=0.45\textwidth]{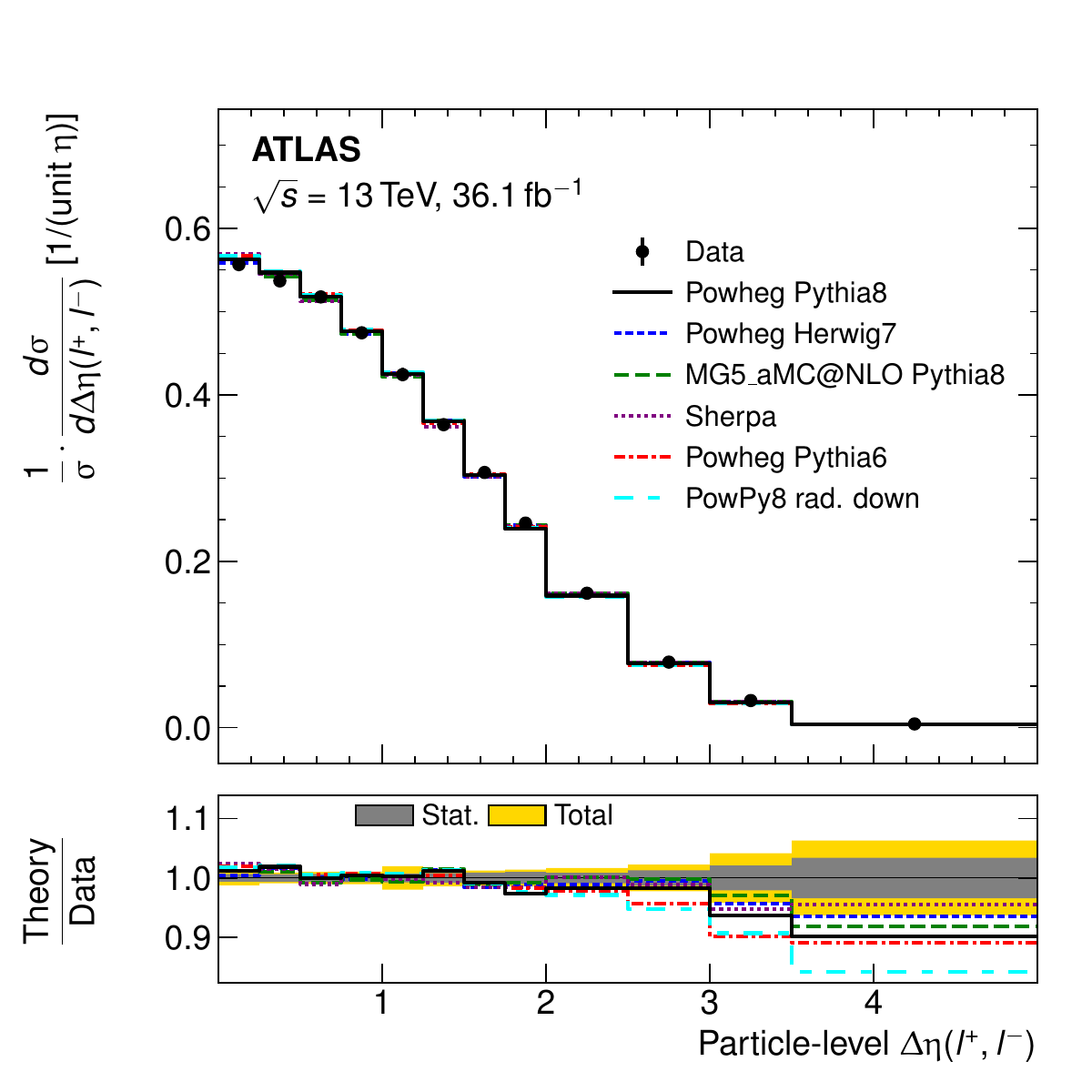}}\\
\end{tabular}
\caption{The fiducial particle-level differential cross-sections compared to predictions from \POWHEG, \MGMCatNLO\ and \SHERPA:  \emph{(top)}, absolute (a) $\Delta \phi$ and (b) $\Delta \eta$
 and \emph{(bottom)}, normalised (c) $\Delta \phi$ and (d) $\Delta \eta$
, using the inclusive selection.}
\label{fig:results_nlo_NoNW}
\end{figure}

The unfolded absolute and normalised particle-level cross-sections for $\Delta\phi$ and $\Delta\eta$ are presented in Figure~\ref{fig:results_nlo_NoNW} and the overall data--MC agreement is very close to that observed at parton level. As with the parton-level results, the normalised uncertainties are significantly smaller than the absolute uncertainties, and signal modelling uncertainties are dominant. The size of the overall uncertainties are similar between fiducial particle and full phase-space parton level for the normalised cross-sections, indicating that the extrapolation to the full phase-space that is modelled by the NLO generators used in the parton-level results is not detrimental. 

\FloatBarrier

%% file: tables/results_dphi_ll_parton.tex
\begin{table}[htb]
\caption{Summary of the parton-level absolute and normalised differential cross-sections as a function of $\Delta\phi(\ell^{+},\ell^{-})$, with statistical and systematic uncertainties in each bin.}
\vspace{3mm}
\setlength\extrarowheight{4pt}
\centering
\begin{tabular}{c c c c c c c c}
\toprule
$\Delta\phi(l^{+},l^{-})$: parton & Cross-section & stat. & syst. & Normalised  & stat. & syst. \\ \relax 
 [rad/$\pi$] & [pb/(rad/$\pi$)] &  &  & [1/(rad/$\pi$)] &  &  \\ 
\midrule
0.0 -- 0.1 & 16.9 & $\pm$ 0.2 & $\pm$ 1.0 & 0.863 & $\pm$ 0.009 & $\pm$ 0.007 \\ 
0.1 -- 0.2 & 17.1 & $\pm$ 0.2 & $\pm$ 1.0 & 0.874 & $\pm$ 0.008 & $\pm$ 0.009 \\ 
0.2 -- 0.3 & 17.2 & $\pm$ 0.2 & $\pm$ 1.1 & 0.879 & $\pm$ 0.008 & $\pm$ 0.019 \\ 
0.3 -- 0.4 & 17.9 & $\pm$ 0.2 & $\pm$ 1.1 & 0.917 & $\pm$ 0.008 & $\pm$ 0.008 \\ 
0.4 -- 0.5 & 18.8 & $\pm$ 0.2 & $\pm$ 1.1 & 0.962 & $\pm$ 0.008 & $\pm$ 0.008 \\ 
0.5 -- 0.6 & 19.6 & $\pm$ 0.2 & $\pm$ 1.2 & 1.001 & $\pm$ 0.008 & $\pm$ 0.019 \\ 
0.6 -- 0.7 & 20.4 & $\pm$ 0.2 & $\pm$ 1.1 & 1.043 & $\pm$ 0.008 & $\pm$ 0.012 \\ 
0.7 -- 0.8 & 21.7 & $\pm$ 0.2 & $\pm$ 1.4 & 1.111 & $\pm$ 0.008 & $\pm$ 0.013 \\ 
0.8 -- 0.9 & 22.6 & $\pm$ 0.2 & $\pm$ 1.4 & 1.156 & $\pm$ 0.008 & $\pm$ 0.009 \\ 
0.9 -- 1.0 & 23.4 & $\pm$ 0.2 & $\pm$ 1.4 & 1.194 & $\pm$ 0.008 & $\pm$ 0.013 \\ 
\bottomrule 
\end{tabular}
\label{tab:results_dphi_ll_parton}
\end{table}

%% file: tables/results_deta_ll_parton.tex
\begin{table}[htb]
\caption{Summary of the parton-level absolute and normalised differential cross-sections as a function of $\Delta\eta(\ell^{+},\ell^{-})$, with statistical and systematic uncertainties in each bin.}
\vspace{3mm}
\setlength\extrarowheight{4pt}
\centering
\begin{tabular}{c c c c c c c c}
\toprule
$\Delta\eta(\ell^{+},\ell^{-})$: parton & Cross-section & stat. & syst. & Normalised  & stat. & syst. \\ \relax 
 [unit $\eta$] & [pb/(unit $\eta$)] &  &  & [1/(unit $\eta$)] &  &  \\ 
\midrule
0.0 -- 0.25 & 9.04 & $\pm$ 0.06 & $\pm$ 0.47 & 0.463 & $\pm$ 0.003 & $\pm$ 0.011 \\ 
0.25 -- 0.5 & 8.81 & $\pm$ 0.06 & $\pm$ 0.48 & 0.451 & $\pm$ 0.003 & $\pm$ 0.008 \\ 
0.5 -- 0.75 & 8.65 & $\pm$ 0.06 & $\pm$ 0.58 & 0.443 & $\pm$ 0.003 & $\pm$ 0.004 \\ 
0.75 -- 1.0 & 8.10 & $\pm$ 0.06 & $\pm$ 0.46 & 0.415 & $\pm$ 0.003 & $\pm$ 0.007 \\ 
1.0 -- 1.25 & 7.48 & $\pm$ 0.06 & $\pm$ 0.57 & 0.383 & $\pm$ 0.003 & $\pm$ 0.007 \\ 
1.25 -- 1.5 & 6.68 & $\pm$ 0.06 & $\pm$ 0.38 & 0.342 & $\pm$ 0.003 & $\pm$ 0.004 \\ 
1.5 -- 1.75 & 5.94 & $\pm$ 0.06 & $\pm$ 0.33 & 0.304 & $\pm$ 0.003 & $\pm$ 0.004 \\ 
1.75 -- 2.0 & 5.15 & $\pm$ 0.05 & $\pm$ 0.37 & 0.264 & $\pm$ 0.003 & $\pm$ 0.003 \\ 
2.0 -- 2.5 & 3.85 & $\pm$ 0.04 & $\pm$ 0.28 & 0.197 & $\pm$ 0.002 & $\pm$ 0.005 \\ 
2.5 -- 3.0 & 2.42 & $\pm$ 0.03 & $\pm$ 0.20 & 0.124 & $\pm$ 0.002 & $\pm$ 0.005 \\ 
3.0 -- 3.5 & 1.46 & $\pm$ 0.03 & $\pm$ 0.15 & 0.075 & $\pm$ 0.002 & $\pm$ 0.005 \\ 
3.5 -- 5.0 & 0.47 & $\pm$ 0.02 & $\pm$ 0.06 & 0.024 & $\pm$ 0.001 & $\pm$ 0.002 \\ 
\bottomrule 
\end{tabular}
\label{tab:results_deta_ll_parton}
\end{table}

%% file: tables/results_dphi_ll_vs_ttbar_m_parton.tex
\begin{table}[htb]
\caption{Summary of the parton-level absolute and normalised differential cross-sections as a function of $\Delta\phi(\ell^{+},\ell^{-})$ in four regions of $m_{t\bar{t}}$, with statistical and systematic uncertainties in each bin.}
\vspace{3mm}
\setlength\extrarowheight{4pt}
\centering
\begin{tabular}{c c c c c c c c}
\toprule
$\Delta\phi(\ell^{+},\ell^{-})$: parton & Cross-section & stat. & syst. & Normalised  & stat. & syst. \\ \relax 
 [rad/$\pi$] & [pb/(rad/$\pi$)] &  &  & [1/(rad/$\pi$)] &  &  \\ 
\midrule
\multicolumn{7}{l}{$m_{t\bar{t}} < 450$ GeV} \\ 
\midrule
0.0 -- 0.2 & 8.99 & $\pm$ 0.15 & $\pm$ 0.71 & 1.099 & $\pm$ 0.016 & $\pm$ 0.035 \\ 
0.2 -- 0.4 & 8.73 & $\pm$ 0.14 & $\pm$ 0.71 & 1.068 & $\pm$ 0.015 & $\pm$ 0.031 \\ 
0.4 -- 0.6 & 8.25 & $\pm$ 0.13 & $\pm$ 0.66 & 1.009 & $\pm$ 0.014 & $\pm$ 0.028 \\ 
0.6 -- 0.8 & 7.89 & $\pm$ 0.12 & $\pm$ 0.60 & 0.965 & $\pm$ 0.014 & $\pm$ 0.024 \\ 
0.8 -- 1.0 & 7.03 & $\pm$ 0.12 & $\pm$ 0.52 & 0.860 & $\pm$ 0.013 & $\pm$ 0.039 \\ 
\midrule
\multicolumn{7}{l}{$450 \leq m_{t\bar{t}} < 550$ GeV} \\ 
\midrule
0.0 -- 0.3 & 4.10 & $\pm$ 0.08 & $\pm$ 0.39 & 0.781 & $\pm$ 0.012 & $\pm$ 0.032 \\ 
0.3 -- 0.6 & 5.17 & $\pm$ 0.07 & $\pm$ 0.40 & 0.986 & $\pm$ 0.011 & $\pm$ 0.031 \\ 
0.6 -- 0.8 & 5.92 & $\pm$ 0.08 & $\pm$ 0.56 & 1.128 & $\pm$ 0.014 & $\pm$ 0.034 \\ 
0.8 -- 1.0 & 6.42 & $\pm$ 0.08 & $\pm$ 0.59 & 1.223 & $\pm$ 0.015 & $\pm$ 0.024 \\ 
\midrule
\multicolumn{7}{l}{$550 \leq m_{t\bar{t}} < 800$ GeV} \\ 
\midrule
0.0 -- 0.4 & 2.91 & $\pm$ 0.07 & $\pm$ 0.32 & 0.665 & $\pm$ 0.013 & $\pm$ 0.024 \\ 
0.4 -- 0.6 & 4.08 & $\pm$ 0.10 & $\pm$ 0.47 & 0.932 & $\pm$ 0.020 & $\pm$ 0.049 \\ 
0.6 -- 0.8 & 5.42 & $\pm$ 0.09 & $\pm$ 0.52 & 1.237 & $\pm$ 0.019 & $\pm$ 0.055 \\ 
0.8 -- 1.0 & 6.57 & $\pm$ 0.09 & $\pm$ 0.65 & 1.500 & $\pm$ 0.020 & $\pm$ 0.031 \\ 
\midrule
\multicolumn{7}{l}{$m_{t\bar{t}} \geq 800$ GeV} \\ 
\midrule
0.0 -- 0.8 & 0.99 & $\pm$ 0.03 & $\pm$ 0.12 & 0.771 & $\pm$ 0.012 & $\pm$ 0.028 \\ 
0.8 -- 1.0 & 2.46 & $\pm$ 0.07 & $\pm$ 0.27 & 1.917 & $\pm$ 0.046 & $\pm$ 0.105 \\ 
\bottomrule 
\end{tabular}
\label{tab:results_dphi_ll_vs_ttbar_m_parton}
\end{table}

%% file: sections/results_spin.tex
\section{Spin correlation results}
\label{sec:results_spin}

The level of spin correlation observed in data is (traditionally) assessed by quantifying it in relation to the amount of correlation expected in the SM~\cite{TOPQ-2011-11,TOPQ-2013-01,TOPQ-2013-10,CMS-TOP-13-003,TOPQ-2014-07,CMS-TOP-13-015,CMS-TOP-14-023,TOPQ-2015-13}. This \emph{fraction of SM-like spin correlation} ($f_{\textrm{SM}}$) is extracted using hypothesis templates that are fitted to the parton-level, unfolded normalised cross-sections from data. Two hypotheses are used: dileptonic \ttbar events with SM spin correlation (the nominal \ttbar sample) and dileptonic events where the effect of spin correlation has been removed (the nominal \ttbar sample where the top quarks are decayed using \MADSPIN with spin correlations disabled), as described in Section~\ref{sec:data}. In each observable, a binned maximum-likelihood fit is performed using \textsc{MINUIT}~\cite{MINUIT}. The predicted normalised cross-section in bin $i$, $x_{i}$, is determined as a function of $f_{\textrm{SM}}$ using the expression:
\begin{equation}
	x_{i} = f_{\textrm{SM}} \cdot x_{\textrm{spin, }i} + (1 - f_{\textrm{SM}}) \cdot x_{\textrm{nospin, }i}~\text{,} \nonumber
\end{equation}
where $x_{\textrm{spin}}$ and $x_{\textrm{nospin}}$ are the expected normalised cross-sections under the SM spin hypothesis and the uncorrelated hypothesis, respectively. The negative logarithm of a likelihood function is minimised in order to determine $f_{\textrm{SM}}$. The extraction of $f_{\textrm{SM}}$ is performed in five observables: the inclusive $\Delta\phi$; and $\Delta\phi$ in each of the four regions of $m_{\ttbar}$. The total number of bins used in the extraction depends upon the region of $m_{\ttbar}$. 

The statistical uncertainty on $f_{\textrm{SM}}$ is determined using ensemble tests. Ten thousand pseudo-data sets are constructed by Poisson-smearing the observed number of events in each bin of the detector-level distribution.  Each of these data samples are unfolded in the usual manner, and fitted to extract $f_{\textrm{SM}}$. The RMS of the resulting distribution of $f_{\textrm{SM}}$ values gives the statistical uncertainty on this quantity.

Systematic uncertainties on $f_{\textrm{SM}}$ are determined using the same procedure as for the unfolded differential cross-sections, considering the same sources as those described in Section~\ref{sec:systematics}. Monte Carlo samples with different sources of systematic uncertainty are unfolded, as described in Section~\ref{sec:unfolding}, and the unfolded spectra are used as pseudo-data. The templates are fitted to this pseudo-data and the difference between the systematic $f_{\textrm{SM}}$ and the nominal (i.e. $f_{\textrm{SM}}=1$) is taken as the systematic uncertainty on $f_{\textrm{SM}}$ due to that source. The dominant uncertainties are summarised in Table~\ref{tab:fsm_syst_new}; the largest sources of systematic uncertainty arise due to the modelling of the \ttbar process. 

\input{tables/fsm_syst_new}

\begin{figure}[htb]
  \centering
  \includegraphics[width=0.8\textwidth]{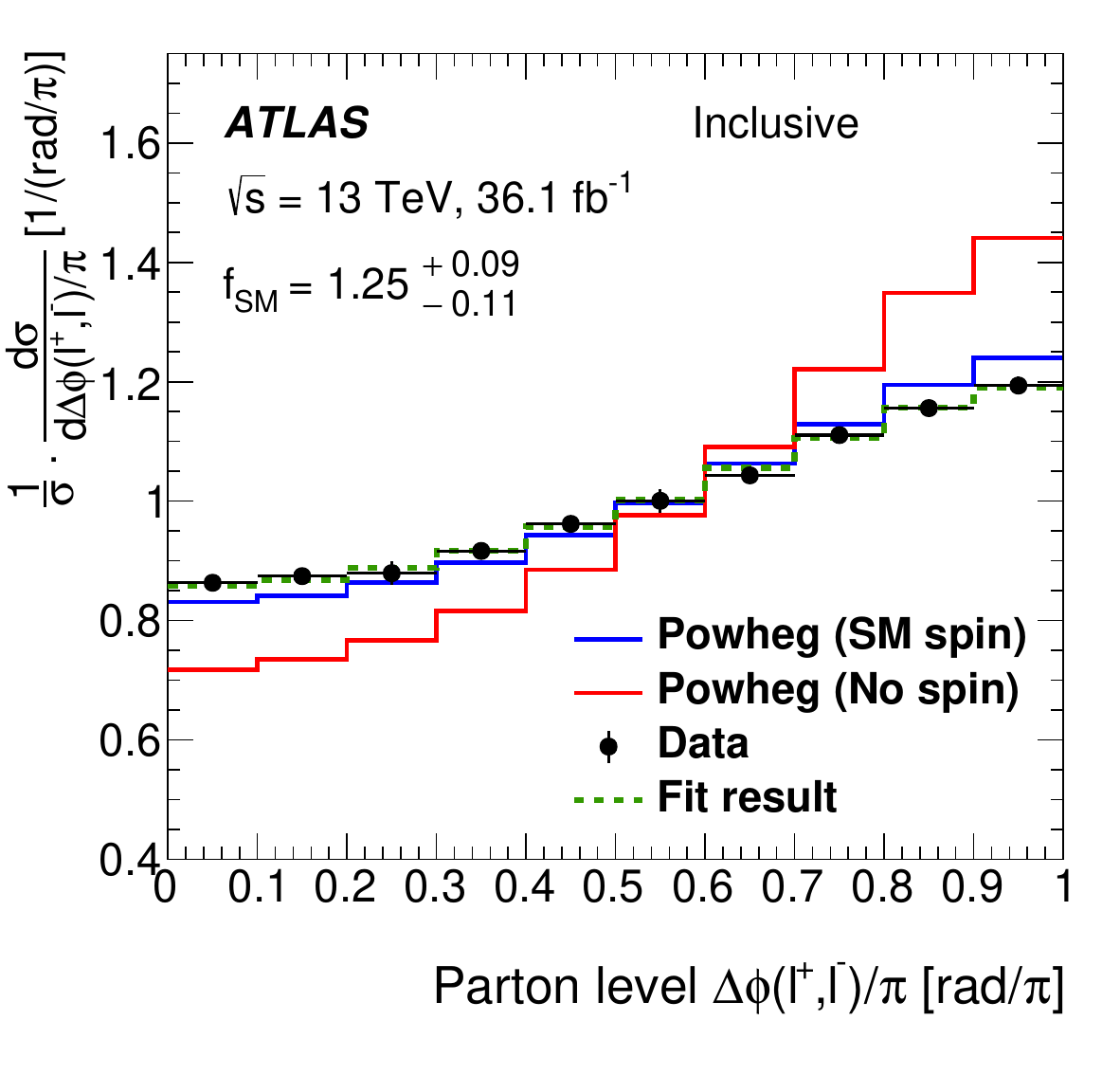}
  \caption{Results of the fit of hypothesis templates to the unfolded data showing the $\Delta\phi$ distribution for the inclusive selection. The hypothesis templates are described in Section~\ref{sec:data}.}
  \label{fig:results_fsm}
\end{figure}

\begin{figure}[htb]
\centering
\begin{tabular}{cc}
\subfloat[\label{fig:results_fsm_450}]{\includegraphics[width=0.48\textwidth]{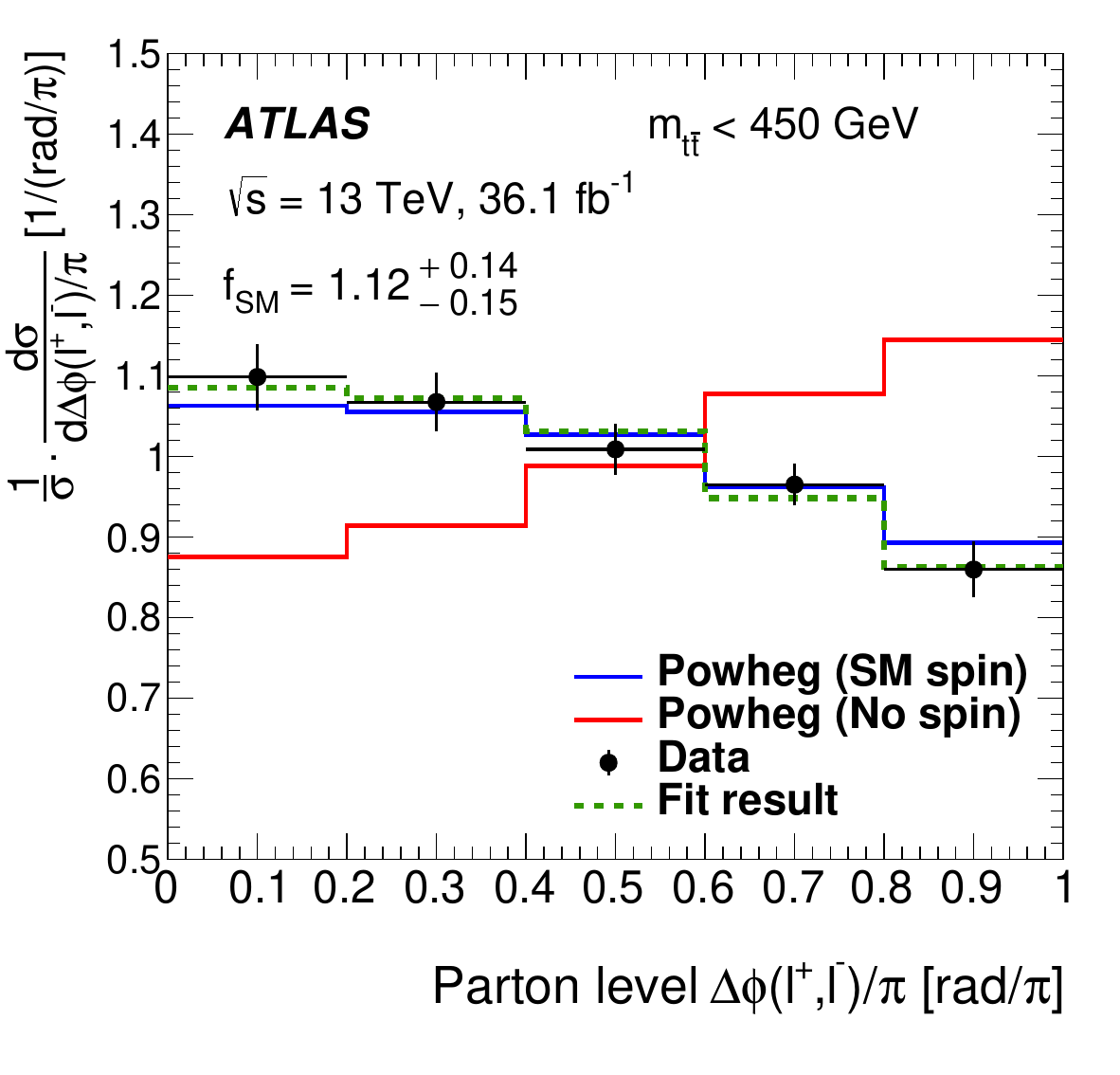}}&
\subfloat[\label{fig:results_fsm_550}]{\includegraphics[width=0.48\textwidth]{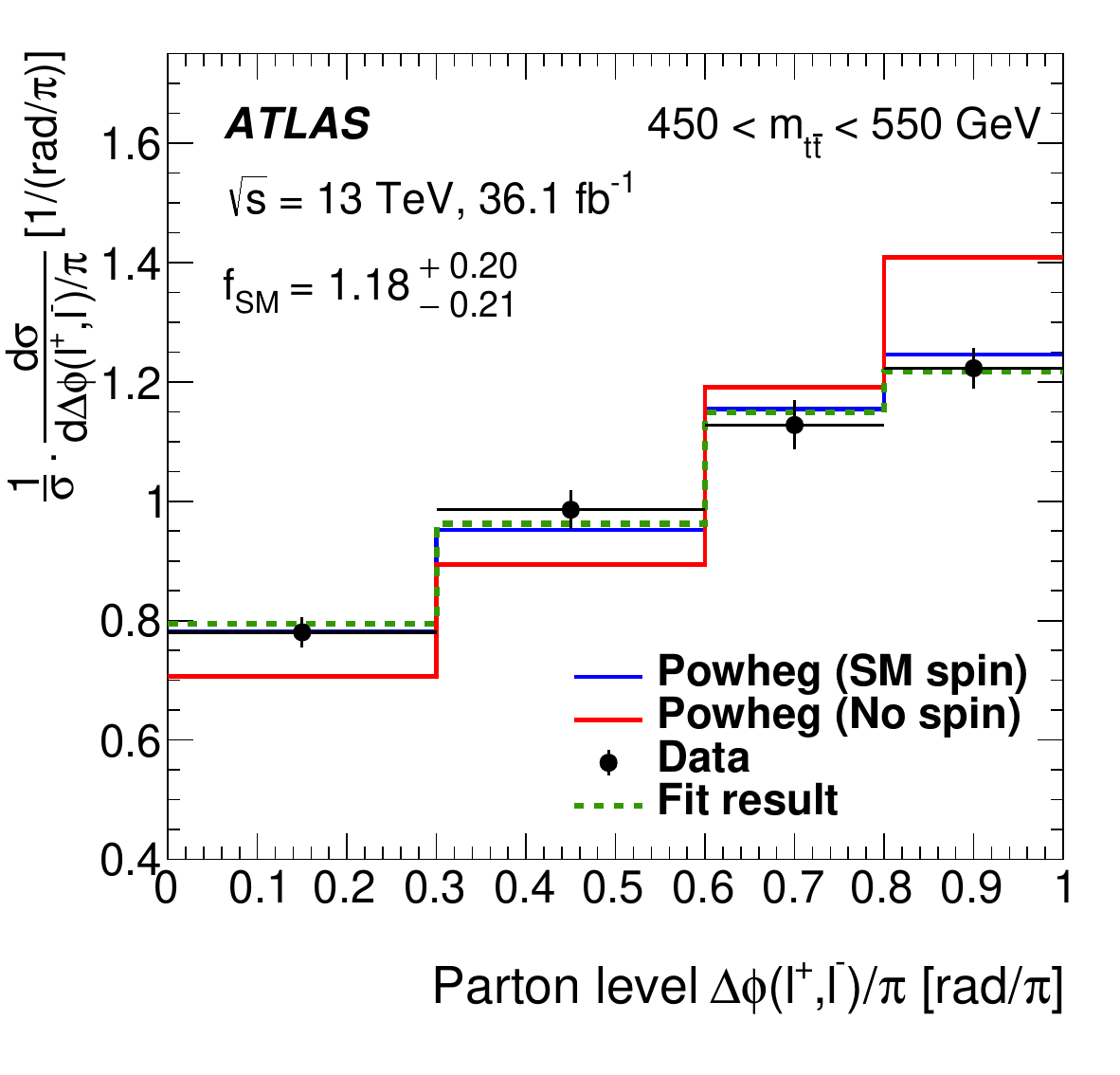}}\\
\subfloat[\label{fig:results_fsm_800}]{\includegraphics[width=0.48\textwidth]{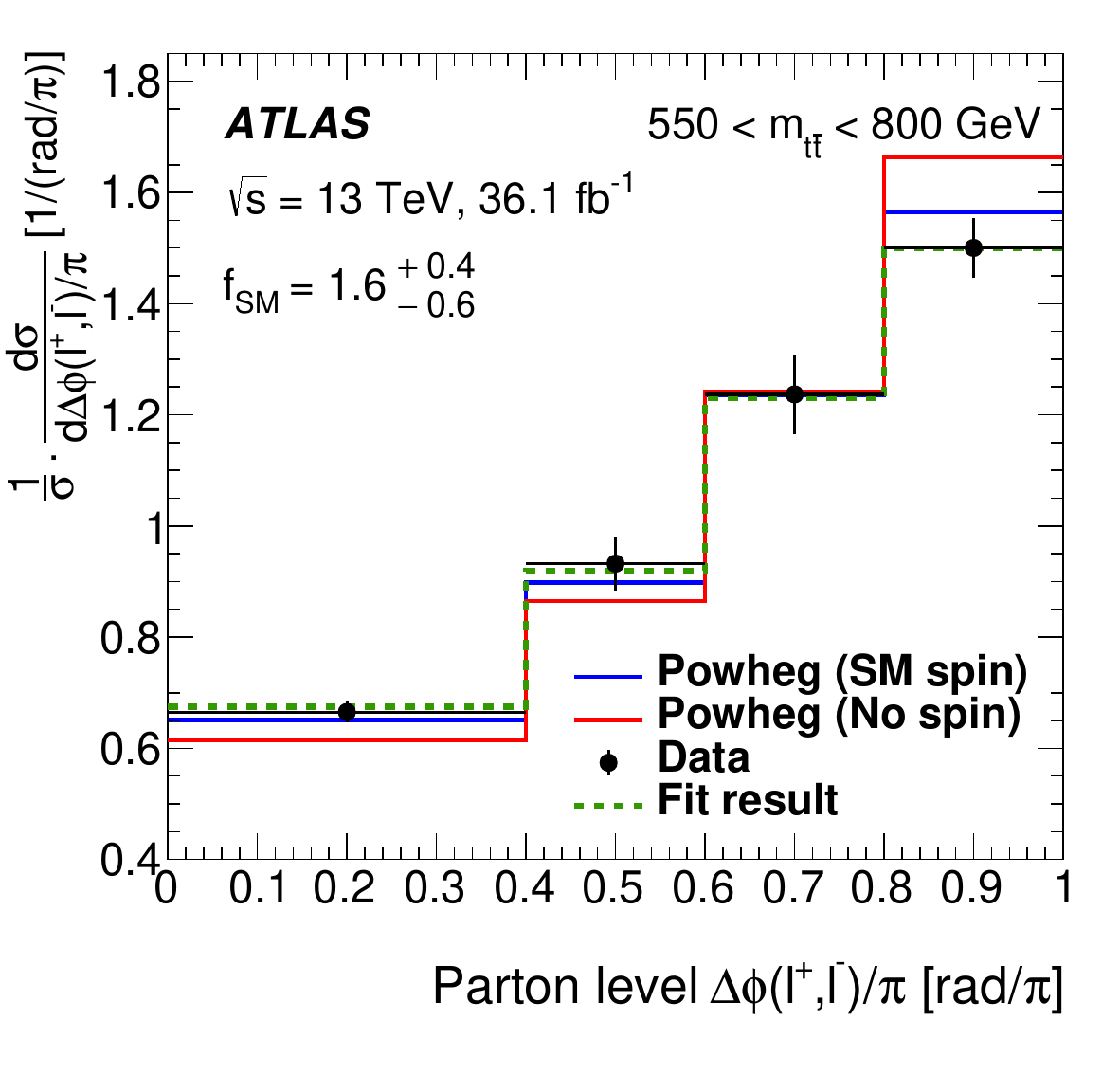}}&
\subfloat[\label{fig:results_fsm_max}]{\includegraphics[width=0.48\textwidth]{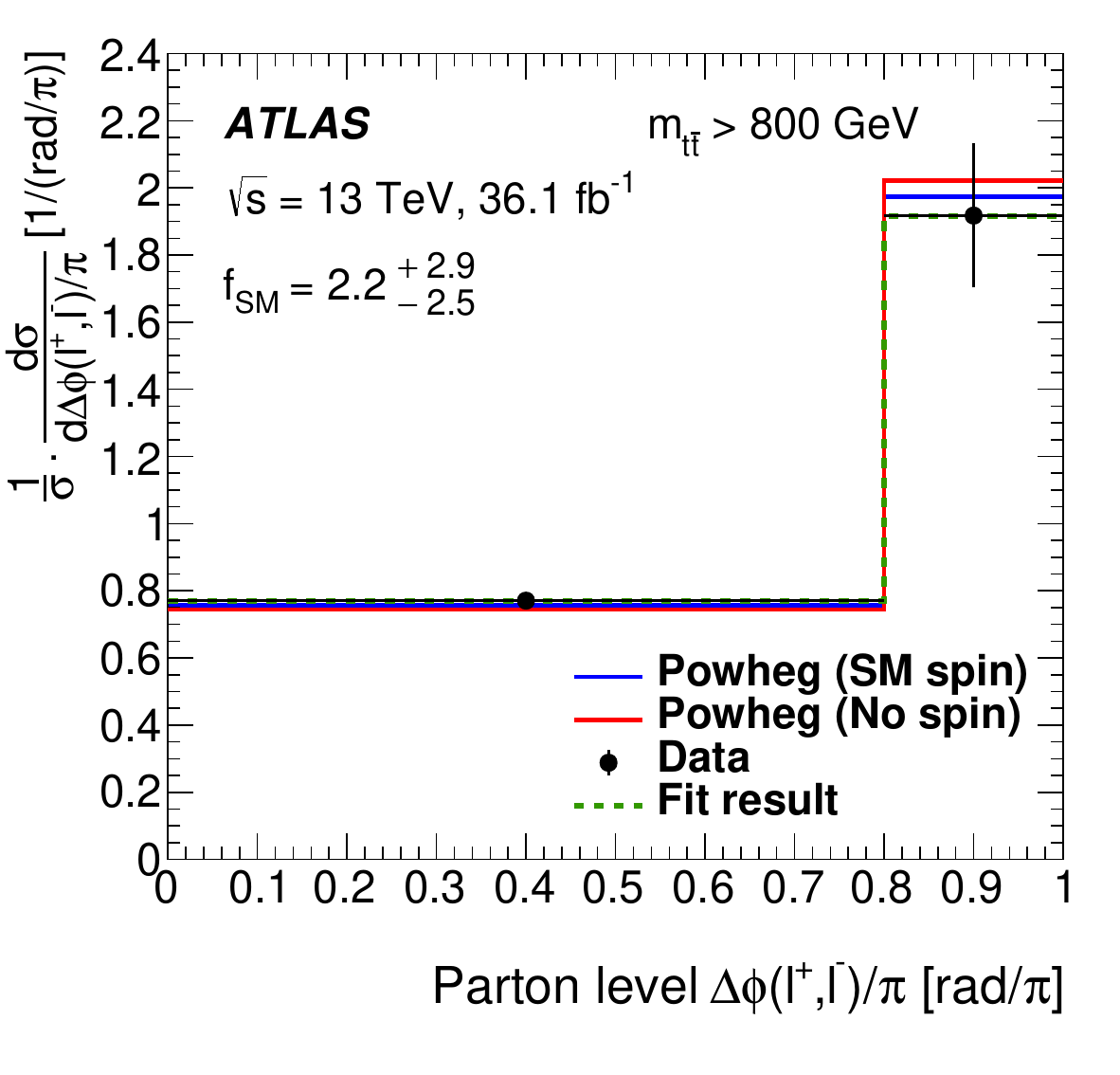}}\\
\end{tabular}
\caption{Results of the fit of hypothesis templates to the unfolded data showing the $\Delta\phi$ distributions in $m_{\ttbar}$ regions for the reconstructed selection:
(a) $m_{\ttbar}<450$~\GeV,
(b) $450\leq m_{\ttbar}<550$~\GeV,
(c) $550\leq m_{\ttbar}<800$~\GeV, and
(d) $m_{\ttbar}\geq800$~\GeV. The hypothesis templates are described in Section~\ref{sec:data}.}
\label{fig:results_fsm_diff}
\end{figure}

The hypothesis templates for each observable, the unfolded data, and the resulting fit are presented in Figure~\ref{fig:results_fsm} and Figure~\ref{fig:results_fsm_diff}. The $f_{\textrm{SM}}$ extracted from each observable and the significance with respect to the SM hypothesis are presented in Table~\ref{tab:results_fsm_chi2}. Two cases are considered: first, only the uncertainties on the unfolded measurement are taken into account, and second, theoretical uncertainties on the hypothesis templates are included. These theoretical contributions include factorisation and renormalisation scale shifts as well as PDF uncertainties\footnote{30 eigenvector variations from the PDF4LHC recommendation~\cite{Butterworth:2015oua}.} and are distinct from the radiation uncertainties (which also include scale variations) that are already included in the unfolded differential cross-section uncertainties.  An additional template uncertainty is considered, which takes into account the difference between the nominal \POWHEG+\PYTHIA\!8 \ttbar sample and the alternative \POWHEG+\PYTHIA\!8 sample in which \MADSPIN\ handles the decays, as described in Section~\ref{sec:data}.

For the inclusive result, the spin correlation extracted from the unfolded data is somewhat higher than the SM expectation, with a significance of 2.2 standard deviations when including theoretical uncertainties on the hypothesis templates, and 3.8 standard deviations without those uncertainties.
Previous measurements from ATLAS and CMS have also observed $f_{\textrm{SM}} > 1$ but the uncertainties were such that the results were consistent with the prediction, even without template uncertainties included~\cite{TOPQ-2011-11,TOPQ-2013-01,TOPQ-2013-10,CMS-TOP-13-003, TOPQ-2014-07,CMS-TOP-13-015,CMS-TOP-14-023,TOPQ-2015-13}. The central $f_{\textrm{SM}}$ value as a function of $m_{\ttbar}$ is found to increase as a function of $m_{\ttbar}$; however, the uncertainties on $f_{\textrm{SM}}$ are much larger than in the inclusive case and none of the results deviate substantially from the SM expectation. 

\input{tables/results_fsm_chi2_newsig3}

\begin{figure}[htb]
  \centering
  \begin{tabular}{c}
  \includegraphics[width=0.6\textwidth]{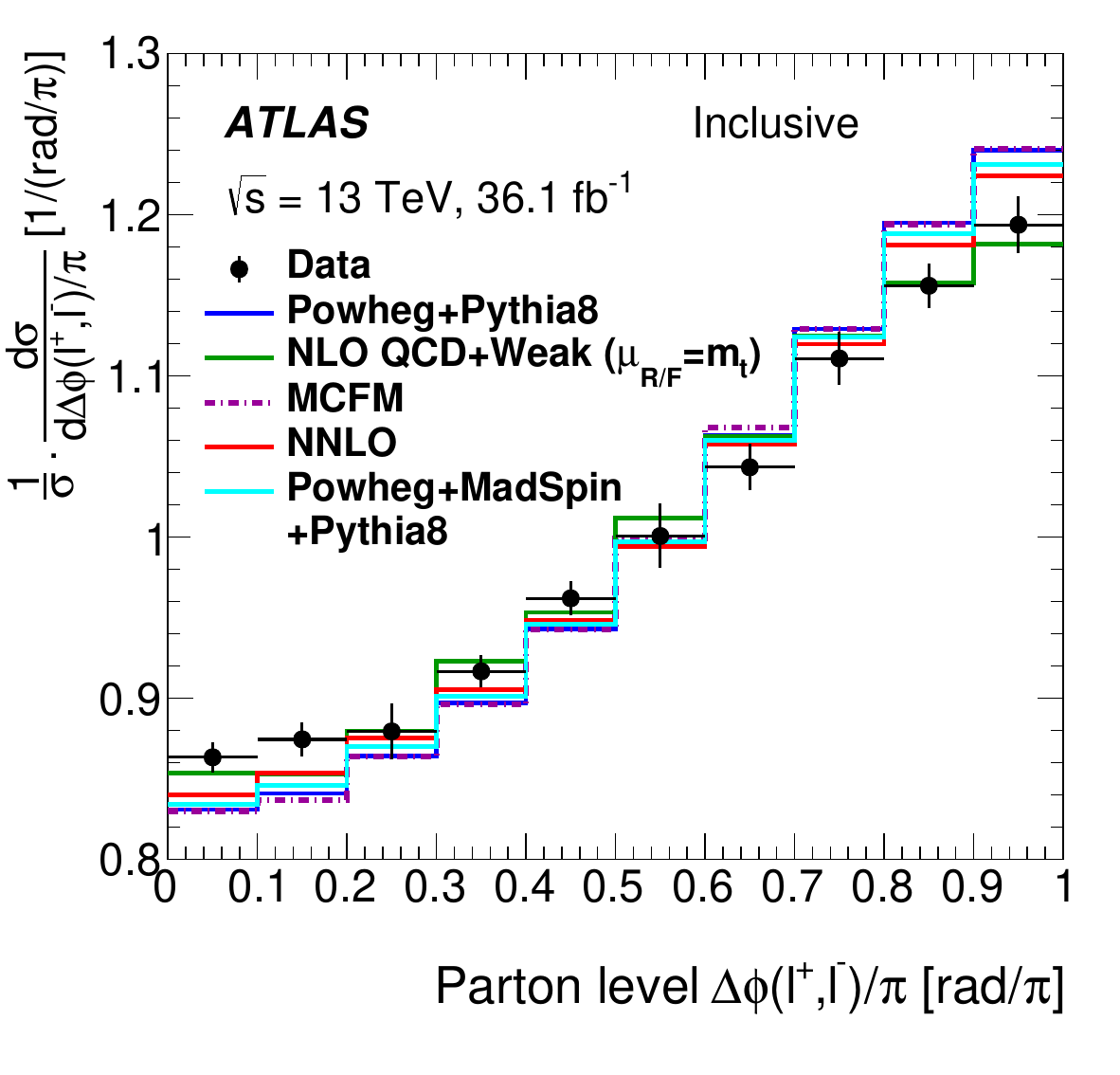}\\
  \end{tabular}
  \caption{Comparison of the unfolded $\Delta\phi$ distribution with theoretical predictions for the normalised cross-section with the inclusive selection. Each prediction is discussed in the text.}
  \label{fig:results_fsm_withpredictions}
\end{figure}

A number of cross-checks were performed to attempt to understand the results in terms of either the limitations of the modelling of the \ttbar system or by experimental effects not covered by the systematic uncertainty prescription described above. The NLO generators used in this analysis model \ttbar production at NLO in QCD (hereafter simply referred to as NLO) but do not fully include NLO effects in the decays of the top quarks, nor do they directly consider the effects of interference between the initial and final states. The production and decay of the top quarks are factorised using the narrow-width approximation (NWA). The MCFM generator \cite{MCFM} can provide fixed-order predictions for \ttbar production and decay at full NLO in the dilepton channel under the NWA. The effect of the spin analysing power of the lepton itself also changes from unity at LO to 0.998 at NLO \cite{Jezabek:1988ja} and this is also not considered in the nominal hypothesis templates. Alternative hypothesis templates were generated using MCFM (using the same scale and PDF settings as the nominal \POWHEG +  \PYTHIA\!8 sample), illustrated in Figure~\ref{fig:results_fsm_withpredictions}, and the prediction is remarkably close to the prediction from \POWHEG + \PYTHIA\!8. It is concluded that the LO decays of the top quarks used in the nominal hypothesis templates have little effect on the measurement and do not explain the observed differences between data and predictions.

The effect of removing the NWA can not be directly tested in the phase-space of this measurement. Without the NWA and with both NLO in production and in decay, it becomes unphysical to separate the \ttbar and the contribution of the $tW$ processes. In this analysis the $tW$ process is directly subtracted as a background from the data, preventing a direct comparison to calculations that do not include the NWA and simulate the full \ttbar + $tW$ process. However, the effect on the $\Delta\phi$ observable was investigated in an inclusive \ttbar + $tW$ ($b\ell^{-}\nu_{\ell}\bar{b}\ell^{+}\bar{\nu_{\ell}}$) phase-space using the \textsc{Powheg-Box-Res} bb4l process \cite{bb4l} and compared to the nominal \ttbar + $tW$ set-up and no significant differences were observed. It is therefore assumed that, in the \ttbar phase-space of this measurement, the NWA in the templates is not a limiting factor and does not explain the observed differences. 

Alternative templates for $f_{\textrm{SM}}$ extraction may be constructed from samples used to evaluate systematic uncertainties, such as the radiation variation of \POWHEG + \PYTHIA\!8, or from alternative generator set-ups, such as \POWHEG + \HERWIG\!7 and \MGMCatNLO + \PYTHIA\!8, or by changing the scales and PDF settings. In each case, the no-spin template is derived by scaling the prediction of the alternative model (with spin included) by the ratio of the no-spin and spin templates in the \POWHEG + \PYTHIA\!8 setup. The results of using different hypothesis templates are presented in Table~\ref{tab:fsm_alternate}. With the exception of the highest $m_{\ttbar}$ bin, which has large statistical and systematic uncertainties, the $f_{\textrm{SM}}$ values remain above 1 for all alternative templates.

The effect of higher orders in the production (NNLO) was investigated by reweighting the $p_{\textrm{T}}(t)$ spectra in \POWHEG + \PYTHIA\!8 to NNLO fixed-order predictions and to observed detector-corrected data spectra~\cite{TOPQ-2016-01}. The reweighting reduced the observed deviation somewhat but was consistent with the scale uncertainties that are already considered in the uncertainties on the hypothesis templates. Fixed-order NNLO predictions recently became available for the observables in this paper~\cite{Behring:2019iiv}. The results of these predictions are illustrated in Figure~\ref{fig:results_fsm_withpredictions} for the default renormalisation ($\mu_{\textrm{R}}$) and factorisation ($\mu_{\textrm{F}}$) scale choice of $\mu_{\textrm{R}}=\mu_{\textrm{F}}=H_T/4$, where $H_T = \sqrt{m_t^2+p_{T,t}^2}+\sqrt{m_t^2+p_{T,\tbar}^2}$.  They are closer to the data than the NLO predictions, but still do not agree fully. The effect of higher orders in $t\bar{t}$ production is therefore assumed to be well-covered by the theoretical uncertainties on the templates and also does not fully explain the observed value of $f_{\textrm{SM}}$. Fiducial predictions are also available with the same NNLO calculation; however, the definition of the particles used to construct the fiducial region (specifically the $b$-jets) are not identical. This results in a somewhat different fiducial region compared to the measurements presented in Section~\ref{sec:results_diff}, and therefore a direct comparison is not made. 

Finally, an alternative differential prediction, made specifically for these observables~\cite{Bernreuther:2010ny,Bernreuther:2013aga,Bernreuther:2015yna}, is used as a template.  It is calculated at NLO in the strong and weak gauge couplings, using an expansion of the normalised differential distribution in powers of the couplings, with
fixed renormalisation and factorisation scales equal to the top mass $\mu_{\textrm{R}}=\mu_{\textrm{F}}=m_t$.
This prediction also has a dedicated no-spin template. The prediction agrees better with the data but has significant scale uncertainties, leading to an $f_{\textrm{SM}}=1.03\pm0.07\text{(stat)}^{+0.10}_{-0.14} \text{(scale)}$, and is consistent both with the result from using the \POWHEG + \PYTHIA\!8 templates and with the SM expectation of $f_{\textrm{SM}}=1$. The value of $f_{\textrm{SM}}$ is consistent with that extracted from a measurement of $\Delta\phi$ by CMS~\cite{CMS-TOP-18-006}, using the same calculation. The predictions of Ref.~\cite{Behring:2019iiv} have also been calculated using the expansion technique of Refs.~\cite{Bernreuther:2010ny,Bernreuther:2013aga,Bernreuther:2015yna}; the NLO expansion with $\mu_{\textrm{R}}=\mu_{\textrm{F}}=m_t$ leads to comparable results, again with significant scale uncertainties.  When the calculation is extended to NNLO in the same framework, it lies further from the data and is consistent with the NNLO prediction without expansion of the normalised cross-section~\cite{Behring:2019iiv}.

\begin{figure}[htb]
  \centering
  \begin{tabular}{c}
  \subfloat[\label{fig:results_fsm_pred2}]{\includegraphics[width=0.6\textwidth]{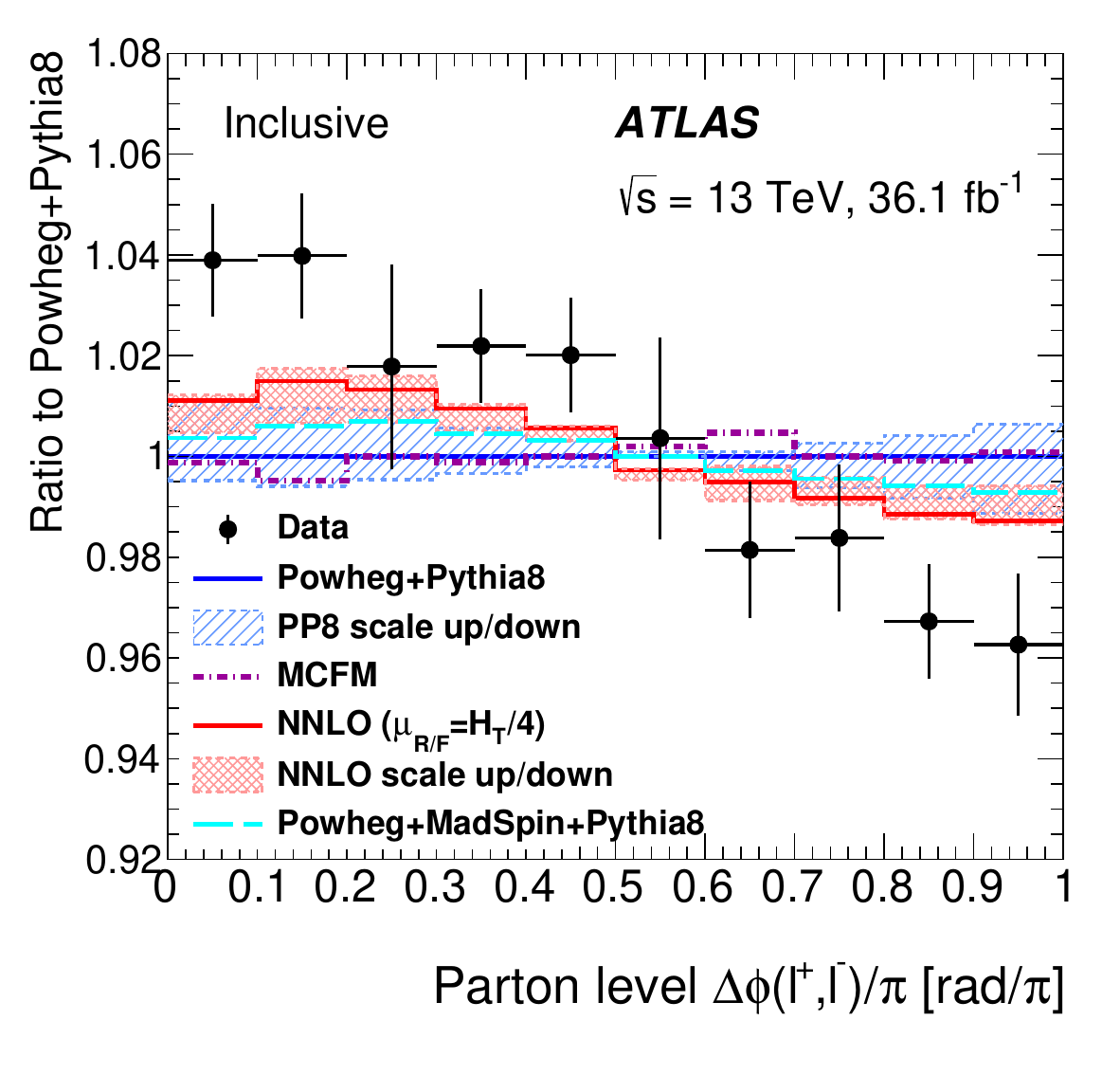}}\\
  \subfloat[\label{fig:results_fsm_pred3}]{\includegraphics[width=0.6\textwidth]{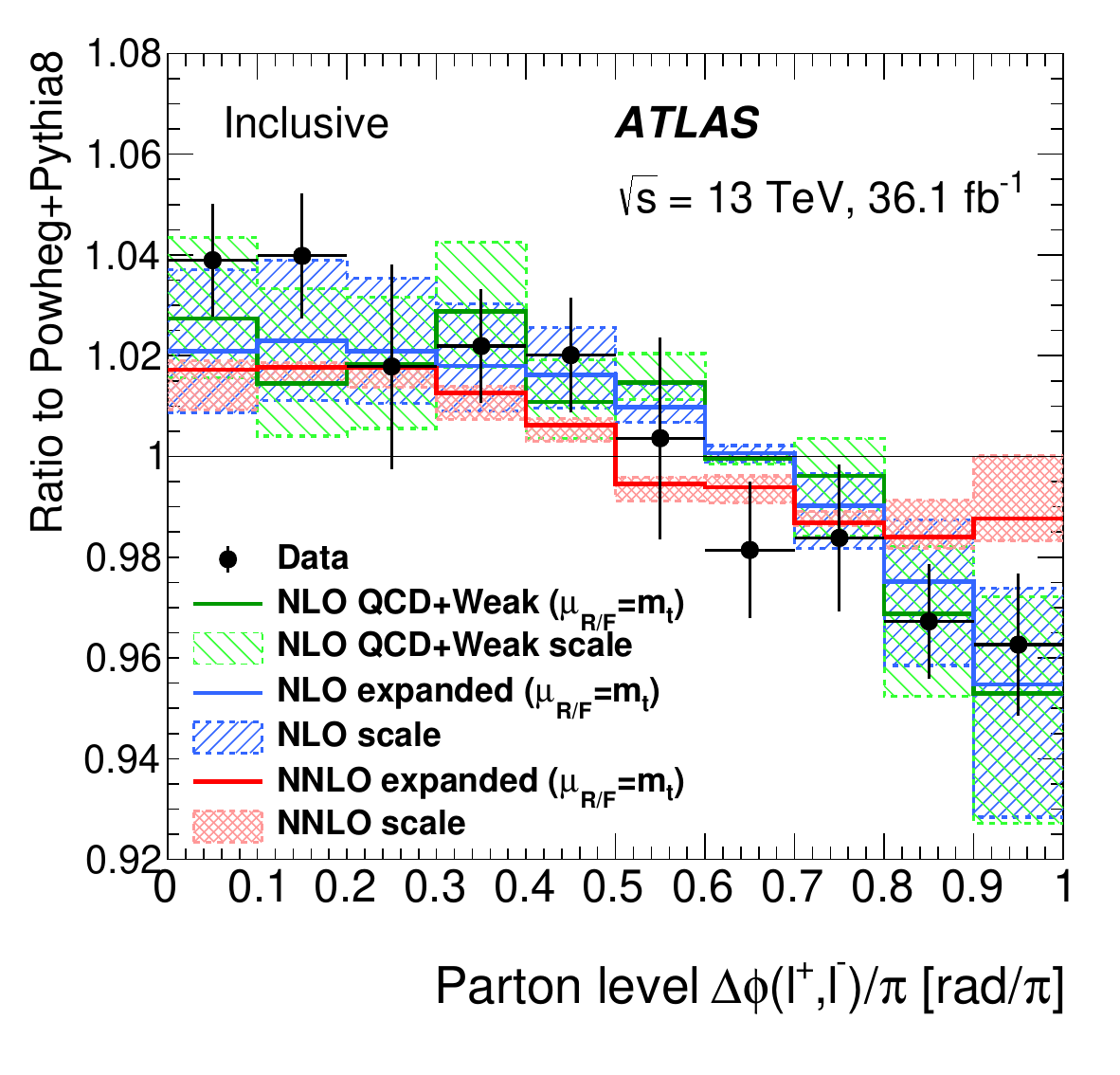}}\\
  \end{tabular}
  \caption{Comparison of the unfolded $\Delta\phi$ distribution with theoretical predictions for the inclusive selection: ratio as compared with \POWHEG + \PYTHIA\!8 for (a) NLO generators and NNLO fixed-order predictions, (b) NLO and NNLO theoretical predictions expanded in the normalised cross-section ratio. Each prediction is discussed in the text.}
  \label{fig:results_fsm_withpredictions_ratio}
\end{figure}

The comparison between data and the various SM predictions is illustrated in Figures~\ref{fig:results_fsm_withpredictions} and~\ref{fig:results_fsm_withpredictions_ratio}. The disagreement between the data and the NLO predictions from MCFM and \POWHEG + \PYTHIA\!8 can be clearly observed in Figure~\ref{fig:results_fsm_pred2}. The NNLO fixed-order prediction agrees better with the data but still differs significantly. Finally, the expanded NLO predictions agree with the data within their large scale uncertainties, as shown in Figure~\ref{fig:results_fsm_pred3}, but the NNLO prediction using the same expansion does not.

\begin{landscape}
\begin{table}
\caption{Summary of the extracted spin correlation values in the inclusive $\Delta\phi$ observable using different hypothesis templates.}

\centering
\begin{tabular}{l c c c c c}
\toprule
Generator & Inclusive & $m_{t\bar{t}} < 450$~\GeV & $450 \leq m_{t\bar{t}} < 550$~\GeV  & $550 \leq m_{t\bar{t}} < 800$~\GeV  & $m_{t\bar{t}} \geq 800$~\GeV  \\
\midrule 
\multicolumn{6}{l}{$f_{\textrm{SM}}$ values} \\
\midrule
\POWHEG + \PYTHIA\!8                                           & 1.25 & 1.12 & 1.18 & 1.65 & 2.2 \\
\POWHEG + \PYTHIA\!8 (2.0~$\mu_{\text{F}},~2.0~\mu_{\textrm{R}}$) & 1.29 & 1.14 & 1.23 & 1.79 & 2.0 \\
\POWHEG + \PYTHIA\!8 (0.5~$\mu_{\text{F}},~0.5~\mu_{\textrm{R}}$) & 1.18 & 1.09 & 1.11 & 1.40 & 1.3 \\
\POWHEG + \PYTHIA\!8 (PDF variations)                          & 1.26 & 1.13 & 1.25 & 1.76 & 2.2 \\
\POWHEG + \PYTHIA\!8 RadLo tune                                & 1.29 & 1.15 & 1.23 & 1.79 & 2.0 \\
\POWHEG + \HERWIG\!7                                           & 1.32 & 1.17 & 1.25 & 1.79 & 2.0 \\
\MGMCatNLO + \PYTHIA\!8                                        & 1.20 & 1.06 & 1.18 & 1.40 & 0.7 \\
\midrule
\POWHEG + \MADSPIN + \PYTHIA\!8                                & 1.20 & 1.06 & 1.09 & 1.46 & 1.3 \\
\midrule
NNLO QCD, $\mu_{\textrm{R,F}}=H_T/4$ \cite{Behring:2019iiv} & 1.16 & - & - & - & - \\
NLO (QCD + Weak expanded), $\mu_{\textrm{R,F}}=m_t$ \cite{Bernreuther:2010ny,Bernreuther:2013aga,Bernreuther:2015yna} & 1.03 & - & - & - & - \\
NLO expanded, $\mu_{\textrm{R,F}}=m_t$  \cite{Behring:2019iiv} & 1.04 & - & - & - & - \\
NNLO expanded, $\mu_{\textrm{R,F}}=m_t$ \cite{Behring:2019iiv} & 1.13 & - & - & - & - \\
\bottomrule
\end{tabular}
\label{tab:fsm_alternate}
\end{table}
\clearpage
\end{landscape}

\FloatBarrier

%% file: tables/fsm_syst_new.tex
\begin{table}
\caption{Summary table of the effect of experimental systematic uncertainties on the $f_{\textrm{SM}}$ extraction. Uncertainties which are smaller than the precision shown are included in the totals and the $f_{\textrm{SM}}$ significance calculations.}
\vspace{3mm}
\setlength\extrarowheight{5pt}
\centering
\begin{tabular}{l c c c c c}
\toprule
 & & \multicolumn{4}{c}{$m_{\ttbar}$ range [GeV]} \\
Systematic & Inclusive  & \small{$m_{\ttbar} < 450$} & \small{$450 \leq m_{\ttbar} < 550$} & \small{$550 \leq m_{\ttbar} < 800$} & \small{$m_{\ttbar} \geq 800$} \\
\midrule 
\small{Matrix element}                            & $\pm 0.006$         & $\pm 0.11$       & $\pm 0.064$           & $\pm 0.01$        & $\pm 0.3$ \\
\small{Parton shower and hadronisation}           & $\pm 0.010$         & $\pm 0.02$       & $\pm 0.005$           & $\pm 0.01$        & $\pm 1.4$ \\
\small{Radiation and scale settings}              & $\pm 0.055$         & $\pm 0.05$       & $\pm 0.061$           & $\pm 0.23$        & $< 0.1$ \\
\small{PDF}                                       & $\pm 0.002$         & $<0.01$          & $\pm 0.003$           & $\pm 0.01$        & $< 0.1$ \\
\small{Background modelling}                      & $\pm 0.009$         & $\pm 0.01$       & $^{+0.014}_{-0.015}$    & $\pm 0.01$        & $\pm 0.1$ \\
\small{Lepton ID and reconstruction}              & $\pm 0.008$         & $\pm 0.01$       & $^{+0.030}_{-0.036}$    & $^{+0.03}_{-0.10}$  & $^{+0.5}_{-0.2}$ \\
\small{$b$-tagging}                               & $^{+0.004}_{-0.003}$  & $\pm 0.01$       & $\pm 0.025$           & $^{+0.04}_{-0.02}$  & $^{+0.1}_{-0.2}$ \\
\small{Jet ID and reconstruction}                 & $^{+0.014}_{-0.017}$  & $^{+0.02}_{-0.05}$ & $^{+0.076}_{-0.093}$    & $^{+0.17}_{-0.26}$  & $^{+1.7}_{-0.6}$ \\
\small{\met reconstruction}                       & $<0.001$            & $^{+0.01}_{-0.02}$ & $^{+0.042}_{-0.034}$    & $^{+0.12}_{-0.14}$  & $^{+0.9}_{-0.7}$ \\
\small{Pile-up effects}                            & $^{+0.013}_{-0.010}$  & $<0.01$          & $^{+0.015}_{-0.019}$    & $^{+0.07}_{-0.04}$  & $^{+0.2}_{-0.4}$ \\
\small{Luminosity}                                & $\pm 0.001$         & $<0.01$          & $^{+0.002}_{-0.000}$      & $<0.01$          & $< 0.1$ \\
\small{MC statistical uncertainty}                & $\pm0.005$          & $<0.01$          & $\pm0.007$            & $\pm0.03$        & $\pm0.05$ \\
\midrule                                                                                                                               
Total systematics                         & $\pm 0.061$         & $^{+0.12}_{-0.13}$ & $^{+0.13}_{-0.14}$      & $^{+0.31}_{-0.41}$ & $^{+2.5}_{-1.7}$ \\
\midrule 
\bottomrule
\end{tabular}
\label{tab:fsm_syst_new}
\end{table}

%% file: tables/results_fsm_chi2_newsig3.tex
\begin{table}
\caption{Summary of extracted $f_{\textrm{SM}}$ values for each explored region with total uncertainties as well as the significance of the result with respect to the SM hypothesis. The significance with respect to the SM hypothesis is calculated using the statistical and systematic uncertainties on the data under a Gaussian assumption, together with the theory uncertainties on the hypothesis templates.  The latter include the effect of scale variations and PDF uncertainties, as well as the uncertainty arising from including top quark decays via \MADSPIN\ in the nominal \ttbar\ MC. The values in brackets exclude the effect of theoretical uncertainties on the hypothesis templates and only include the statistical and systematic uncertainties on the data.}
\vspace{3mm}
\setlength\extrarowheight{5pt}
\centering
\begin{tabular}{c c c}
\toprule
Region                        & $f_{\textrm{SM}} \pm \textrm{(stat.,syst.,theory)}$ & Significance (excl. theory) \\
\midrule
Inclusive                       & $1.249 \pm 0.024~\pm 0.061~^{+0.067}_{-0.090} $    &  2.2 (3.8) \\[1ex]
\midrule
$m_{\ttbar} < 450$ GeV          & $1.12  \pm 0.04~^{+0.12}_{-0.13}~^{+0.06}_{-0.07} $  & 0.78 (0.87) \\[0.5ex]
$450 \leq m_{\ttbar} < 550$ GeV & $1.18  \pm 0.08~^{+0.13}_{-0.14}~^{+0.13}_{-0.15} $  & 0.84 (1.1)  \\[0.5ex]
$550 \leq m_{\ttbar} < 800$ GeV & $1.65  \pm 0.19~^{+0.31}_{-0.41}~^{+0.26}_{-0.33} $  & 1.2  (1.4)  \\[0.5ex]
$m_{\ttbar} \geq 800$ GeV       & $2.2   \pm 0.9~^{+2.5}_{-1.7}~^{+1.2}_{-1.5}      $ & 0.49 (0.61) \\[1ex]

\bottomrule
\end{tabular}
\label{tab:results_fsm_chi2}
\end{table}

%% file: sections/results_susy.tex
\section{SUSY interpretation}
\label{sec:results_susy}

\begin{figure}[htb]
\vspace*{-8mm}
\centering         
\begin{tabular}{cc} 
\subfloat[\label{fig:susy_example_dphi_1}]
{\includegraphics[width=0.40\textwidth]{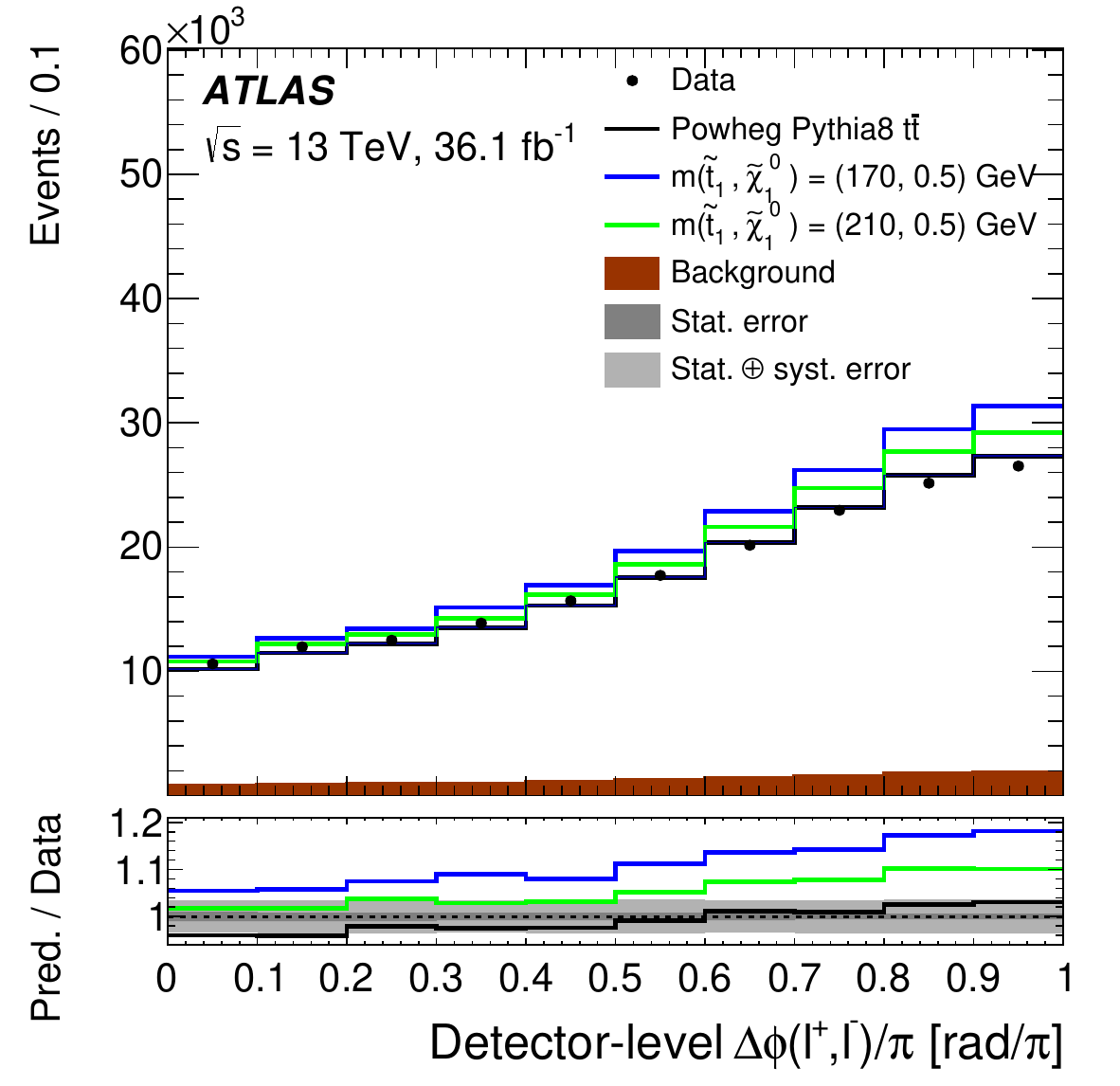}}&
\subfloat[\label{fig:susy_example_deta_1}]
{\includegraphics[width=0.40\textwidth]{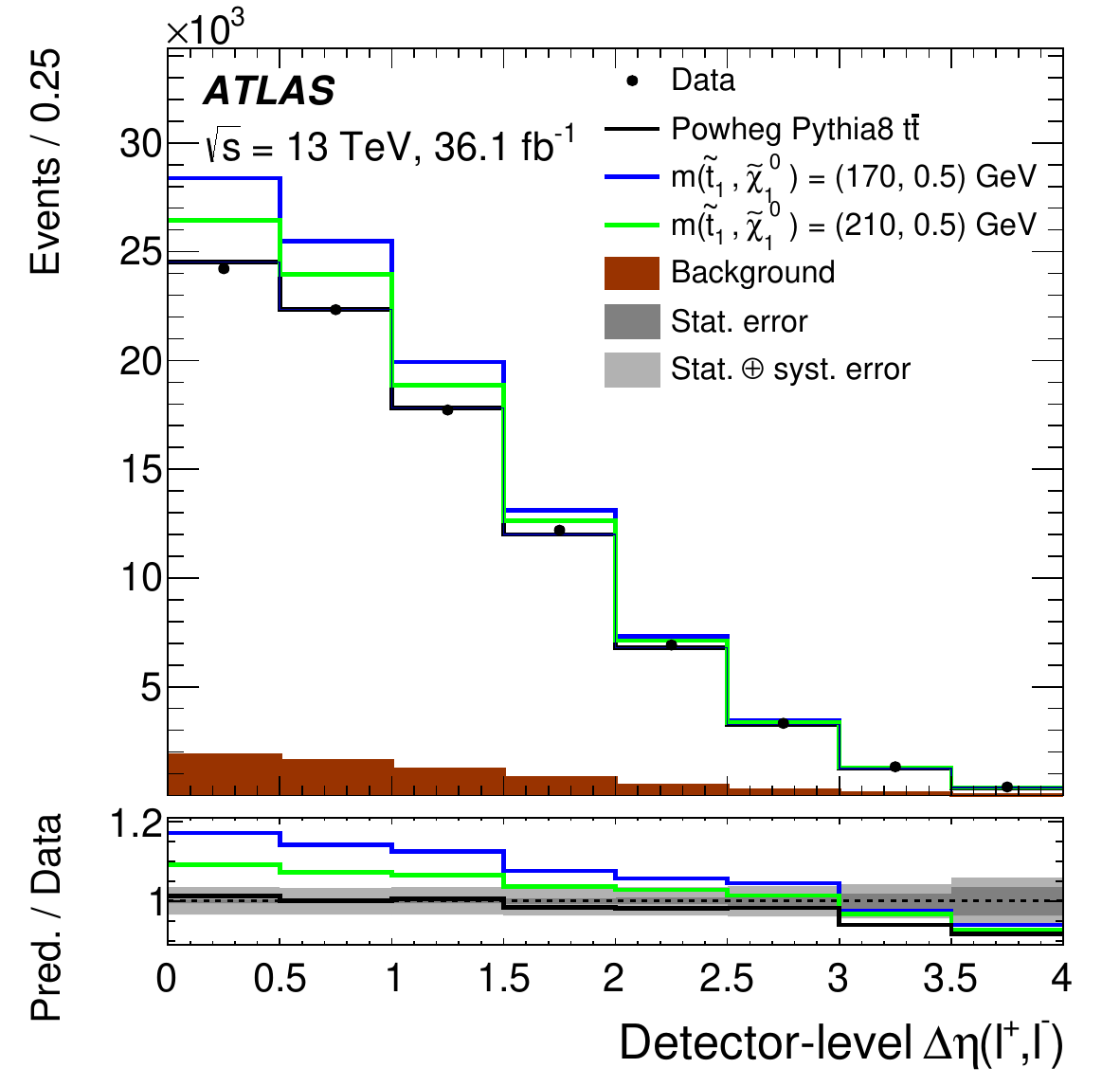}}\\
\subfloat[\label{fig:susy_example_dphi_deta_1}]
{\includegraphics[width=0.40\textwidth]{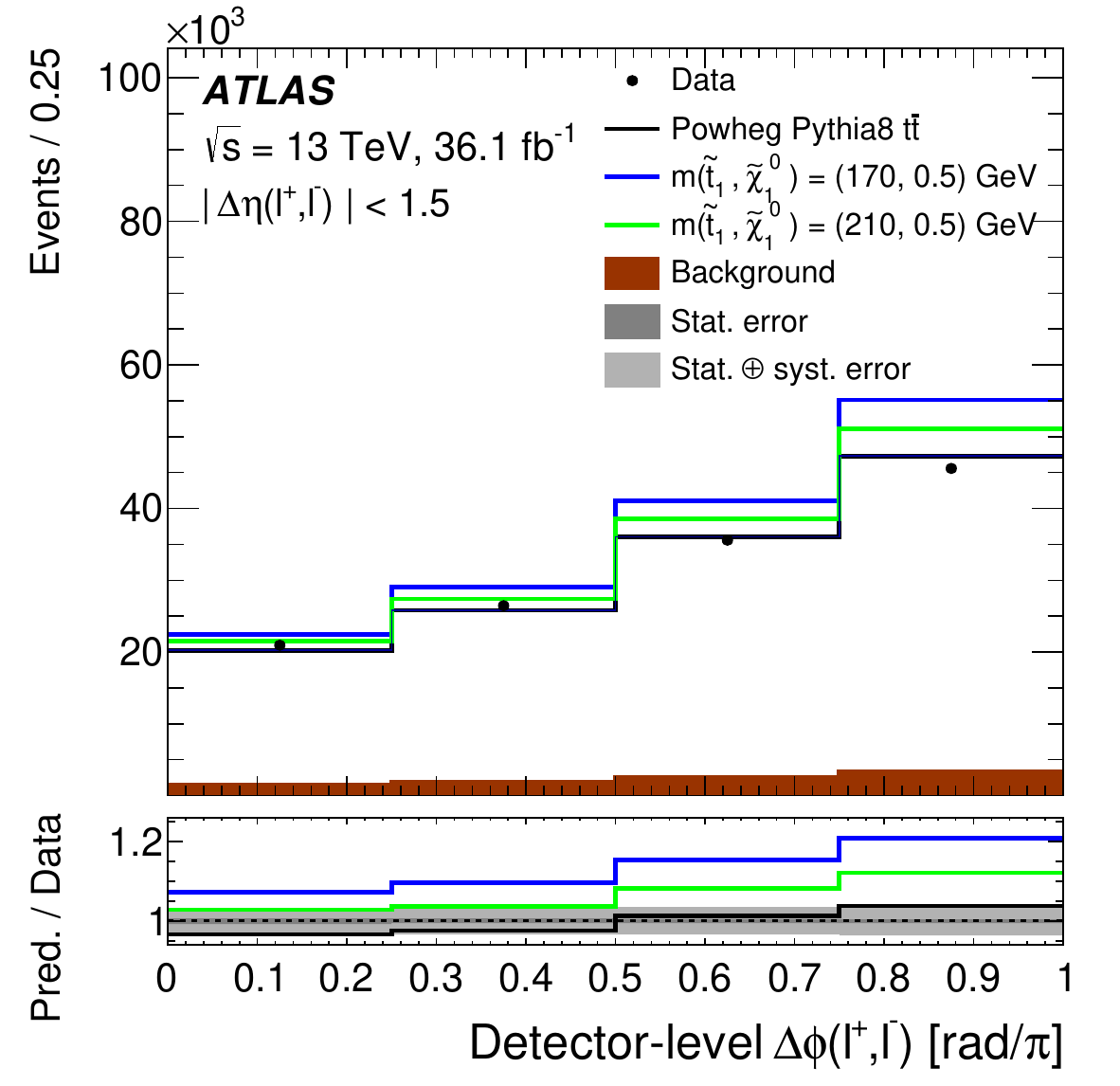}} &
\subfloat[\label{fig:susy_example_dphi_deta_2}]
{\includegraphics[width=0.40\textwidth]{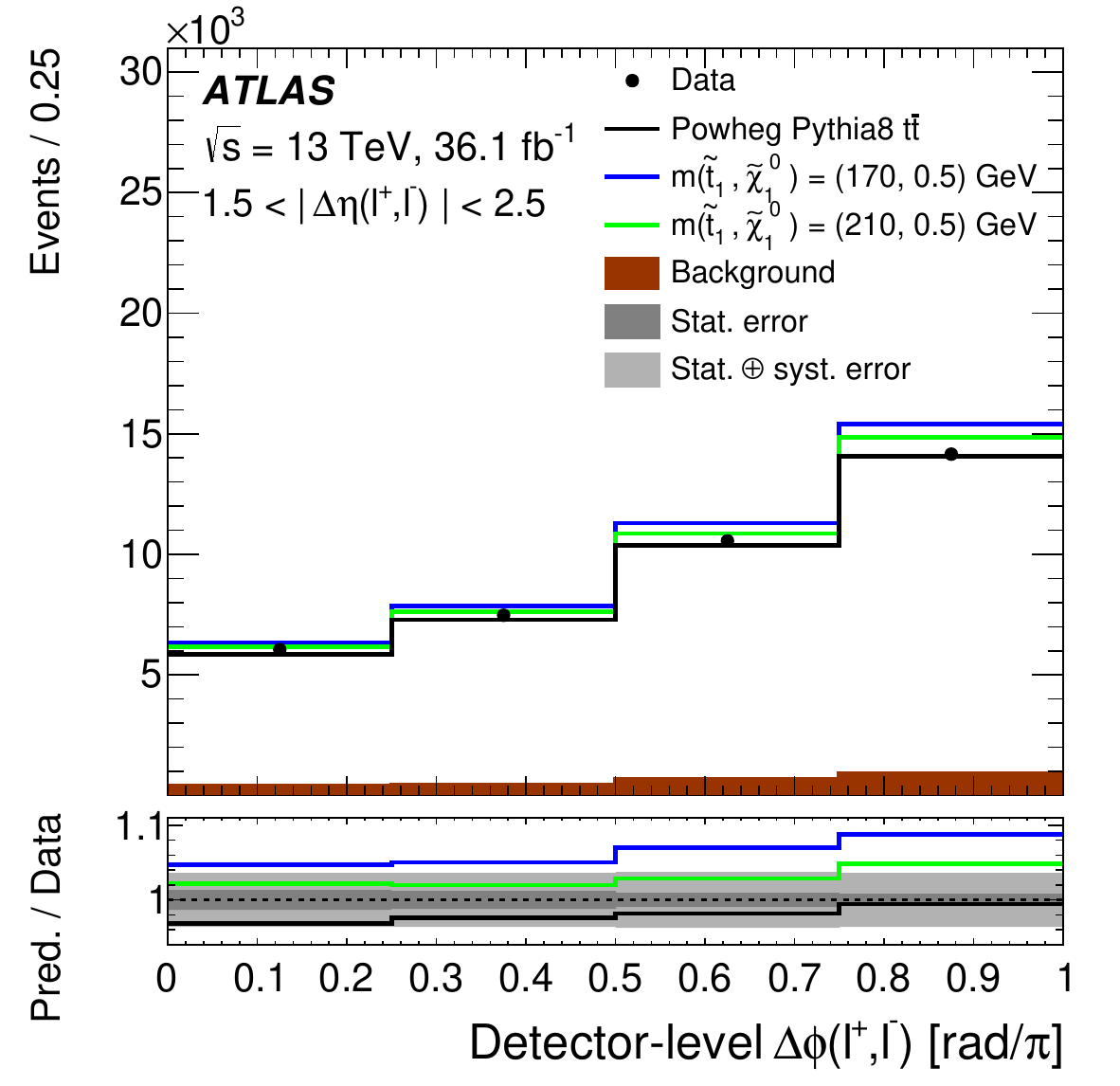}} \\
\subfloat[\label{fig:susy_example_dphi_deta_3}]
{\includegraphics[width=0.40\textwidth]{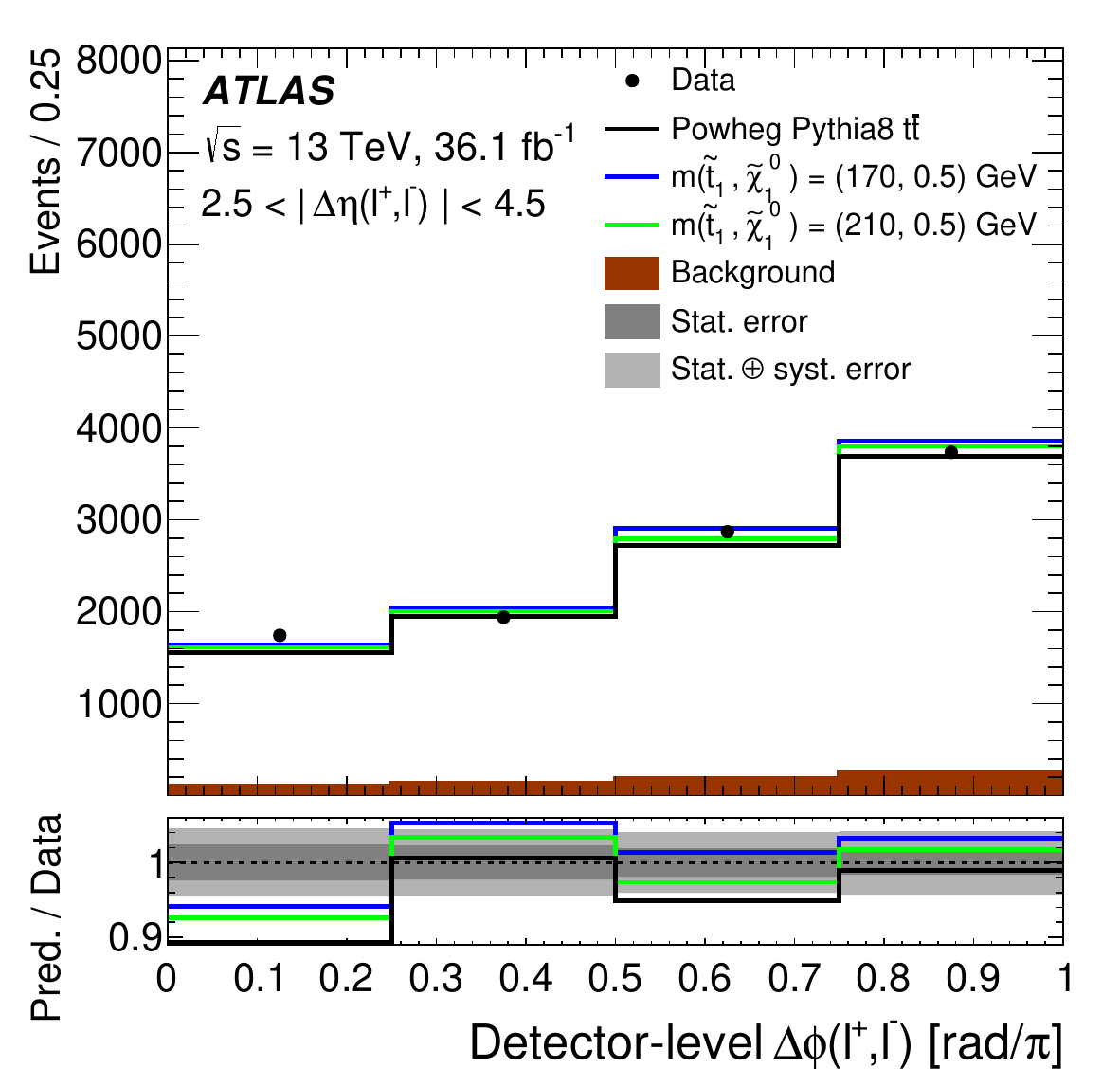}} & \\
\end{tabular}
\caption{The inclusive (a)  $\Delta\phi$ and (b)  $\Delta\eta$ distributions compared to the sum of the SM and SUSY predictions, for $m_{\tilde{t}_{1}} = 170$~GeV and $210$~GeV, and $m_{\tilde{\chi}^{0}_{1}} = 0.5$~GeV as well as the $\Delta\phi$ in regions of $\Delta\eta$: (c)  $|\Delta\eta  | <1.5$, (d) $1.5< |\Delta\eta | <2.5$, and (e) $2.5 <| \Delta\eta | <4.5 $. The dark uncertainty bands in the ratio plots represent the statistical uncertainties while the light uncertainty bands represent the statistical and systematic uncertainties added in quadrature. The systematic uncertainties include contributions from leptons, jets, missing transverse momentum, background modelling, pile-up modelling and luminosity, but not PDF or \ttbar modelling uncertainties.}
\label{fig:susy_example_reco}
\end{figure}

The detector-level $\Delta\phi$ and $\Delta\eta$ observables using the inclusive selection described in Section~\ref{sec:selection} are used to search for supersymmetric top squark pair production (\StSt) with $\tilde{t}_{1} \rightarrow t\tilde{\chi}^{0}_{1}$ decays. Naturalness arguments suggest that SUSY models with light top squarks may provide a solution to the hierarchy problem; however, a light top squark with a mass nearly degenerate with that of the top quark (so called ``stealth stops'') are challenging to detect using direct searches. It has been shown that leptonic spectra, such as $\Delta\eta$, can differentiate between \StSt and SM \ttbar production~\cite{Han:2012fw} and in previous searches ATLAS exploited differences between the expected spin correlations in the $\Delta\phi$ observable to set limits on \StSt production at a centre-of-mass energy of 8~TeV~\cite{TOPQ-2014-07}.  In this analysis the sensitivity of both of these observables is exploited simultaneously to maximise the sensitivity to stealth stop scenarios.  Double-differential distributions of $\Delta\phi$ in ranges of $\Delta\eta$ of $|\Delta\eta| < 1.5$, $1.5 < |\Delta\eta| < 2.5$, and $2.5 < |\Delta\eta| < 4.5$ are constructed as this was found to provide the optimal sensitivity to a wide variety of \StSt scenarios, where the majority of the expected sensitivity comes from the $|\Delta\eta|$ component. The MC samples used to simulate left-handed \StSt production are described in Section~\ref{sec:data}, and are generated at LO in perturbative QCD. Figure~\ref{fig:susy_example_reco} shows the effect on the expected $\Delta\phi$ and $\Delta\eta$ distributions individually, and on the double-differential distributions, from the inclusion of \StSt signal with $m_{\tilde{\chi}^{0}_{1}} = 0.5$~GeV and $m_{\tilde{t}_{1}} = 170$~GeV or $m_{\tilde{t}_{1}} = 210$~GeV compared to the data and SM \ttbar background.  In the $\Delta\phi$ distribution alone, $f_{\textrm{SM}}$ varies from approximately $0.85$ to $0.99$ between $m_{\tilde{t}_{1}} = 170$~GeV and $m_{\tilde{t}_{1}} = 240$~GeV.

Observed and expected limits are set on the \StSt production cross-section by simultaneously fitting the SM prediction to the observed data in the three double-differential distributions and varying the supersymmetric signal strength parameter $\mu$.  Limits are determined using a profile likelihood ratio in the asymptotic limit, using nuisance parameters to account for sources of systematic uncertainties. The limits are extracted at the $95\%$ confidence level (CL) using the CL$_\text{s}$  prescription~\cite{CLs}. The absolute cross-sections are used in the limits and the $t\bar{t}$ cross-section is set to its SM value (as described in Section~\ref{sec:data}) but allowed to vary as a nuisance parameter within the theoretical uncertainties. All experimental and modelling systematic uncertainties that are considered for the differential cross-section results and for the spin correlation measurement are also considered here as nuisance parameters in the fit. In addition, these sources of experimental uncertainty are also considered on the \StSt samples. In contrast to the $t\bar{t}$ prediction, SUSY modelling uncertainties such as the choice of factorisation and renormalisation scales, the merging and matching scale, and the \PYTHIA tune are found to produce a negligible contribution on the shape of the distributions.  The limits are found to be insensitive to these systematic uncertainties, which are therefore neglected. Finally, an additional uncertainty is included to account for the spread of predictions observed in Figure~\ref{fig:results_fsm_withpredictions}. The difference between the NLO background $t\bar{t}$ predictions from \POWHEG + \PYTHIA\!8 and the NLO QCD + Weak expanded predictions is taken, symmetrised, as a theoretical model uncertainty in order to account for the differences observed in the data compared to the prediction as discussed in Section~\ref{sec:results_spin}. This encompasses the range of current NLO predictions of the $t\bar{t}$ background process, particularly for the $\Delta\phi$ observable. The fit was found to be very sensitive to correlations between the additional radiation and MC generator uncertainties, therefore these uncertainties were split into individual nuisance parameters per bin to break the correlations. The largest sources of systematic uncertainty on the limits are the \ttbar cross-section uncertainty, the systematic uncertainties due to the choice of radiation settings and MC generator in the \ttbar simulation, and the theoretical model uncertainty on the $t\bar{t}$ background. Without this final systematic uncertainty, the observed limits would become stronger due to the poorer description of the background compared to the data and the difference between the expected and observed limits would also become stronger. The limits are also found to be stable when the $t\bar{t}$ prediction is reweighted to match the NLO QCD + Weak expanded prediction.

\begin{figure}
  \centering
  \includegraphics[width=0.6\textwidth]{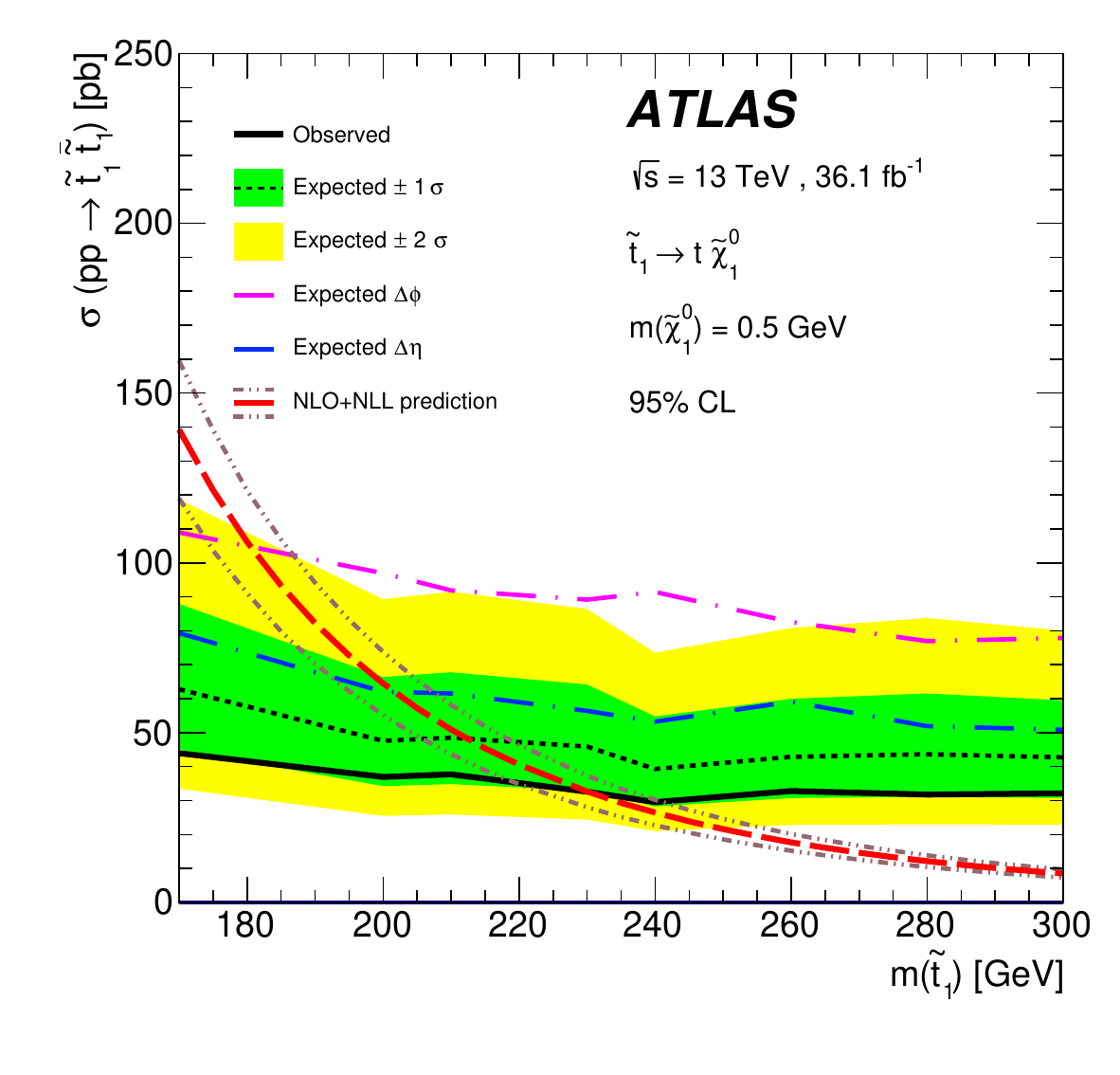}
  \caption{Expected and observed limits at $95\%$ CL on the top squark pair production cross-section as a function of $m_{\tilde{t}_{1}}$ assuming a $100\%$ branching ratio for $\tilde{t}_{1} \rightarrow t\tilde{\chi}^{0}_{1}$ decays with $m_{\tilde{\chi}^{0}_{1}} = 0.5$~GeV. The dashed line shows the expected limit with $\pm 1$ and $\pm 2$ standard deviation bands. The dashed line shows the theoretical cross-section with uncertainties. The solid line gives the observed limit. Also shown are the expected limits using the $\Delta\phi$ and $\Delta\eta$ distributions separately.}
  \label{fig:susy_dphi_1d_excl}
\end{figure}

\begin{figure}
  \centering
  \includegraphics[width=1.0\textwidth]{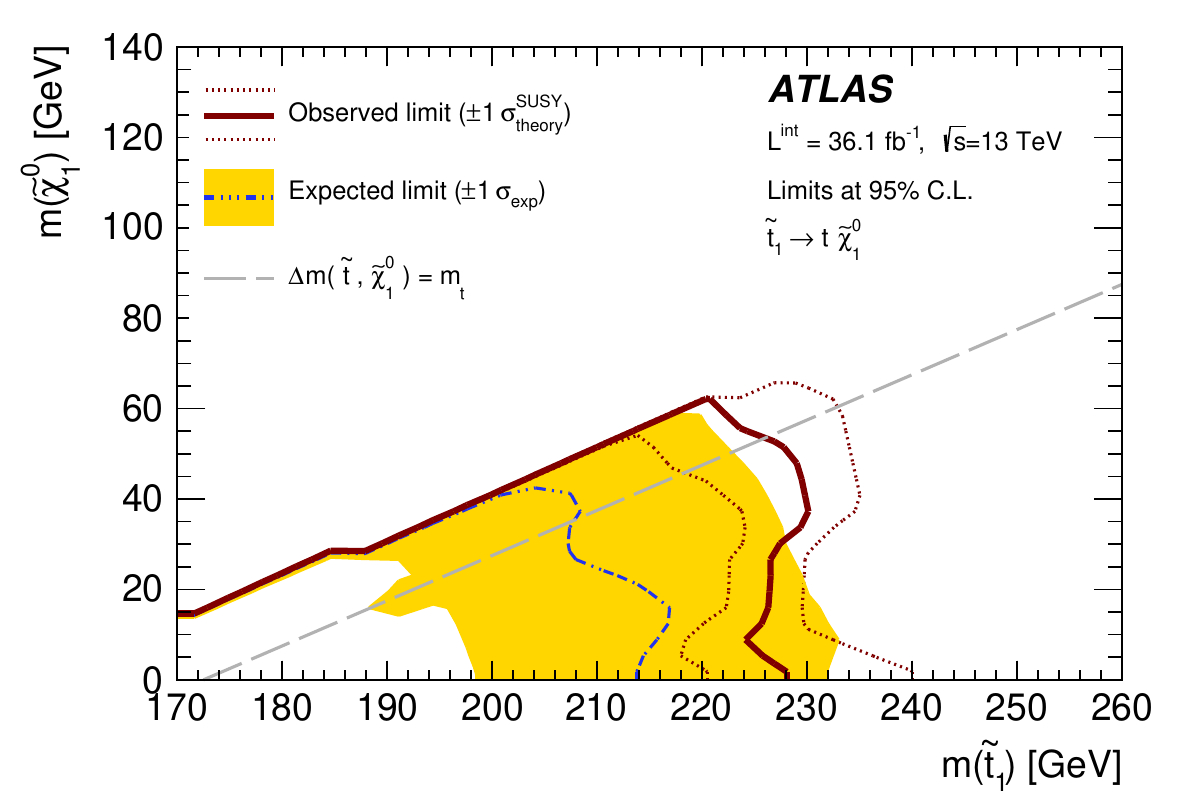}
\caption{Expected and observed limits at $95\%$ CL on the top squark pair production cross-section as a function of $m_{\tilde{\chi}^{0}_{1}}$ and $m_{\tilde{t}_{1}}$ assuming a $100\%$ branching ratio for $\tilde{t}_{1} \rightarrow t\tilde{\chi}^{0}_{1}$ decays. The dashed line shows the expected limit with $\pm 1$ standard deviation band.  The solid line shows the observed limit with the $\pm 1\sigma$ (dotted) SUSY cross-section theoretical uncertainties.}
  \label{fig:susy_dphi_2d_excl}
\end{figure}

SUSY production for a given $m_{\tilde{t}_{1}},m_{\tilde{\chi}^{0}_{1}}$ is considered to be excluded when the observed limit is below the expected SUSY cross-section, the theoretical uncertainty on which is 15\% (from PDF and scale uncertainties~\cite{Borschensky:2014cia}). For a neutralino mass $m_{\tilde{\chi}^{0}_{1}} = 0.5$~GeV, Figure~\ref{fig:susy_dphi_1d_excl} shows the observed (expected) limit, where top squarks with a mass between $170~(170)$~GeV and $230~(213)$~GeV are excluded with respect to the background generator prediction. Figure~\ref{fig:susy_dphi_2d_excl} shows the observed (expected) limit as functions of both $m_{\tilde{\chi}^{0}_{1}}$ and $m_{\tilde{t}_{1}}$ assuming the expected SUSY cross-sections. Observed (expected) limits are set on top squarks with masses between $170~(170)$~GeV and $230~(217)$~GeV for different values of $m_{\tilde{\chi}^{0}_{1}}$, and stop production with neutralinos with masses below $62~(42)$~GeV is excluded for different values of $m_{\tilde{t}_{1}}$. Figure~\ref{fig:susy_dphi_1d_excl} also shows the expected limits derived using only the single-differential distributions of $\Delta\phi$ and $\Delta\eta$. The limits on top squark production are dominated by the $\Delta\eta$ observable, and are relatively insensitive to the contribution of the $\Delta\phi$ distribution and its modelling at NLO.

If the $t\bar{t}$ cross-section normalisation were a free parameter of the fit and not constrained within its theory prediction uncertainty, the expected cross-section limit would increase, giving a lower mass limit of $m_{\tilde{t}_{1}} = 205$~GeV.  If, on the other hand, the shape information of $\Delta\phi$ and $\Delta\eta$ were not used in the fit, the expected cross-section limit would increase, giving a lower mass limit of just below $m_{\tilde{t}_{1}} = 170$~GeV.  This shows that the shape information drives the limit, in contrast with the same study performed at 8 TeV~\cite{TOPQ-2014-07}.

The presence of low-mass top squarks could cause a bias in the determination of the top quark mass; this would cause an underestimation of the top quark mass in direct measurements. In the presence of two-body top squark decays, the top production cross-section used in the fit would be too high, resulting in more stringent limits being set on top squarks near and above the top quark mass. The limits would be most affected for $m_{\tilde{t}_{1}} = m_{t}$ and the effect would disappear with increasing top squark mass. Equivalently, three-body top squark decays would affect limits near and below the top quark mass; however, as this analysis does not consider masses below $m_{\tilde{t}_{1}} = 170$~GeV this effect is small.  The magnitude of the expected bias on the top quark mass and the top quark pair production cross-section due to two-body top squark decays is estimated in Refs.~\cite{Eifert:2014kea, Cohen:2019ycc}.  For $m_{\tilde{\chi}^{0}_{1}} = 1$~GeV, the maximum bias is approximately $\Delta m_{t} = 2.0~(0.4)$~GeV for $m_{\tilde{t}_{1}} = 170~(200)$~GeV, resulting in a cross-section shift of 5.3\% (1.1\%). In addition, for $m_{\tilde{\chi}^{0}_{1}} = 20$~GeV, the maximum bias is approximately $\Delta m_{t} = 1.5~(0.5)$~GeV, resulting in a cross-section shift of 3.9\% (1.3\%).  The fits are performed again assuming these biases, interpolated for the region of $170 \leq m_{\tilde{t}_{1}} \leq 200$~GeV and $m_{\tilde{\chi}^{0}_{1}} \leq 20$~GeV.  This results in an increase in the observed top squark cross-section limit at $m_{\tilde{t}_{1}} = 170$~GeV of $8$~pb, and an extension of the $1\sigma$ error band in Figure~\ref{fig:susy_dphi_2d_excl} by $10$~GeV towards lower values of $m_{\tilde{t}_{1}}$.  However, neither the expected nor the observed upper limits set by the analysis are altered.

Some of the excluded phase-space has already been excluded by existing direct measurements, however, these results are more stringent than the existing limits~\cite{ATL-PHYS-PUB-2018-046} in a kinematically challenging region (where $m_{\tilde{t}} \sim m_{t}$) not currently excluded by direct searches~\cite{SUSY-2016-16,CMS-SUS-18-003}. The entire phase-space excluded by this analysis is shown for completeness.

\FloatBarrier

%% file: sections/conclusion.tex
\clearpage
\section{Conclusion}
\label{sec:conclusion}

Absolute and normalised differential cross-sections have been measured as a function of the azimuthal angle difference, $\Delta\phi$, and the pseudorapidity difference, $\Delta\eta$, between the two charged leptons in the $e\mu$ decay channel of top quark pairs using 36.1 fb$^{-1}$ of data recorded by the ATLAS detector in proton--proton collisions at $\sqrt{s} = 13$~\TeV\ during 2015 and 2016 at the LHC. The $\Delta\phi$ differential cross-section is also measured as a function of the \ttbar invariant mass. None of the studied generators are able to reproduce the normalised $\Delta\phi$ distribution within the experimental errors. A comparison was made with fixed-order predictions at NNLO in QCD and with the expansion at NLO in QCD and Weak couplings using a fixed scale choice. The former slightly improves the description of the data, while the latter describes the data but with large scale uncertainties. 

An extraction of spin correlation was performed using the normalised parton-level $\Delta\phi$ observable. The spin correlation was found to be higher than that predicted by the SM as implemented in NLO MC generators with a significance of 2.2 standard deviations. The measured value of spin correlation agrees well with the prediction by the expansion at NLO in QCD and Weak couplings, but is less consistent with NNLO predictions, with or without expansion in the normalised cross-section. The spin correlation was also found to increase slightly as a function of the invariant mass of the \ttbar system but no individual bin indicates a discrepancy above 1.2 standard deviations, due to the larger statistical and systematic uncertainties in these regions.  

A search for \StSt production was also performed using double-differential distributions of $\Delta\phi$ in ranges of $\Delta\eta$. In the absence of a SUSY signal in data, limits were set on top squark and neutralino production, taking into account the current limitations of the signal and background modelling. Top squarks with masses between 170 and 230 GeV are excluded for most kinematically allowed values of the neutralino mass, compared to expected limits of 170 and 217 GeV.

%% file: acknowledgements/Acknowledgements.tex


We thank CERN for the very successful operation of the LHC, as well as the
support staff from our institutions without whom ATLAS could not be
operated efficiently.

We acknowledge the support of ANPCyT, Argentina; YerPhI, Armenia; ARC, Australia; BMWFW and FWF, Austria; ANAS, Azerbaijan; SSTC, Belarus; CNPq and FAPESP, Brazil; NSERC, NRC and CFI, Canada; CERN; CONICYT, Chile; CAS, MOST and NSFC, China; COLCIENCIAS, Colombia; MSMT CR, MPO CR and VSC CR, Czech Republic; DNRF and DNSRC, Denmark; IN2P3-CNRS and CEA-DRF/IRFU, France; SRNSFG, Georgia; BMBF, HGF and MPG, Germany; GSRT, Greece; RGC and Hong Kong SAR, China; ISF and Benoziyo Center, Israel; INFN, Italy; MEXT and JSPS, Japan; CNRST, Morocco; NWO, Netherlands; RCN, Norway; MNiSW and NCN, Poland; FCT, Portugal; MNE/IFA, Romania; MES of Russia and NRC KI, Russia Federation; JINR; MESTD, Serbia; MSSR, Slovakia; ARRS and MIZ\v{S}, Slovenia; DST/NRF, South Africa; MINECO, Spain; SRC and Wallenberg Foundation, Sweden; SERI, SNSF and Cantons of Bern and Geneva, Switzerland; MOST, Taiwan; TAEK, Turkey; STFC, United Kingdom; DOE and NSF, United States of America. In addition, individual groups and members have received support from BCKDF, CANARIE, Compute Canada and CRC, Canada; ERC, ERDF, Horizon 2020, Marie Sk{\l}odowska-Curie Actions and COST, European Union; Investissements d'Avenir Labex, Investissements d'Avenir Idex and ANR, France; DFG and AvH Foundation, Germany; Herakleitos, Thales and Aristeia programmes co-financed by EU-ESF and the Greek NSRF, Greece; BSF-NSF and GIF, Israel; CERCA Programme Generalitat de Catalunya and PROMETEO Programme Generalitat Valenciana, Spain; G\"{o}ran Gustafssons Stiftelse, Sweden; The Royal Society and Leverhulme Trust, United Kingdom.

The crucial computing support from all WLCG partners is acknowledged gratefully, in particular from CERN, the ATLAS Tier-1 facilities at TRIUMF (Canada), NDGF (Denmark, Norway, Sweden), CC-IN2P3 (France), KIT/GridKA (Germany), INFN-CNAF (Italy), NL-T1 (Netherlands), PIC (Spain), ASGC (Taiwan), RAL (UK) and BNL (USA), the Tier-2 facilities worldwide and large non-WLCG resource providers. Major contributors of computing resources are listed in Ref.~\cite{ATL-SOFT-PUB-2020-001}.


%% file: atlas_authlist.tex
 
\begin{flushleft}
\hypersetup{urlcolor=black}
{\Large The ATLAS Collaboration}

\bigskip

\AtlasOrcid{M.~Aaboud}$^\textrm{\scriptsize 35d}$,    
\AtlasOrcid[0000-0002-6665-4934]{G.~Aad}$^\textrm{\scriptsize 102}$,    
\AtlasOrcid[0000-0002-5888-2734]{B.~Abbott}$^\textrm{\scriptsize 129}$,    
\AtlasOrcid{D.C.~Abbott}$^\textrm{\scriptsize 103}$,    
\AtlasOrcid{O.~Abdinov}$^\textrm{\scriptsize 13,*}$,    
\AtlasOrcid[0000-0002-2788-3822]{A.~Abed~Abud}$^\textrm{\scriptsize 71a,71b}$,    
\AtlasOrcid[0000-0002-2987-4006]{D.K.~Abhayasinghe}$^\textrm{\scriptsize 94}$,    
\AtlasOrcid[0000-0002-8496-9294]{S.H.~Abidi}$^\textrm{\scriptsize 167}$,    
\AtlasOrcid[0000-0002-8279-9324]{O.S.~AbouZeid}$^\textrm{\scriptsize 40}$,    
\AtlasOrcid{N.L.~Abraham}$^\textrm{\scriptsize 156}$
% ...
\bigskip
\\

% ...
$^{13}$Institute of Physics, Azerbaijan Academy of Sciences, Baku; Azerbaijan.\\
% ...
$^{35}$$^{(a)}$Facult\'e des Sciences Ain Chock, R\'eseau Universitaire de Physique des Hautes Energies - Universit\'e Hassan II, Casablanca;$^{(b)}$Facult\'{e} des Sciences, Universit\'{e} Ibn-Tofail, K\'{e}nitra;$^{(c)}$Facult\'e des Sciences Semlalia, Universit\'e Cadi Ayyad, LPHEA-Marrakech;$^{(d)}$Facult\'e des Sciences, Universit\'e Mohamed Premier and LPTPM, Oujda;$^{(e)}$Facult\'e des sciences, Universit\'e Mohammed V, Rabat; Morocco.\\
% ...
$^{40}$Niels Bohr Institute, University of Copenhagen, Copenhagen; Denmark.\\
% ...
$^{71}$$^{(a)}$INFN Sezione di Pavia;$^{(b)}$Dipartimento di Fisica, Universit\`a di Pavia, Pavia; Italy.\\
% ...
$^{94}$Department of Physics, Royal Holloway University of London, Egham; United Kingdom.\\
% ...
$^{102}$CPPM, Aix-Marseille Universit\'e, CNRS/IN2P3, Marseille; France.\\
$^{103}$Department of Physics, University of Massachusetts, Amherst MA; United States of America.\\
% ...
$^{129}$Homer L. Dodge Department of Physics and Astronomy, University of Oklahoma, Norman OK; United States of America.\\
% ...
$^{156}$Department of Physics and Astronomy, University of Sussex, Brighton; United Kingdom.\\
% ...
$^{167}$Department of Physics, University of Toronto, Toronto ON; Canada.\\
% ...
$^{a}$ Also at Borough of Manhattan Community College, City University of New York, New York NY; United States of America.\\
$^{b}$ Also at Centre for High Performance Computing, CSIR Campus, Rosebank, Cape Town; South Africa.\\
$^{c}$ Also at CERN, Geneva; Switzerland.\\
$^{d}$ Also at CPPM, Aix-Marseille Universit\'e, CNRS/IN2P3, Marseille; France.\\
$^{e}$ Also at D\'epartement de Physique Nucl\'eaire et Corpusculaire, Universit\'e de Gen\`eve, Gen\`eve; Switzerland.\\
% ...
$^{*}$ Deceased

\end{flushleft}
